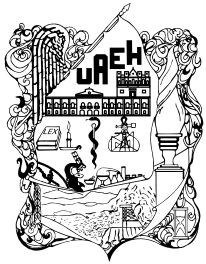

# Universidad Autónoma del Estado de Hidalgo

## Instituto de Ciencias Básicas e Ingeniería

### Área Académica de Matemáticas y Física

---

# Soluciones exactas para la interacción de materiales de Dirac anisótropos con campos eléctricos y magnéticos

---

T E S I S

Que para obtener el título de

Maestro en Física y Tecnología Avanzada

P R E S E N T A

Julio Armando Mojica Zárate

Directores

Dr. Erik Díaz Bautista
Dr. Omar Pedraza Ortega

Pachuca de Soto, Septiembre de 2025

# Índice general

# Agradecimientos

A toda mi familia, incluidos mis hermanos mayores Omar, Luis y César, y mis hermanos menores Rubí, Esme y Alex, sepan que siempre voy a apoyarlos en lo que necesiten. A mis padres, Armando Mojica Ruiz y Adelita Zárate Castro, gracias por su apoyo incondicional en todos los sentidos; no podría estar más feliz con los padres que tengo. A mis tíos, que son como mis otros padres, Domingo Mojica Ovando y Mariana Guevara Acosta, gracias por su cuidado y apoyo como si fuera un hijo más. A mis tíos Irais Castro y Marcelo, por su ayuda y por hacer mi estancia en Veracruz más amena. A mis tíos Cecilia Guevara, Eluterio, y mi padrino Carlos, gracias por apoyarme y darme su cariño como familia.

Quiero agradecer también a mis "pseudohermanas" Mitzi y Lupita, quienes me dieron consejos y alegraron mi vida durante este trayecto. A mis sobrinos, en particular a Abril, siempre contará con mi apoyo; se les quiere y aprecia a todos.

A mis abuelos, que en paz descansen, todo lo que soy se lo debo a ellos: mi forma de ser y mi filosofía de vida. Dondequiera que estén, sé que me cuidan y se alegran por mis metas alcanzadas.

A mis compañeros de maestría, en especial a mi hermana Andrea, gracias por estar ahí apoyándome y por compartir penas y glorias en nuestra vida académica y personal.

A mis amigos Alfredo, Joe, Rafa, Harold, Fernanda, Arturo, Aurora y Andi Montiel, gracias por apoyarme incondicionalmente y soportarme en esta etapa. Se les tiene un gran cariño y aprecio. Aunque ahora viene el doctorado, no sé si estaré peor o mejor, pero sé que seguirán conmigo.

A mis medicas Jera, Melisa y Jesi, gracias por su amistad y por esas charlas con café que hicieron más amena mi última etapa en la maestría.

A todos los doctores del área académica de matemáticas y física, en particular:

- Dra. Victoria Cerón, por su apoyo emocional y por motivarme a ser mejor persona y seguir adelante en el doctorado.

- Dr. Luis, por su clase y tutoría, y por las palabras que me ayudaron a ser mejor persona e investigador.

- Dr. Roberto Noriega Papaqui, por darme un espacio de trabajo en mis primeros semestres, por su amistad, consejos y por transmitirme su pasión por la física, de igual manera a su esposa la lic. Gabi, por sus consejos y amistad, y a su hijo Jequi, que me dio varias sonrisas y me recordó mi niñez.

- Dr. Pedro Miranda, por su apoyo en clase, su amistad y consejos.

- Mtra. Andrea, por su amistad y respaldo durante mi proceso de maestría.

- Dr. Mario Pérez, amigo y hermano, por estar siempre ahí para mí y apoyarme en todos los sentidos.

- Dr. Francisco, otro hermano, por sus consejos, su filosofía de la física y por escucharme cuando lo necesitaba.

- Dr. Lao, por sus clases, su visión distinta de la física y por su amistad.

- Dr. Donado, un amigo y un tío para mí, por su apoyo en los momentos difíciles, su consejo y por transmitirme la ilusión y alegría por el experimento.

- Dr. Gerardo, Dr. Adolfo y a la Dra. Alba, por su amistad y consejos que enriquecieron mi formación.

- Dr. Roberto Ávila Pozos, por su apoyo en todos los sentidos durante mi proceso y estancia en la maestría, ademas de su amistad desde estaba en la licenciatura.

- Dr. Selim Gómez Ávila, por su apoyo en mi formación y por estar siempre al tanto de mi trayectoria.

- Dr. Criollo, por su amistad, sus consejos y por escucharme en la etapa final de la maestría.

- Dr. Benjamín y Dra. Erika, por sus consejos y amistad.

- Mtro. Roger, por su apoyo emocional, académico y personal durante toda la maestría, pero sobre todo por su amistad.

A mis sinodales, muchos de los cuales ya he mencionado, gracias por apoyarme en su revisión y por su paciencia conmigo.
A mis asesores:

- Dr. Omar Pedraza Ortega, quien siempre será como un padre para mí. Agradezco mucho conocerlo; siempre será el mejor asesor y guía que pude haber escogido. Valoro y aprecio todas sus enseñanzas, su tiempo y paciencia, que me ayudaron a formarme y a ser mejor persona e investigador.

- Dr. Erik Díaz, mi otro padre académico, por enseñarme una manera diferente de ver la física y el mundo, y por integrarme al grupo de QM, lo que amplió enormemente mi visión. Gracias por su tiempo, dedicación, consejos y por el gran ser humano que es.

AL Dr. Daniel Ortiz Campa, por ayudarme a iniciar mi etapa como investigador, por su apoyo constante y por su amistad.

De igual forma, agradezco a todas las personas y amigos que estuvieron a lo largo de todo mi trayecto.

Finalmente, agradezco a mí mismo por estos dos años de perseverancia y sacrificio.

# 1 | Introducción

El carbono es la base de la vida tal como la conocemos, gracias a sus propiedades químicas únicas que permiten formar diferentes estructuras llamadas alótropos. Entre ellos, el diamante es conocido por ser el material más duro que existe. Otro alótropo destacado es el grafeno, el cual puede ser considerado el material más delgado jamás descubierto y que fue aislado por primera vez en 2004 por Andre Geim y Konstantin Novoselov [Novoselov et al., 2004]. Sin embargo, el término *grafeno* fue introducido previamente por Hans-Peter Boehm, y comenzó a cobrar relevancia en la literatura científica a finales del siglo XX, impulsado por los estudios de Saito, Wang y otros investigadores [Saito et al., 1998, Wang et al., 2000]. La descripción teórica del grafeno fue abordada inicialmente por Wallace en 1947 [Wallace, 1947]. Sin embargo, la teoría que describe este material en términos de una ecuación de Dirac sin masa fue desarrollada de forma independiente por Semenoff, y por DiVincenzo y Mele en 1984 [DiVincenzo and Mele, 1984, Semenoff, 1984].

El grafeno es, quizá, el material bidimensional (2D) más conocido [Castro Neto et al., 2009]. Se trata de un material compuesto por átomos de carbono dispuestos en una red con estructura de panal de abeja. Cerca del nivel de Fermi, sus propiedades pueden describirse mediante la ecuación de Dirac-Weyl (2+1)D sin masa. Por esta razón, varios fenómenos de la mecánica cuántica relativista pueden observarse en el grafeno, como el tunelamiento de Klein y el Zitterbewegung [Katsnelson et al., 2006, Kamal and Jellal, 2021, Martinez et al., 2010]. Dado que la celda unitaria de la red de grafeno está formada por dos átomos de carbono, la función de onda asociada a los electrones en ella está descrita por espinores de dos componentes. Este grado de libertad se conoce como pseudoespín (o grado de libertad de subred), ya que se comporta de manera análoga al espín real en el espacio recíproco. Además, gracias a las simetrías presentes en la red hexagonal, en particular la simetría rotacional de orden 6 ($C_6$) y la equivalencia entre las subredes triangulares A y B en que puede dividirse la red cristalina del grafeno, la relación de dispersión es lineal alrededor del nivel de Fermi. Esto da lugar a estructuras energéticas en las bandas de valencia y de conducción, conocidas como conos de Dirac, que se tocan en un único punto, lo cual hace que el grafeno sea considerado un semimetal. La presencia de estos conos de Dirac en el régimen energético de bajas energías alrededor del nivel de Fermi origina un grado de libertad adicional conocido como *índice de valle*, el cual abre la puerta a aplicaciones en la nanoelectrónica para codificar y manipular información utilizando los llamados estados de valle, con potencial en transistores avanzados, sensores y dispositivos cuánticos.

Si bien el enfoque en el grafeno se debe a sus propiedades únicas, entre ellas un transporte electrónico eficiente, una alta conductividad térmica y propiedades mecánicas notables, en los últimos años también ha crecido el interés por encontrar otros sistemas donde los portadores de carga se comporten como fermiones de Dirac sin masa [Wehling et al., 2014]. Estos materiales, conocidos como *materiales de Dirac*, presentan propiedades universales como una elevada capacidad calorífica, una alta conductividad óptica y una notable susceptibilidad magnética, entre otras. La linealidad de su relación de dispersión a bajas energías es consecuencia directa de la preservación de simetrías fundamentales en el sistema, como la inversión temporal y la inversión espacial. Esta característica confiere a los materiales de Dirac una identidad física única, diferenciándolos claramente de los semiconductores y metales convencionales. Entre ellos, el grafeno, el borofeno y los aislantes topológicos, destacan por sus inusuales características electrónicas, derivadas del comportamiento relativista de sus portadores de carga. Estas peculiaridades los convierten en candidatos ideales para aplicaciones avanzadas en nanoelectrónica y tecnologías cuánticas. Adicionalmente, el estudio de los materiales de Dirac anisótropos se ha consolidado como un tema emergente en la física del estado sólido. Un aspecto particularmente relevante es el análisis de su respuesta óptica, ya que permite entender cómo estos materiales interactúan con campos electromagnéticos. Desde esta perspectiva, se han investigado propiedades como la conductividad óptica, el peso de Drude y la conductividad Hall en sistemas específicos, como el boro $8 - Pmmn$ y el grafeno tipo quinoide (caracterizado por la separación de sus conos de Dirac), revelando efectos únicos derivados de su anisotropía [Mojarro et al., 2021]. Existen también trabajos en los que se estudian los efectos combinados de campos magnéticos y electrostáticos en monocapas de borofeno $8B-$ y $2BH - Pmmn$ [Díaz-Bautista, 2022].

Por otro lado, el estudio de soluciones exactas de la ecuación de Dirac-Weyl (2+1)D en campos inhomogéneos ha atraído mucha atención debido a sus posibles aplicaciones en la monocapa y la bicapa de grafeno, pues permiten comprender las propiedades electrónicas de cintas, nanocables y nanotubos. En los últimos años, se han realizado varios estudios sobre la interacción de los electrones en grafeno con campos magnéticos perpendiculares al plano de la muestra, considerando en algunos casos la presencia adicional de campos electrostáticos paralelos a la superficie del material. El objetivo principal de estas investigaciones ha sido encontrar mecanismos para confinar las cargas en el material. Dado que la obtención de niveles de energía para estados confinados o los coeficientes de transmisión de estados dispersivos suele requerir cálculos complejos, en muchos de estos estudios se han utilizado métodos numéricos para resolver el problema. Uno de los primeros trabajos donde se estudiaron las soluciones exactas con interacción de campos magnéticos fue el trabajo de Kuru, Negro y Nieto [Ş. Kuru et al., 2009]. En él se explora la interacción de electrones en grafeno con campos magnéticos perpendiculares a la superficie del material, enfocándose en soluciones analíticas exactas de la ecuación de Dirac-Weyl.

Siguiendo esta línea de pensamiento, este trabajo de tesis se enfocará en obtener soluciones exactas de la ecuación de Dirac-Weyl que describa a los electrones en presencia de campos externos inhomogéneos, con el fin de estudiar el comportamiento electrónico en materiales de Dirac anisótropos bajo condiciones de confinamiento. Este enfoque resulta particularmente relevante en sistemas como el grafeno, el borofeno y otros materiales bidimensionales, donde los portadores de carga se comportan como fermiones de Dirac sin masa, y las condiciones de frontera, la geometría del sistema y el perfil del campo aplicado influyen fuertemente en las propiedades electrónicas.

Una de las motivaciones principales de este estudio es comprender cómo el transporte electrónico se ve afectado cuando se modifica el entorno efectivo mediante diferentes condiciones de confinamiento espacial. Para ello, inicialmente se analizarán configuraciones en las que los campos externos presenten perfiles dependientes de la coordenada espacial $x$ y tengan una extensión sobre toda la superficie del material de Dirac, que se asume infinitamente extendido. Seguido de lo anterior, los perfiles de los campos externos considerados se modificarán mediante la introducción de un parámetro identificado como $z_0$, el cual permitirá controlar explícitamente el dominio físico de las interacciones. Este enfoque permitirá evaluar con precisión el impacto de estos *perfiles ajustables* sobre las soluciones de la ecuación de Dirac-Weyl, es decir, tanto en el espectro de energía como en las densidades de probabilidad y de corriente, que constituyen herramientas fundamentales para describir el transporte electrónico. Adicionalmente, este tipo de análisis proporcionará herramientas teóricas valiosas para el diseño de dispositivos basados en materiales de Dirac anisótropos para futuras aplicaciones en nanoelectrónica y tecnologías cuánticas.

En este trabajo se utilizan dos perfiles de campo para describir las interacciones electrónicas de forma simple. El primero es un perfil *singular*, que concentra la inter-
acción en puntos muy localizados de la red y refleja fenómenos de corto alcance donde
los electrones sienten una atracción o repulsión muy intensa en un sitio específico.
El segundo es un perfil de *decaimiento exponencial*, en el que la intensidad de la
interacción disminuye progresivamente al alejarse, representando el hecho de que
la influencia de un electrón no se extiende indefinidamente sino que se atenúa con
la distancia. Cada perfil se analiza de manera independiente, permitiendo explorar
diferentes regímenes de interacción. Esta aproximación permite capturar de manera
más fiel las propiedades electrónicas de materiales de Dirac, lo que resulta esencial
para comprender fenómenos de transporte cuántico con aplicaciones potenciales en
el diseño de dispositivos $2D$ como transistores de alto desempeño, sensores, entre otras.

Por todo lo anterior, el contenido de esta tesis se organiza de la siguiente manera:

El Capítulo 2 se centra en la descripción general de los materiales de Dirac, con
énfasis en los sistemas bidimensionales como el grafeno. Se detalla la construcción del
Hamiltoniano efectivo tipo Dirac mediante el modelo de amarre fuerte (*tight-binding*),
haciendo uso de las matrices de solapamiento y transferencia. Se profundiza en los
llamados puntos y conos de Dirac, fundamentales para describir la dispersión lineal de
los portadores de carga en estos materiales. Además, se discute la existencia de otros
materiales bidimensionales similares al grafeno, para los cuales también es posible
construir modelos efectivos análogos.

En el Capítulo 3, se retoma el Hamiltoniano que describe materiales de Dirac aniso-
trópos y se consideran interacciones externas de tipo magnético y electrostático. El
sistema se estudia bajo la influencia de campos eléctricos y magnéticos que dependen
de la posición, considerando perfiles con decaimiento exponencial y con singularidades
en el origen. Se resuelve la ecuación de Dirac correspondiente a cada tipo de perfil,
obteniendo el espectro de energía, las eigenfunciones del sistema, y se analizan las
distribuciones de densidad de probabilidad y densidad de corriente asociadas.

El Capítulo 4 introduce el método de iteración asintótica, primero desde un enfoque
general y luego bajo un tratamiento perturbativo. Este método se aplica a los modelos
estudiados en el Capítulo 3, pero ahora modificando las condiciones de frontera: el
dominio espacial del sistema se restringe al intervalo $(-\infty, x_0]$ o $(0, x_0]$, según sea el
caso. A través del método perturbativo de iteración asintótica, se obtiene el espectro
de energía a primer orden y se determinan las eigenfunciones correspondientes a los
primeros estados cuánticos ($n = 0$ y $n = 1$). Finalmente, se presentan las gráficas del espectro de energía, así como las distribuciones de densidad de probabilidad y de corriente para cada caso.

En el Capítulo 5, se presentan las conclusiones del trabajo.

# 2 | Materiales de Dirac

Los materiales de Dirac comprenden a todos los sistemas de la física de la materia condensada en los que la energía de sus portadores de carga sigue una relación lineal respecto al momentum, por lo que pueden describirse mediante una ecuación de Dirac efectiva en ciertos régimenes energéticos. El estudio de los materiales de Dirac ha crecido exponencialmente en los últimos años [Wehling et al., 2014, Wang et al., 2015] debido a sus potenciales aplicaciones en la tecnología y en la ciencia básica. Dicho crecimiento, se refleja en el hallazgo reciente de diversos sistemas que presentan este comportamiento, desde fluidos cuánticos exóticos hasta sólidos cristalinos [Wehling et al., 2014].

Algunos materiales de Dirac comparten la misma estructura cristalina, como se muestra en la Tabla 1 de la Referencia [Wehling et al., 2014]. Otros, como los llamados aislantes topológicos (TIs), presentan una característica dual: en su volumen (o *bulk*) exhiben una brecha energética como aislantes convencionales, pero en sus superficies o bordes emergen estados electrónicos protegidos topológicamente[1] con dispersión tipo Dirac. Por otro lado, si se toma en cuenta la simetría quiral del sistema[2], la ecuación que describe a los materiales de Dirac tridimensionales adopta la forma de la ecuación de Weyl; tales materiales se conocen en la literatura como semimetales de Weyl (WSMs). Mientras tanto, se les denomina semimetales de Dirac tridimensionales (DSMs) a aquellos en los que se mantiene la degeneración de bandas en el *punto de*

---

[1]Dentro del contexto de materia condensada, un estado de la materia cuyas propiedades persisten a pesar de estar sometido a ciertas perturbaciones o imperfecciones del material se dice que está topológicamente protegido. Dicho de otra manera, las propiedades del material no son sensibles a modificaciones locales en la estructura del material, a menos que la estructura global de este se modifique a un estado topológicamente diferente.

[2]La simetría quiral (también llamada de subred) implica la existencia de un operador $\mathcal{S}$ tal que $\{\mathcal{S}, H(\mathbf{k})\} = 0$. En redes bipartitas, como el grafeno, $\mathcal{S}$ intercambia las subredes A y B (ver Figura 2.1.4). Esta anticonmutación fuerza al espectro a ser simétrico alrededor de $E = 0$ y protege estados de energía nula frente a perturbaciones que no rompan dicha simetría. La simetría de red en el grafeno se expresa como $\sigma_z H(\mathbf{k})\sigma_z = -H(\mathbf{k})$.

*Dirac* sin que se abra una brecha de energía.

En síntesis, los materiales de Dirac exhiben una notable diversidad estructural, manifestándose en sistemas de dimensiones variables. A pesar de estas diferencias y sus distintas composiciones químicas, todos comparten una característica fundamental: sus espectros de energía a bajas energías siguen un mismo patrón, lo que sugiere que hay un principio fundamental que los organiza. Este principio está relacionado con las simetrías del sistema, ya que son ellas las que determinan si los puntos de Dirac pueden formarse o desaparecer, como la simetría de inversión temporal[3] en los aislantes topológicos o la simetría de subred en el grafeno. En general, la presencia de puntos de Dirac hace que el número de estados accesibles para las excitaciones de energía cero se reduzca drásticamente, y esta reducción está gobernada por las simetrías del sistema en el régimen de baja energía.

## 2.1 | **Una descripción breve del grafeno**

El carbono es el sexto elemento de la tabla periódica y tiene dos isótopos estables, $^{12}C$ y $^{13}C$, con diferentes propiedades nucleares. En su estado fundamental, el átomo de carbono tiene una configuración electrónica $2s^2 2p^2$ y un momento angular total $J = 0$. Su primer estado excitado, $^3P_1$, muestra un multiplete $J = 1$ con una energía de excitación igual a 2 meV, aproximadamente [Radzig and Smirnov, 1985]. La formación de enlaces covalentes, los cuales son fundamentales para la estabilidad molecular, maximiza la superposición de funciones de onda entre átomos, influenciada por la disposición espacial de los átomos vecinos. Esta maximización se busca especialmente en direcciones específicas en relación con el átomo central.

Como resultado de lo anterior, el carbono exhibe diversas formas estructurales conocidas como alótropos. Entre estos, un alótropo particularmente importante es el diamante, el cual posee una estructura tetraédrica (ver Figura 2.1.1b). Por su parte, el grafito presenta orbitales dispuestos de manera planar entre capas con acoplamientos débiles (ver Figura 2.1.1c).

La obtención de estos materiales a partir del carbono depende de la distribución de sus electrones y su estructura, lo que involucra procesos de hibridización. La hibridización implica cambiar la base orbital, es decir, realizar combinaciones lineales de estos orbitales. La configuración electrónica del carbono en la base de los orbitales es $1s^2\,2s^2\,2p^2$, mientras que, como se aprecia en la Figura 2.1.1, los orbitales híbridos presentan una estructura tetraédrica y planar.

---

[3]En el caso del grafeno, la simetría de inversión temporal, dentro del modelo de amarre fuerte y en ausencia el grado de libertad de espin, se escribe como $H(\mathbf{k}) = H^*(-\mathbf{k})$ en el espacio de momentos. Adicionalmente, la simetría de inversión temporal envía el valle $K_+$ a $K_-$ y, por lo tanto, intercambia los dos conos de Dirac (ver Figura 2.1.4).

La familia $sp^2$ basada en carbono muestra una amplia variedad de alotropías, desde fullerenos y nanotubos hasta la monocapa del grafeno y multicapas apiladas. En particular, la monocapa de grafeno es la unidad fundamental de la cual se derivan las demás formas: las cintas de grafeno (cuasi-unidimensionales), los nanotubos y el grafito. Este material fue descubierto en 2004 [Novoselov et al., 2004] y presenta una estructura de una sola capa de átomos de carbono dispuestos en una red hexagonal. Esta disposición le confiere propiedades electrónicas únicas, como una excepcional conductividad eléctrica y térmica, así como una notable resistencia mecánica [Castro Neto et al., 2009].

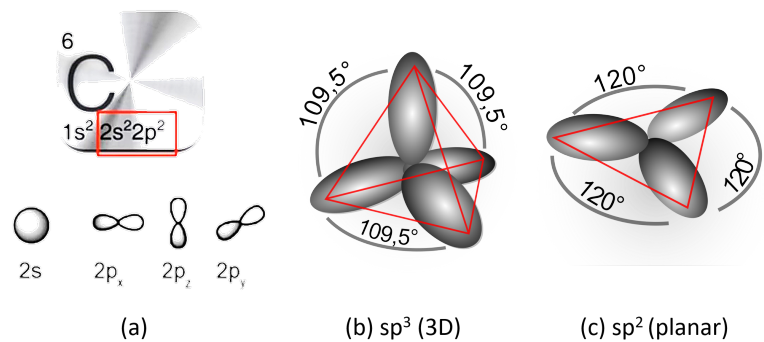

**Figura 2.1.1:** (a) Configuración electrónica del carbono. (b) Hibridaciones $sp^3$. (c) Hibridación $sp^2$. Imagen adaptada de [De Sousa Ferreira, 2020].

### Estados $\pi$ en el grafeno

La resistencia mecánica del grafeno se atribuye a los enlaces $\sigma$, mientras que sus notables propiedades eléctricas son resultado de los enlaces $\pi$ (ver Figura 2.1.2). Las interacciones $\pi$ entre átomos de carbono refuerzan su robustez mecánica, proporcionando una estructura sorprendentemente resistente. Por otro lado, la hibridización de orbitales favorece la movilidad de los electrones, lo que confiere al grafeno una buena conductividad eléctrica.

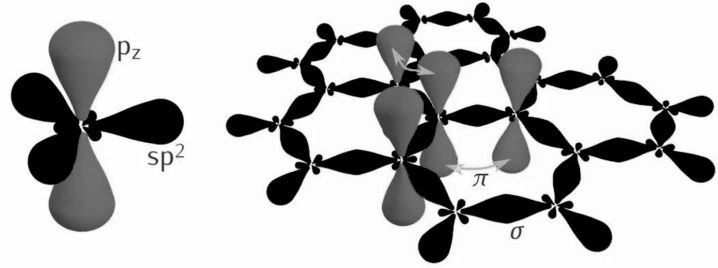

**Figura 2.1.2:** La estructura hexagonal de carbono en el grafeno y el átomo de carbono con hibridación $sp^2$. Los orbitales $p_z$ se hibridan, permitiendo la deslocalización de electrones por encima y por debajo del plano de los átomos de carbono. Estos orbitales forman las llamadas bandas $\pi$ y $\pi^*$. Imagen adaptada de [De Paoli, 2023].

El grafeno no solo tiene aplicaciones prácticas en la industria, sino que también sirve como un campo de estudio para comprender conceptos fundamentales como la fase de Berry [Zhang et al., 2005], los modos cero topológicamente protegidos [Nair et al., 2008], el tunelamiento de Klein [Katsnelson et al., 2006] y la reconstrucción en vacío [Zhang et al., 2009]. Las propiedades electrónicas básicas del grafeno, como la simetría electrón-hueco y la dispersión lineal de banda cerca de los vértices de la zona de Brillouin (punto de Dirac), se conocen desde el trabajo seminal de Wallace en 1947 [Wallace, 1947]. Sin embargo, el interés en las propiedades de materiales basados en grafeno se reavivó después del descubrimiento de los nanotubos de carbono por Iijima en 1991 (para una revisión más amplia, consulte [Charlier et al., 2007]).

Como se mencionó anteriormente, la forma más común del grafeno consiste en una red hexagonal bidimensional de átomos de carbono, como en la Figura 2.1.3. Esta red hexagonal está dividida en dos subredes triangulares, A y B. Por su parte la celda unitaria convencional está definida por los vectores de red primitivos (ver Figura 2.1.4a):

$$\mathbf{a}_1 = \frac{a}{2}\left(\sqrt{3}, 1\right), \quad \mathbf{a}_2 = \frac{a}{2}\left(\sqrt{3}, -1\right), \tag{2.1.1}$$

donde $a \approx 2.46$ Å es la distancia entre dos átomos de carbono adyacentes. Cada átomo de la subred A esta rodeado por tres átomos de la subred B y viceversa (red bipartita) [Katsnelson, 2020]. Como la red hexagonal tiene dos subredes triangulares, la distancia a los átomos vecinos más cercanos viene dada por la norma de los tres vectores $\delta_l$ con $l = 1, 2, 3$ (ver Figura 2.1.3):

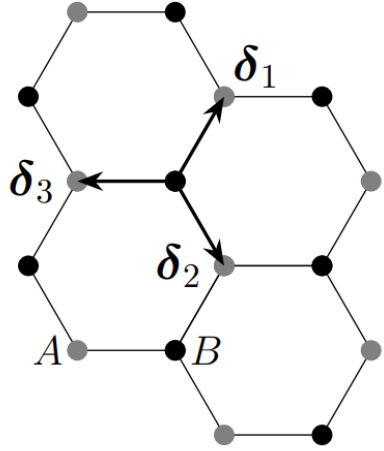

**Figura 2.1.3:** Los átomos más cercanos a un átomo B se denominan vecinos más próximos. Los vectores de posición relativa de estos vecinos más próximos se denotan como $\boldsymbol{\delta}_l$, y sus normas son $|\boldsymbol{\delta}_l| = a_l \approx 1.42$ Å. Imagen adaptada de [García Muñoz, 2022].

$$\delta_{\mathbf{1}} = \frac{a}{2}\left(\frac{1}{\sqrt{3}}, 1\right), \quad \delta_{\mathbf{2}} = \frac{a}{2}\left(\frac{1}{\sqrt{3}}, -1\right), \quad \delta_{\mathbf{3}} = a\left(0, -\frac{1}{\sqrt{3}}\right). \tag{2.1.2}$$

El espacio recíproco, mostrado en la Figura 2.1.4b, permite representar la periodicidad cristalina en el espacio de momentos. En el caso del grafeno, la red recíproca también presenta simetría hexagonal, definida por vectores que reflejan la periodicidad de la estructura real, dados por

$$\mathbf{b_1} = \frac{2\pi}{a}\left(\frac{1}{\sqrt{3}}, 1\right), \quad \mathbf{b_2} = \frac{2\pi}{a}\left(\frac{1}{\sqrt{2}}, -1\right). \tag{2.1.3}$$

La primera zona de Brillouin se presenta en la Figura 2.1.4b. En ella, se muestran los puntos especiales de alta simetría $K_+$, $K_-$ (puntos de Dirac) y M, con sus respectivos vectores de onda

$$\mathbf{k} = \left(\frac{2\pi}{3a}, -\frac{2\pi}{3\sqrt{3}a}\right), \quad \mathbf{k'} = \left(\frac{2\pi}{3a}, \frac{2\pi}{3\sqrt{3}a}\right), \quad \mathbf{M} = \left(\frac{2\pi}{3a}, 0\right). \tag{2.1.4}$$

Los estados hibridados $sp^2$ (estados $\sigma$) forman bandas ocupadas y vacías con un espacio entre ellas, mientras que los estados $p_z(\pi)$ forman una sola banda, con un punto cónico de autocruce en $K_+$ (por simetría, este punto también existe en

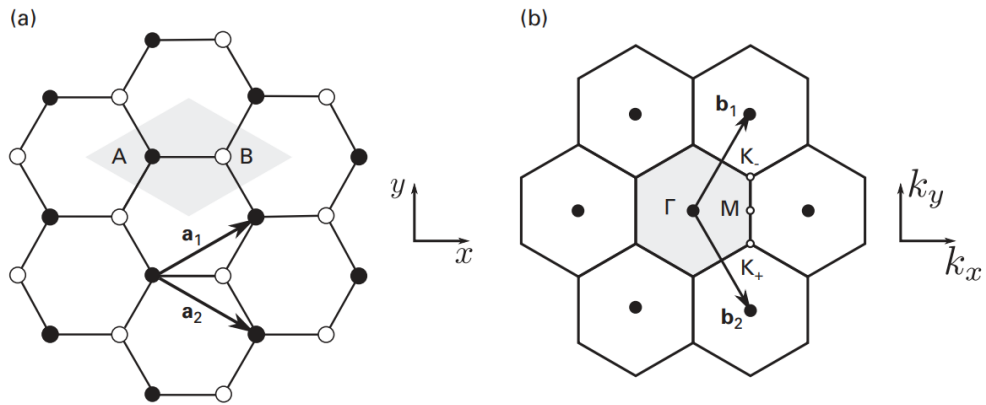

**Figura 2.1.4:** (a) Ilustración de los vectores de base $\mathbf{a}_1$ y $\mathbf{a}_2$ en la red hexagonal del grafeno. Esta red es una red de Bravais triangular que contiene una base de dos átomos: A (representados por puntos sólidos) y B (representados por puntos vacíos). (b) Representación de los puntos de la red recíproca que corresponden a la red de Bravais triangular (puntos sólidos), junto con los vectores de base recíprocos $\mathbf{b}_1$ y $\mathbf{b}_2$. La celda unitaria y la zona de Brillouin están sombreadas en gris en (a) y (b), respectivamente. En (b) también se muestran los puntos de alta simetría, que están etiquetados como $\Gamma$ (centro de la zona), $K_+$, $K_-$ y M. Imagen adaptada de [Foa Torres et al., 2014].

K$_-$) (véase la Figura 2.1.5). Este punto cónico es una característica de la peculiar estructura electrónica del grafeno y el origen de sus propiedades electrónicas únicas. Fue obtenido por primera vez por Wallace en 1947 [Wallace, 1947] en el marco de un modelo simple de *tight-binding*, el cual fue desarrollado por McClure en 1957 [McClure, 1957] y Slonczewski y Weiss en 1958 (para mayor información consulte [Slonczewski and Weiss, 1958]) .

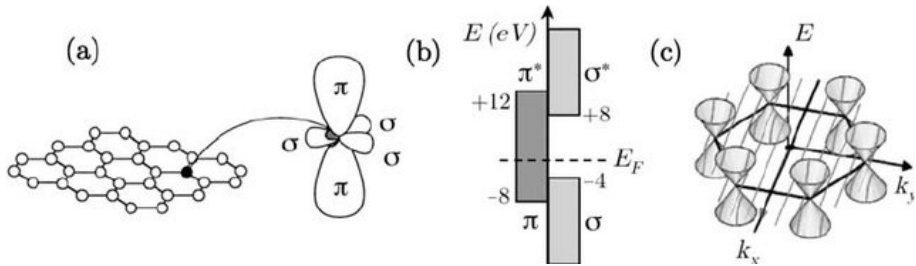

**Figura 2.1.5:** (a) Los tres orbitales $\sigma$ en el grafeno y el orbital $\pi$ perpendicular a la hoja de grafeno. Los enlaces $\sigma$ en la red hexagonal de carbono unen fuertemente a los átomos de carbono y son responsables de la energía de enlace y de las propiedades estructurales del grafeno. Los enlaces $\pi$ son perpendiculares a la superficie de la hoja. (b) Las bandas de enlace y anti-enlace $\sigma$ correspondientes están separadas por una brecha de energía igual a 12 eV, aproximadamente, mientras que los estados de enlace y anti-enlace $\pi$ están cerca del nivel de Fermi ($E_F$). Por lo tanto, los enlaces $\sigma$ a menudo se ignoran al predecir las propiedades electrónicas del grafeno alrededor de la energía de Fermi. (c) Zona de Brillouin donde se representan los puntos de Dirac y cada enlace $p_z$. Imagen adaptada de [Dubois et al., 2009].

**Modelo de *tight-binding***

El modelo de amarre fuerte (o *tight-binding*) es una herramienta útil para describir la estructura electrónica de sólidos cristalinos como el grafeno. Este modelo permite identificar la matriz Hamiltoniana $H$ [Saito et al., 1998, McCann and Koshino, 2013, Foa Torres et al., 2014] que describe las interacciones entre los electrones en la red cristalina. El modelo describe las interacciones entre los electrones en la red cristalina, considerando su localización en sitios discretos. Para capturar su carácter no localizado, se emplean las funciones de Bloch $\phi(\mathbf{k}, \mathbf{r})$, donde $\mathbf{k}$ es el vector de onda y $\mathbf{r}$ es el vector de posición, estas funciones combinan una factor de onda plana con una función periódica de la red. Estas funciones permiten construir la matriz del Hamiltoniano $H$, describiendo tanto la energía en cada sitio como la capacidad de los electrones de desplazarse entre ellos.

Una implicación importante de este enfoque es que, debido a la simetría del sistema, la superposición entre los orbitales $p_z$ y los electrones $s$, o entre los orbitales $p_x$ y $p_y$, es estrictamente cero. Esta característica implica que los electrones $p_z$ que forman los enlaces $\pi$ en el grafeno pueden tratarse de forma independiente de los otros electrones de valencia [Foa Torres et al., 2014].

Hablando matemáticamente, la condición anterior significa que existen funciones de onda $\phi$ para el grafeno que deben cumplir la misma simetría translacional [Cohen-Tannoudji et al., 2019]. Esto implica que en el modelo de *tight-binding* para el grafeno, las funciones de onda deben mostrar un comportamiento periódico. Dicho de otro modo, cualquier función de onda de la red, $\phi$, debería satisfacer el teorema de Bloch:

$$T_{\alpha_j}\phi = \mathrm{e}^{i\mathbf{k}\cdot\alpha_j}\phi. \tag{2.1.5}$$

Aquí, $T_{\alpha_j}$ representa un vector de traslación, $\phi$ es una función que está siendo trasladada además de tener la misma periodicidad que la red cristalina, y $\mathbf{k}$ es un vector de onda. Las soluciones naturales para (2.1.5) son las ondas planas, ya que al ser trasladadas solo adquieren un factor de fase $e^{i\mathbf{k}\cdot\alpha_j}$. Sin embargo, debido a que el grafeno, al igual que cualquier cristal, posee un potencial periódico, las funciones de onda no pueden describirse únicamente como ondas planas, sino como una combinación infinita de ellas. Por esta razón, y de acuerdo con el teorema de Bloch, se recurre al uso de funciones de Bloch expresadas como

$$\phi_j(\mathbf{k},\mathbf{r}) = \frac{1}{\sqrt{N}}\sum_{\mathbf{R}_{j,m}} e^{i\mathbf{k}\cdot\mathbf{R}_{j,m}}\,\psi_j(\mathbf{r}-\mathbf{R}_{j,m}), \quad j=1,2,\dots,n. \tag{2.1.6}$$

donde $\phi_j(\mathbf{k},\mathbf{r})$ es la función de onda de Bloch asociada al orbital $j$ en el cristal; $\mathbf{k}$ es el vector de onda cristalina, $\mathbf{r}$ es el vector de posición dentro de la celda unitaria y $n$ es el número total de funciones base consideradas por cada vector de onda $\mathbf{k}$ en la zona de Brillouin. La función $\psi_j(\mathbf{r}-\mathbf{R}_{j,m})$ representa el orbital atómico $j$ localizado en la celda unitaria $m$, desplazado por el vector de red $\mathbf{R}_{j,m}$, el índice $m=1,\dots,N$ recorre todas las celdas unitarias del cristal.

El teorema de Bloch establece que dentro de un potencial periódico, como el de una red cristalina, las soluciones a la ecuación de Schrödinger toman la forma de ondas de Bloch, las cuales verifican lo siguiente:

$$
\begin{aligned}
\phi_j(\mathbf{k}, \mathbf{r} + \alpha) &= \frac{1}{\sqrt{N}} \sum_{R}^{N} \mathrm{e}^{i\mathbf{k}\cdot\mathbf{R}_{j,m}} \psi_j(\mathbf{r} + \alpha - \mathbf{R}_{j,m}), \\
&= \frac{1}{\sqrt{N}} \sum_{R-\alpha}^{N} \mathrm{e}^{i\mathbf{k}\cdot\alpha} \mathrm{e}^{i\mathbf{k}\cdot(\mathbf{R}_{j,m}-\alpha)} \psi_j(\mathbf{r} - (\mathbf{R}_{j,m} - \alpha)), \\
&= \mathrm{e}^{i\mathbf{k}\cdot\alpha} \phi_j(\mathbf{k}, \mathbf{r}).
\end{aligned}
\tag{2.1.7}
$$

Estas ondas de Bloch forman un conjunto completo de funciones base para las soluciones de la ecuación de Schrödinger en el cristal. Por lo tanto, cualquier función de onda $\Phi_i(\mathbf{k}, \mathbf{r})$ puede ser expresada como una combinación lineal de estas ondas de Bloch (ver Figura 2.1.6):

$$
\Phi_i(\mathbf{k}, \mathbf{r}) = \sum_{j=1}^{n} C_{ij}(\mathbf{k})\, \phi_j(\mathbf{k}, \mathbf{r}), \quad i = 1, \ldots, n,
\tag{2.1.8}
$$

donde $C_{ij}(\mathbf{k})$ son los coeficientes de expansión que determinan la contribución de cada función de Bloch $\phi_j$ al estado electrónico $\Phi_i$.

Según la mecánica cuántica, los eigenvalores de energía $E_j$ del operador $H$ pueden calcularse como los valores esperados

$$
E_j = \frac{\langle \Phi_j |\, H\, | \Phi_j \rangle}{\langle \Phi_j | \Phi_j \rangle},
\tag{2.1.9}
$$

y en el grafeno estos valores corresponden a los niveles de energía de los electrones descritos por las funciones de Bloch en la red cristalina.

Sustituyendo la ecuación (2.1.8) en (2.1.9) y realizando un cambio de subíndices, se obtiene la siguiente expresión

$$
E_j = \frac{\sum\limits_{l,m=1}^{n} C_{jl}^{*} C_{jm} \langle \phi_l |\, H\, | \phi_m \rangle}{\sum\limits_{l,m=1}^{n} C_{jl}^{*} C_{jm} \langle \phi_l | \phi_m \rangle},
\tag{2.1.10}
$$

donde $C^{*}$ denota el conjugado complejo de $C \in \mathbb{C}$. Es necesario optimizar los coeficientes $C_{ij}$ para minimizar la energía lo que garantiza que el estado construido se acerque al eigenvalor real del Hamiltoniano, reflejando la energía física del electrón en la red de grafeno. Por lo tanto, se toma una derivada parcial de $E_j$ con respecto a $C_{jl}^*$, lo cual conduce a lo siguiente:

$$\frac{\partial E_j}{\partial C_{jl}^*} = \frac{1}{\sum_{l,m=1}^{n} C_{jl}^* C_{jm} \langle \phi_l | \phi_m \rangle} \left( \sum_{m=1}^{n} C_{jm} \langle \phi_l | H | \phi_m \rangle - E_j \sum_{m=1}^{n} C_{jm} \langle \phi_l | \phi_m \rangle \right),$$
(2.1.11)

en donde se consideró la relación $\frac{\partial C_{jl}^*}{\partial C_{jl'}^*} = \delta_{l,l'}$ y la ecuación (2.1.10). Al igualar la ecuación (2.1.11) a cero, se obtiene

$$\sum_{m=1}^{n} C_{jm} \langle \phi_l | H | \phi_m \rangle = E_j \sum_{m=1}^{n} C_{jm} \langle \phi_l | \phi_m \rangle.$$
(2.1.12)

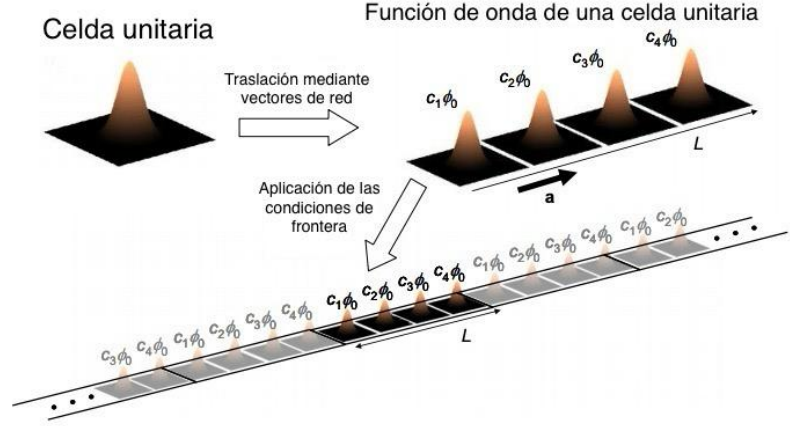

**Figura 2.1.6:** Representación de los estados de Bloch para una red unidimensional. Un átomo aislado en una celda unitaria de una red cristalina es descrito mediante la función de onda $\phi_0$, la cual es trasladada mediante los vectores de red, $T_{\alpha_j}\phi_0 = e^{i\mathbf{k}\cdot\alpha_j}\phi_0$, para obtener la función de onda de una celda unitaria de longitud $L$. Luego, la función de onda de la red cristalina $\Phi_i$ se construye como combinación lineal de funciones de onda de $N$ celdas unitarias. Imagen adaptada de [Baldo, 2011].

Como $\langle \phi_l | H | \phi_m \rangle$ y $\langle \phi_l | \phi_m \rangle$ son elementos de matriz, es posible definir las matrices de tamaño $n \times n$:

$$(H)_{ij} = \langle \phi_i | H | \phi_j \rangle, \quad (S)_{ij} = \langle \phi_i | \phi_j \rangle.$$
(2.1.13)

Las matrices $(H)_{ij}$ y $(S)_{ij}$ son conocidas como las matrices de integrales de transferencia y de solapamiento, respectivamente [Saito et al., 1998]. Dado que el índice $j$ está fijo en la ecuación (2.1.12), los coeficientes $C_{ij}$ pueden considerarse como vectores

columna

$$C_j = \begin{pmatrix} C_{j1} \\ C_{j2} \\ \vdots \\ C_{jn} \end{pmatrix}. \tag{2.1.14}$$

Las dos últimas ecuaciones nos permiten reescribir (2.1.12) de la siguiente manera:

$$HC_j = E_j S C_j. \tag{2.1.15}$$

Las energías de banda $E_j$ se determinan a partir de la ecuación de valores propios generalizada (2.1.15) tras resolver la ecuación secular

$$\det(H - E_j S) = 0, \tag{2.1.16}$$

la cual tiene $n$ soluciones y que son los autovalores de energía de $H$.

**Hamiltoniano efectivo para una monocapa de grafeno**

En esta sección se aplicará el modelo de amarre fuerte al grafeno. Teniendo en cuenta un orbital $2p_z$ por sitio atómico, y dado que hay dos átomos en la celda unitaria de la monocapa de grafeno como se muestra en la Figura 2.1.4, se incorporan dos orbitales ($n = 2$) por celda unitaria etiquetados como $j = $ A, B. Así, las funciones de Bloch se simplifican a:

$$\phi_j(\mathbf{k}, \mathbf{r}) = \frac{1}{\sqrt{N}} \sum_{\mathbf{R}_{j,m}} \mathrm{e}^{i\mathbf{k}\cdot\mathbf{R}} \psi_j(\mathbf{r} - \mathbf{R}_{j,m}), \quad m = 1, \dots, N. \tag{2.1.17}$$

Los elementos diagonales de la matriz $H$ se calculan como

$$\langle \phi_j | H | \phi_j \rangle = \frac{1}{N} \sum_{m,l=1}^{N} \mathrm{e}^{i\mathbf{k}\cdot(\mathbf{R}_{j,l} - \mathbf{R}_{j,m})} \langle \psi_j(\mathbf{r} - \mathbf{R}_{j,m}) | H | \psi_j(\mathbf{r} - \mathbf{R}_{j,l}) \rangle. \tag{2.1.18}$$

En esta ecuación, se considera la interacción entre el $j$-ésimo átomo en la $m$-ésima celda unitaria y los $j$-ésimos átomos en las $l$-ésimas celdas unitarias. Sin embargo, utilizando la aproximación a primeros vecinos, solo se consideran las interacciones del $j$-ésimo átomo consigo mismo, ya que su contribución dominante es significativa [McCann and Koshino, 2013]. Entonces, la ecuación (2.1.13) se reduce a aproximadamente:

$$\langle \phi_j | H | \phi_j \rangle \approx \frac{1}{N} \sum_{j=1}^{N} \langle \psi_j | H | \psi_j \rangle. \tag{2.1.19}$$

Esto puede considerarse como una suma del parámetro $\epsilon_j = \langle \phi_j | H | \phi_j \rangle$ sobre todas las celdas unitarias. Tal cantidad consiste en la energía atómica del átomo libre más cualquier potencial de la capa de grafeno donde toma el mismo valor en cada celda unitaria. Por lo tanto, el elemento de la matriz puede expresarse simplemente como $(H)_{jj} = \epsilon_j$.

Para el cálculo de los elementos diagonales de la matriz de solapamiento $S$, se procede de la misma manera que el de $H$, con el traslape de un orbital consigo mismo igual a la unidad:

$$
\begin{aligned}
\langle \phi_j | S | \phi_j \rangle &= \frac{1}{N} \sum_{m,l=1}^{N} \mathrm{e}^{i\mathbf{k} \cdot (\mathbf{R}_{j,l} - \mathbf{R}_{j,m})} \langle \psi_j(\mathbf{r} - \mathbf{R}_{j,m}) | \psi_j(\mathbf{r} - \mathbf{R}_{j,l}) \rangle, \\
&\approx \frac{1}{N} \sum_{m=1}^{N} \langle \psi_j | \psi_j \rangle = 1. \tag{2.1.20}
\end{aligned}
$$

Por otro lado, el elemento $H_{\mathrm{AB}}$ fuera de la diagonal de la matriz de integrales de transferencia $H$ representa la probabilidad de salto entre orbitales en los sitios A y B. Al sustituir la función de Bloch (2.1.18) en el elemento matricial (2.1.13), se obtiene una suma sobre todos los sitios A y B, en donde se supone que la contribución dominante a la dinámica electrónica surge de los saltos entre átomos vecinos más cercanos. En el modelo de primeros vecinos, cada átomo de carbono interactúa únicamente con sus tres átomos adyacentes en la red hexagonal, esto es, para un sitio A específico, se considera el salto a sus tres sitios B más cercanos (véase Figura 2.1.3) (aproximación a primeros vecinos), cuyos vectores de posición relativa $\delta_l$ están escritos en la ecuación (2.1.2), con $l = 1, 2, 3$:

$$
\begin{aligned}
\langle \phi_{\mathrm{A}} | H | \phi_{\mathrm{B}} \rangle &\approx \frac{1}{N} \sum_{m=1}^{N} \sum_{l=1}^{3} \mathrm{e}^{-i\mathbf{k} \cdot \delta_l} \langle \psi_{\mathrm{A}} (r - \mathbf{R}_{\mathrm{B},l} - \delta_l) | H | \psi_{\mathrm{B}} (r - \mathbf{R}_{\mathrm{B},l}) \rangle, \\
&= \frac{1}{N} \sum_{m=1}^{N} (-\gamma_0) \left( \mathrm{e}^{i\frac{ak_x}{\sqrt{3}}} + \mathrm{e}^{-i\frac{a}{2}\left(\frac{k_x}{\sqrt{3}} + k_y\right)} + \mathrm{e}^{-i\frac{a}{2}\left(\frac{k_x}{\sqrt{3}} - k_y\right)} \right), \\
&= \frac{1}{N} \sum_{m=1}^{N} (-\gamma_0) \left( i\frac{ak_x}{\sqrt{3}} + 2\mathrm{e}^{-i\frac{a}{2}\frac{k_x}{\sqrt{3}}} \cos\left(\frac{a}{2}k_y\right) \right), \\
&= -\frac{1}{N} \sum_{m=1}^{N} \gamma_0 h(\mathbf{k}),
\end{aligned} \tag{2.1.21}
$$

donde el parámetro de salto en el plano $\gamma_0$ se define como:

$$
\gamma_0 = -\langle \phi_{\mathrm{A}}(r - \mathbf{R}_{\mathrm{B},l} - \delta_l) | H | \phi_{\mathrm{B}}(r - \mathbf{R}_{\mathrm{B},l}) \rangle > 0, \tag{2.1.22}
$$

el cual representa la energía de enlace covalente entre dos átomos de carbono adyacentes, y es una cantidad positiva.

Análogamente, para los elementos que están fuera de la diagonal en la matriz $S$ se tiene lo siguiente

$$\langle \phi_A | S | \phi_B \rangle = \frac{1}{N} \sum_{m,l=1}^{N} e^{i\mathbf{k}\cdot(\mathbf{R}_{B,l}-\mathbf{R}_{A,m})} \langle \psi_A(r-\mathbf{R}_{A,m}) | \psi_B(r-\mathbf{R}_{B,l}) \rangle, \qquad (2.1.23)$$

$$\approx \frac{1}{N} \sum_{m=1}^{N} \sum_{l=1}^{3} e^{-i\mathbf{k}\cdot\delta_l} \langle \psi_A(r-\mathbf{R}_{B,l}-\delta_l) | \psi_B(r-\mathbf{R}_{B,l}) \rangle, \qquad (2.1.24)$$

$$= \frac{1}{N} \sum_{m=1}^{N} s_0 \left( i\frac{ak_x}{\sqrt{3}} + 2e^{-i\frac{a}{2}\frac{k_x}{\sqrt{3}}} \cos\left(\frac{a}{2}k_y\right) \right), \qquad (2.1.25)$$

$$= \frac{1}{N} \sum_{m=1}^{N} s_0 h(k) = s_0 h(k), $$

donde

$$s_0 = \langle \psi_A(r-\mathbf{R}_{B,l}-\delta_l) | \psi_B(r-\mathbf{R}_{B,l}) \rangle. \qquad (2.1.26)$$

Este parámetro se introduce para considerar la posibilidad de un solapamiento no nulo entre los orbitales de átomos adyacentes. Se puede observar que las expresiones (2.1.21) y (2.1.23) involucran una función $h(k)$, la cual está dada por

$$h(k) = e^{i\frac{ak}{\sqrt{3}}} + 2e^{-i\frac{a}{2}\frac{k}{\sqrt{3}}} \cos\left(\frac{a}{2}ky\right). \qquad (2.1.27)$$

Por lo tanto, los elementos fuera de la diagonal pueden escribirse como:

$$(H)_{AB} = -\gamma_0 h(k), \quad (S)_{AB} = s_0 h(k). \qquad (2.1.28)$$

Dado que $h(k)$ es una función compleja, y el Hamiltoniano corresponde a una matriz Hermítica, se tiene $H_{BA} = H_{AB}^* \approx -\gamma_0 h^*(k)$ y $S_{BA} = S_{AB}^* \approx s_0 h^*(k)$. Así, los elementos de la matriz de transferencia $H_m$ y de la matriz de integral de solapamiento $S_m$ de la monocapa de grafeno pueden escribirse como:

$$H = \begin{pmatrix} \epsilon_A & -\gamma_0 h(k) \\ -\gamma_0 h^*(k) & \epsilon_B \end{pmatrix}, \quad S = \begin{pmatrix} 1 & s_0 h(k) \\ s_0 h^*(k) & 1 \end{pmatrix}. \qquad (2.1.29)$$

Ahora, es posible resolver la ecuación secular indicada en (2.1.16):

$$\det \begin{pmatrix} \epsilon_A - E & -(\gamma_0 + Es_0)h(k) \\ -(\gamma_0 + Es_0)h^*(k) & \epsilon_B - E \end{pmatrix} = 0, \qquad (2.1.30)$$

lo cual lleva a un polinomio de segundo grado para la energía. La solución a dicho polinomio es:

$$E_{\pm} = \frac{\epsilon_{\mathrm{A}} + \epsilon_{\mathrm{B}} + 2\gamma_0 s_0 |h|^2 \pm \sqrt{(\epsilon_{\mathrm{A}} + \epsilon_{\mathrm{B}})^2 - 4\epsilon_{\mathrm{A}}\epsilon_{\mathrm{B}} + 4\left[\gamma_0^2 + s_0^2 \epsilon_{\mathrm{A}}\epsilon_{\mathrm{B}} + \gamma_0 s_0 (\epsilon_{\mathrm{A}} + \epsilon_{\mathrm{B}})\right]|h|^2}}{2(1 - s_0^2 |h|^2)}.$$
(2.1.31)

Dado que $\epsilon_j$ es la suma de la energía atómica para el átomo libre más cualquier potencial de la capa de grafeno, a partir de este momento se pueden considerar como iguales para cada átomo de carbono en la capa, es decir, $\epsilon_{\mathrm{A}} = \epsilon_{\mathrm{B}} = \epsilon$. Por lo tanto, la ecuación anterior se reduce a:

$$E_{\pm} = \frac{\epsilon \pm \gamma_0 |h(k)|}{1 \mp s_0 |h(k)|}$$
(2.1.32)

Los valores de los parámetros según Saito et al. [Saito et al., 1998] son $\gamma_0 = 3.033$ eV y $s_0 = 0.129$. La función $h(k)$, definida en la ecuación (2.1.27), se anula en las esquinas de la zona de Brillouin. Dos de estas esquinas se conocen como $\mathrm{K}_+$ y $\mathrm{K}_-$, y se ubican en los vectores de onda. Estos puntos reciben el nombre de puntos K o valles, y resultan de gran importancia porque alrededor de ellos la dispersión electrónica es lineal. Para distinguirlos se introduce el llamado índice de valle, denotado por $\nu = \pm 1$, que indica si un electrón se encuentra en el valle $\mathrm{K}_+$ o en el valle $\mathrm{K}_-$.
En estos puntos, las soluciones (2.1.32) son degeneradas, lo que significa que las bandas de conducción y valencia se tocan, resultando en un *gap* de banda nulo.

La matriz de transferencia $H$ se aproxima a un Hamiltoniano tipo Dirac cerca del punto $\mathrm{K}_{\pm}$, describiendo cuasi-partículas quirales sin masa con una relación de dispersión lineal. Estos puntos son especialmente relevantes porque el nivel de Fermi se encuentra cerca de ellos en el grafeno prístino. Cuando la integral de solapamiento $S$ se hace cero, las bandas $\pi$ y $\pi^*$ se vuelven simétricas alrededor de $E = \epsilon$, como se puede entender a partir de la ecuación (2.1.32). Las relaciones de dispersión energética en el caso de $S_0 = 0$ (es decir, en el esquema de Slater-Koster [Papaconstantopoulos and Mehl, 2003]) son comúnmente utilizadas como una aproximación simple para la estructura electrónica de una capa de grafeno:

$$E_{\pm}(\mathbf{k}) = \pm \gamma_0 \sqrt{1 + 4\cos\left(\frac{\sqrt{3}k_x a}{2}\right)\cos\left(\frac{k_y a}{2}\right) + 4\cos^2\left(\frac{k_y a}{2}\right)}.$$
(2.1.33)

Los vectores de onda $\mathbf{k} = (k_x, k_y)$ se eligen dentro de la primera zona de Brillouin hexagonal. Los ceros de $\alpha(\mathbf{k})$ corresponden a las intersecciones entre las bandas

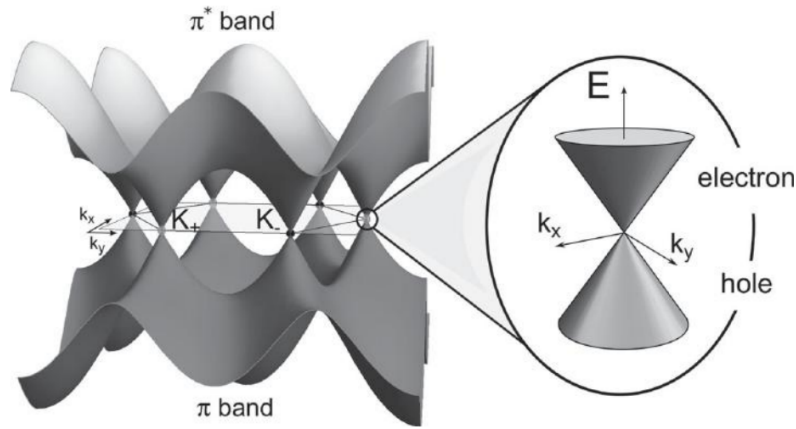

**Figura 2.1.7:** Las bandas electrónicas $\pi$ y $\pi^*$ del grafeno. En este enfoque simple, las bandas $\pi$ y $\pi^*$ son simétricas con respecto a las bandas de valencia y conducción. La relación de dispersión lineal cerca de los puntos $K_+$ (puntos grises claros) y $K_-$ (puntos negros) de la primera zona de Brillouin 2D da lugar a los "conos de Dirac", como se muestra a la derecha. Cabe destacar que cerca de estos conos, $k_x$ y $k_y$ se utilizan para denotar el desplazamiento desde el punto K correspondiente. Imagen adaptada de [Foa Torres et al., 2014].

con signos $+$ y $-$, las cuales ocurren en los puntos $K_+$ y $K_-$. Como se mencionó anteriormente, estos puntos representan las posiciones en el espacio recíproco en donde las bandas se tocan, dando lugar a una degeneración en la energía que es característica de los valles del sistema. Además, con un solo electrón $p_z$ por átomo en el modelo $\pi$–$\pi^*$, las otras tres configuraciones electrónicas ($s$, $p_x$, $p_y$) llenan la banda $\sigma$ de baja energía. Así, la banda $(-)$ (rama de energía negativa) está completamente ocupada, mientras que la banda $(+)$ está vacía, al menos para el grafeno eléctricamente neutro. Por lo tanto, el nivel de Fermi $E_F$ (o punto de neutralidad de carga) es la referencia de energía cero (véase Figura 2.1.7) y la superficie de Fermi está compuesta por el conjunto de puntos $K_+$ y $K_-$. El grafeno muestra un carácter metálico, es decir, de brecha cero. Sin embargo, dado que la superficie de Fermi es de dimensión cero (ya que se reduce a un conjunto discreto y finito de puntos), se suele emplear el término semimetálico o semiconductor de brecha cero. Al expandir la ecuación (2.1.33) para $\mathbf{k}$ en las cercanías de $K_+$ (o $K_-$), $\mathbf{k} = K_+ + \delta\mathbf{k}$ ($\mathbf{k} = K_- + \delta\mathbf{k}$), se obtiene una dispersión lineal para las bandas $\pi$ y $\pi^*$ cerca de estos seis vértices de la zona de Brillouin hexagonal bidimensional:

$$E_\pm(\delta\mathbf{k}) = \pm\hbar v_F|\delta\mathbf{k}| \qquad (2.1.34)$$

con $v_F = \frac{\sqrt{3}\gamma_0 a}{2\hbar}$, siendo la velocidad de grupo electrónica. El grafeno es, por lo tanto,

altamente peculiar debido a esta relación lineal energía-momento y a la simetría electrón-hueco.

Al expandir la matriz de transferencia $H$ en la ecuación (2.1.29) alrededor de $K_+$ y $K_-$, se obtiene una aproximación cercana a esos puntos. Una expansión lineal da:

$$H_{K_+} = \hbar v_F \begin{pmatrix} 0 & k_x - ik_y \\ k_x + ik_y & 0 \end{pmatrix} = v_F(p_x\sigma_x + p_y\sigma_y), \qquad (2.1.35)$$

donde $p_{x(y)} = \hbar k_{x(y)}$ y las matrices de Pauli se definen como sigue:

$$\sigma_x = \begin{pmatrix} 0 & 1 \\ 1 & 0 \end{pmatrix}, \quad \sigma_y = \begin{pmatrix} 0 & -i \\ i & 0 \end{pmatrix}. \qquad (2.1.36)$$

Este Hamiltoniano efectivo también puede escribirse en forma más compacta como:

$$H_{K_+} = v_F\sigma \cdot \mathbf{p}, \qquad (2.1.37)$$

donde $\boldsymbol{\sigma} = (\sigma_x, \sigma_y)$ son las matrices de Pauli, y a través del producto escalar, representan una combinación lineal con los operadores de momento $\mathbf{p} = (p_x, p_y)$ [DiVincenzo and Mele, 1984].

De manera similar, cerca del punto $K_-$, el Hamiltoniano efectivo se expresa como:

$$H_{K_-} = v_F\sigma^* \cdot \mathbf{p}, \qquad (2.1.38)$$

donde $\sigma^* = (-\sigma_x, \sigma_y)$ representa la conjugación de las matrices de Pauli para este valle, y en ambos casos $\mathbf{p} = \hbar\mathbf{k}$ es el operador momento en las cercanías de $K_\pm$.

Esta estructura refleja la simetría entre los dos valles, donde las diferencias en el signo frente a $\sigma_x$ corresponden a la inversión del momento y aseguran la conservación del tiempo en el sistema. En conjunto, los Hamiltonianos $H_{K_-}$ y $H_{K_-}$ describen el comportamiento lineal y relativista de los electrones en grafeno cerca de los puntos de valle, mostrando una analogía directa con los fermiones de Dirac sin masa en dos dimensiones.

## 2.2 | Una nueva familia de materiales bidimensionales

Los cristales puramente bidimensionales son una subclase de nanomateriales que muestran características físicas interesantes debido al confinamiento cuántico de sus electrones. El grafeno fue el primer material en ser aislado mediante la técnica de exfoliación de capas individuales de un cristal de grafito usando cinta adhesiva Scotch [Novoselov et al., 2004]. Este método original y eficiente también se aplicó a otros materiales con estructuras laminadas, creando una nueva familia de cristales atómicamente delgados. En la actualidad, se ha reportado la presencia y estabilidad bajo condiciones normales de más de unas pocas docenas de cristales bidimensionales, que incluyen nitruro de boro hexagonal (h-BN), dicalcogenuros de metales de transición ($MoS_2$, $MoSe_2$, $WS_2$, $WSe_2$, $NbSe_2$, etc.), capas delgadas de óxidos ($TiO_2$, $MoO_3$, $WO_3$, etc.), siliceno, germaneno, fosforeno, borofeno, arseneno, estaneno, entre otros.

Recientemente, la búsqueda de nuevos materiales 2D ha llevado a la predicción y síntesis de varias capas atómicamente delgadas con propiedades electrónicas interesantes. Entre estos materiales se encuentran el siliceno y el germaneno, que tienen estructuras electrónicas prometedoras gracias a sus fuertes acoplamientos espín-órbita, los cuales pueden abrir una brecha de energía en la zona de Brillouin, ofreciendo ventajas sobre el grafeno [Cahangirov et al., 2009]. En 2014, se predijo la existencia de fosforeno azul 2D y otras fases de fósforo 2D [Zhu and Tomanek, 2014, Guan et al., 2014]. En 2015, se sintetizó el estaneno, investigándose sus aspectos topológicos [Zhu et al., 2015], y se crearon polimorfos de boro 2D con características metálicas anisotrópicas en superficies de plata [Mannix et al., 2015].

### Materiales bidimensionales anisótropos

Adicionalmente, existe una familia de sistemas de Dirac que presentan un espectro de energía anisótropo e inclinado con respecto al punto K. Esto significa que la velocidad de las cuasipartículas en estos sistemas depende del ángulo azimutal del momento en el espacio recíproco, y la simetría electrón-hueco no se preserva. En la imagen de la Figura 2.2.1, se presentan 3 espectros de energía de materiales de Dirac: siliceno y germaneno ambos con anisotropía y el grafeno prístino

### Borofeno

El boro es el elemento más ligero que forma redes covalentes extendidas y que tiene una capa exterior trivalente que crea enlaces deslocalizados, resultando en 16 alótropos 3D conocidos [Sergeeva et al., 2014], incluyendo cúmulos icosaédricos $B_{12}$ [Oganov et al., 2009]. La síntesis de materiales 2D sin análogos en alótropos volumétricos es difícil de obtener [Mannix et al., 2018], pero se han logrado sintetizar monocapas de borofeno mediante la deposición controlada de átomos de boro en una superficie de plata [Mannix et al., 2015, Feng et al., 2016]. Por otro lado, una de las estructuras de boro que ha mostrado una gran estabilidad a través de cálculos de primeros principios es el llamado borofeno ortorrómbico $8 - Pmmn$ [Zabolotskiy and Lozovik, 2016] donde su estructura se aprecia en la Figura 2.2.2. Este material en 2D posee propiedades electrónicas similares a las del grafeno en lo que respecta a sus portadores de carga cuando se coloca alrededor de la energía de Fermi. La estructura de bandas en este material se compone de dos conos de Dirac; a diferencia del grafeno, estos conos son anisótropos y están inclinados como se muestra en la Figura 2.2.3, lo que significa que la velocidad de los portadores de carga varía con el ángulo azimutal en el espacio de momentos. Además, las bandas de conducción y de valencia no son simétricas en relación al nivel de Fermi.

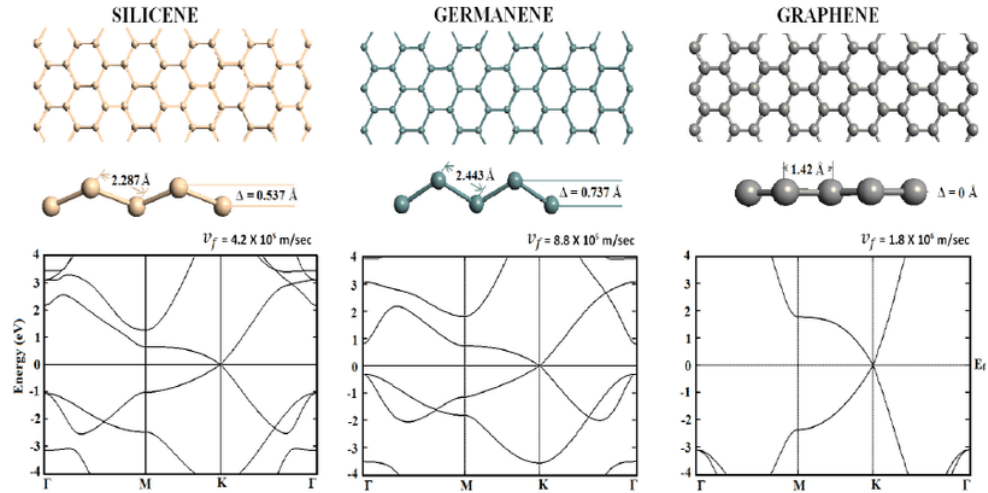

**Figura 2.2.1:** Estructura de bandas con velocidades de Fermi de siliceno, germaneno y grafeno. Imagen adaptada de [Trivedi et al., 2014].

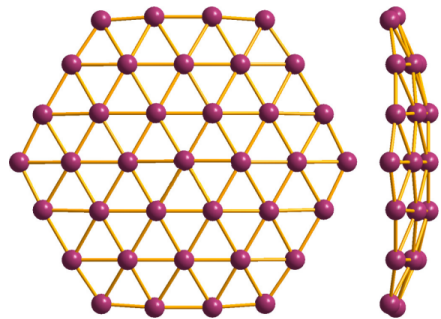

**Figura 2.2.2:** Representación de la red del borofeno ortorrómbico $8 - Pmmn$, donde cada nodo corresponde a un átomo de boro. Imagen adaptada de [Champo and Naumis, 2019].

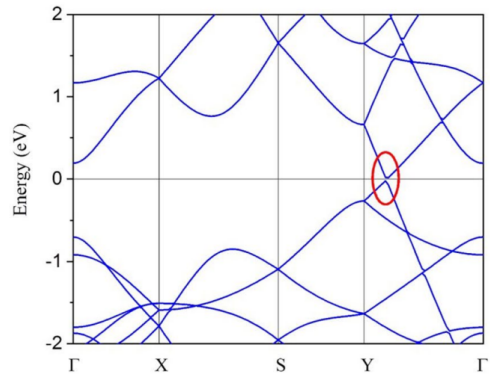

**Figura 2.2.3:** Estructura de bandas donde el punto de Dirac ha sido resaltado con un círculo rojo del borofeno $8 - Pmmn$ prístino. Imagen adaptada de [Chowdhury et al., 2019].

**Grafeno deformado**

El grafeno deformado es una variante del grafeno convencional que adquiere propiedades electrónicas únicas debido a la modificación de su estructura cristalina mediante tensiones mecánicas, pliegues o curvaturas controladas. Estas deformaciones generan cambios significativos en el Hamiltoniano efectivo, especialmente alrededor de los puntos de Dirac, y permiten explorar diferentes propiedades basadas en la interacción entre la deformación y las propiedades electrónicas. Si el grafeno es sometido a tensión uniaxial [Concha et al., 2018], las velocidades $v_x$ y $v_y$ son las componentes diagonales no nulas de la velocidad de Fermi anisótropa a lo largo de los ejes $x$ y $y$. Además, están dadas, a primer orden en la deformación $\varepsilon$ por

$$v_x = v_{\mathrm{F}}(1 - \varrho\,\varepsilon), \quad v_y = v_{\mathrm{F}}(1 + \varrho\,\varepsilon\,\varsigma) \quad \text{(tensión uniaxial a lo largo del eje } x),$$
$$(2.2.1)$$

$$v_x = v_{\mathrm{F}}(1 + \varrho\,\varepsilon\,\varsigma), \quad v_y = v_{\mathrm{F}}(1 - \varrho\,\varepsilon) \quad \text{(tensión uniaxial a lo largo del eje } y),$$
$$(2.2.2)$$

donde $\varrho \approx 2.6$ es la constante de Grüneisen, $\varepsilon$ mide el porcentaje de tensión aplicada y $\varsigma = 0.18$ es el coeficiente de Poisson en el grafeno.

# 3 | Materiales de Dirac con campos eléctricos y magnéticos externos

En 2009, se desarrolló un trabajo teórico en el que se consideró la interacción de una placa de grafeno con campos magnéticos perpendiculares a la superficie del material. Eligiendo una Norma de Landau que permitiera conservar la invariancia traslacional en una dirección, fue posible obtener soluciones analíticas exactas para la ecuación de Dirac-Weyl (para más información, véase [Ş. Kuru et al., 2009]). Este tipo de estudios teóricos se enmarca dentro del creciente interés en el grafeno, motivado por sus propiedades únicas, entre ellas un transporte electrónico balístico capaz de modificarse externamente [Castro Neto et al., 2009, Han et al., 2007], su alta conductividad térmica y propiedades mecánicas.

Por otro lado, varios autores han investigado materiales de Dirac en interacción con campos magnéticos y eléctricos dependientes de la posición [Mojarro et al., 2021, Romero Jorge et al., 2024, Isobe and Nagaosa, 2022, Schulze-Halberg and Roy, 2021]. En estos trabajos se estudia cómo los campos externos modifican la estructura de bandas, la densidad de estados y las propiedades de transporte, lo que abre nuevas posibilidades para aplicaciones en tecnologías y dispositivos electrónicos. Un estudio adicional abordó el borofeno interactuando con un campo eléctrico y magnético de perfil constante [Díaz-Bautista, 2022], además de examinar los estados coherentes en estos materiales en el espacio fase [O-Campa and Díaz-Bautista, 2024].

En este capítulo se describirá el Hamiltoniano efectivo para un material anisótropo de Dirac en presencia de campos magnéticos y eléctricos externos, considerando que la intensidad de estos varía en la dirección $x$ y empleando el *gauge* de Landau que garantice que $[H, p_y] = 0$. Asumiendo, por lo tanto, una solución de onda plana en la dirección $y$, se resolverá la ecuación de eigenvalores que surge para la función de onda de dos componentes $\Psi(x, y) = (\phi_1(x, y), \phi_2(x, y))^{\mathrm{T}}$, donde el superíndice T denota la transpuesta de la matriz. Posteriormente, se estudiarán las soluciones para dos tipos de perfiles de campos magnéticos y eléctricos: el primero con un decaimiento exponencial $e^{-\alpha x}$, y el segundo de la forma $1/x^2$.

Los campos externos con perfil exponencial, $\mathbf{E}, \mathbf{B} \sim e^{-\alpha x}$, representan interacciones que se atenúan rápidamente a medida que aumenta la distancia desde su fuente. Este comportamiento es característico de sistemas donde existe *apantallamiento*, como materiales conductores o medios dieléctricos con cargas libres, que reorganizan su distribución para neutralizar parcialmente el campo. Físicamente, un campo de este tipo restringe la influencia de la fuente a una región limitada, generando potenciales efectivos que pueden inducir la *localización de los estados electrónicos* en materiales tipo Dirac. Esto se puede observar en el comportamiento de la densidades de probabilidad y de corriente en la región en donde el campo es significativo.

Por otro lado, los campos externos con un perfil de tipo singular, $\mathbf{E}, \mathbf{B} \sim 1/x^2$, describen interacciones de largo alcance que disminuyen lentamente con la distancia y que pueden interpretarse como aquellas generadas por *impurezas puntuales* o defectos dentro del material que actúan como centros de dispersión efectivos. Este tipo de campos permite que la influencia de las imperfecciones se extienda a escalas espaciales mayores, afectando la dinámica de los portadores de carga y modificando el comportamiento de las densidades de probabilidad y de corriente de los estados electrónicos.

## 3.1 │ **Material anisótropo de Dirac**

Como se mostró en la Sección 2.2, existe una nueva familia de materiales bidimensionales cuyas propiedades electrónicas difieren notablemente de los sistemas convencionales. En estos materiales, con conos de Dirac en su estructura de bandas, los electrones se comportan como si no tuvieran masa, moviéndose a velocidades efectivas del orden de $c/300$, siendo $c$ la velocidad de la luz en el vacío.

Este comportamiento se manifiesta claramente a bajas energías, cerca del nivel de Fermi, en donde la dinámica de los electrones puede describirse mediante el siguiente Hamiltoniano [Champo and Naumis, 2019]

$$H(\mathbf{k}) = \nu \left( v_x \sigma_x p_x + v_y \sigma_y p_y + v_t \sigma_0 p_y \right), \tag{3.1.1}$$

siendo $\sigma_0$ la matriz identidad $2 \times 2$, $\sigma_x, \sigma_y$ son las matrices de Pauli, que actúan sobre el grado libertad de subred, los términos $v_x$ y $v_y$ son las velocidades de Fermi que están relacionadas con la excentricidad de los conos de Dirac en la red recíproca, y $\nu = \pm 1$ es el índice de valle ($\nu = 1$, hace referencia al valle $K_+$, mientras que $\nu = -1$, al valle $K_-$). La inclinación de los conos de Dirac está codificada en el tercer término en el Hamiltoniano que es proporcional a $\sigma_0$, y que está caracterizada por la velocidad $v_t$. Nótese que si se elige $v_x = v_y = v_F$ y $v_t = 0$, el Hamiltoniano coincide con el del grafeno en la ecuación (2.1.37).

En presencia de un campo magnético $\mathbf{B}$ externo perpendicular a la superficie del material y un campo eléctrico $\mathbf{E}$ externo perpendicular al magnético, y considerando la regla de acoplamiento mínimo, el Hamiltoniano (3.1.1) se puede expresar, en unidades naturales ($\hbar = e = 1$), como:

$$H = \nu \left\{ v_x \left[ p_x + A_x(x,y) \right] \sigma_x + v_y \left[ p_y + A_y(x,y) \right] \sigma_y + v_t \left[ p_y + A_y(x,y) \right] \sigma_0 \right\} - \phi(x)\sigma_0, \tag{3.1.2}$$

donde se ha considerado que la superficie del material es paralelo al plano $xy$. Además, $\vec{A}(x,y) = A_x(x,y)\hat{e}_x + A_y(x,y)\hat{e}_y$ y $\phi(x,y)$ son los potenciales vectorial y escalar, respectivamente, con los cuales se describen el campo magnético y el campo eléctrico:

$$\mathbf{B} = \nabla \times \mathbf{A}(x,y), \tag{3.1.3}$$

$$\mathbf{E} = -\nabla \phi(x,y). \tag{3.1.4}$$

Para continuar con el análisis, las posiciones y momentos se pueden expresar de la siguiente manera, manteniendo los conmutadores canónicos invariantes:

$$x_c = \sqrt{\frac{v_y}{v_x}}x, \quad p_x^c = \sqrt{\frac{v_x}{v_y}}p_x, \quad y_c = \sqrt{\frac{v_x}{v_y}}y, \quad p_y^c = \sqrt{\frac{v_y}{v_x}}p_y. \tag{3.1.5}$$

Así, los potenciales vectorial y escalar se escriben como

$$A_x^c(x_c, y_c) = \sqrt{\frac{v_x}{v_y}}A_x(x,y), \quad A_y^c(x_c, y_c) = \sqrt{\frac{v_y}{v_x}}A_y(x,y), \quad \phi^c(x_c, y_c) = \sqrt{\frac{v_y}{v_x}}\phi(x,y). \tag{3.1.6}$$

Combinando las expresiones (3.1.5) y (3.1.6) con (3.1.2), el Hamiltoniano efectivo toma la siguiente forma:

$$H = \nu\sqrt{v_x v_y}\left\{ \boldsymbol{\sigma} \cdot \left[ \mathbf{p}^c + \mathbf{A}^c(x_c, y_c) \right] + \frac{v_t}{v_y}\left[ p_y^c + A_y^c(x_c, y_c) \right] - \frac{\nu}{v_t}\phi^c(x_c, y_c)\sigma_0 \right\}. \tag{3.1.7}$$

Considerando estados estacionarios, la ecuación a resolver es

$$[H - E\sigma_0]\Psi(x_c, y_c) = 0. \tag{3.1.8}$$

Donde $E$ es la energía física del sistema

Sustituyendo la propuesta matricial de $\Psi$ en (3.1.7), se obtienen las siguientes ecuaciones:

$$\left\{ \frac{v_t}{v_y} \left[ p_y^c + A_y^c - \phi^c \right] \right\} \phi_1 + \left\{ v_x \left[ p_x^c + A_x^c \right] - iv_y \left[ p_y^c + A_y^c \right] \right\} \phi_2 = E\phi_1, \tag{3.1.9}$$

$$\left\{ \frac{v_t}{v_y} \left[ p_y^c + A_y^c - \phi^c \right] \right\} \phi_2 + \left\{ v_x \left[ p_x^c + A_x^c \right] - iv_y \left[ p_y^c + A_y^c \right] \right\} \phi_1 = E\phi_2. \tag{3.1.10}$$

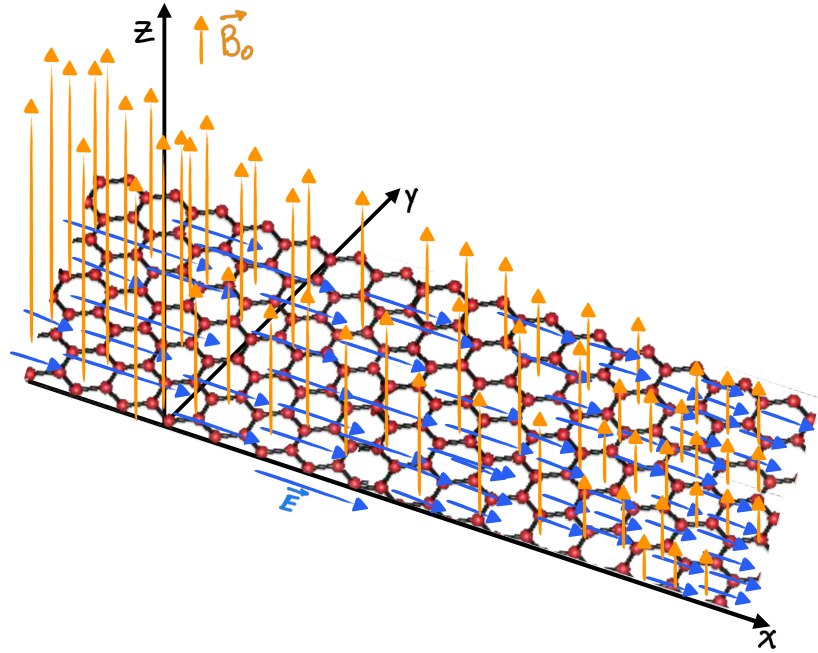

**Figura 3.1.1:** Monocapa de un material de Dirac (red hexagonal) interactuando con un campo eléctrico **E** y un campo magnético **B** externos. Ambos campos externos poseen simetría traslacional en la coordenada $y$.

Se puede observar que las ecuaciones (3.1.9) y (3.1.10) están acopladas, por lo que resolverlas puede ser una tarea no trivial. Una forma de encontrar su solución es asumir que los campos externos eléctrico y magnético poseen simetría traslacional en una dirección particular, por ejemplo, en la dirección del eje $y$, con lo cual las intensidades de los campos son función únicamente de la coordenada $x$. Bajo estas consideraciones y sin pérdida de generalidad, el potencial vectorial en la norma de Landau puede expresarse como $\mathbf{A} = A_y^c(x_c)\hat{e}_y$, con lo cual, el campo magnético toma la siguiente forma $\mathbf{B} = \frac{\mathrm{d}A_y^c}{\mathrm{d}x_c}\hat{e}_z$, mientras que el potencial escalar se puede escribir como $\phi^c = \phi(x_c)$, y esto implica que el campo eléctrico sea $\mathbf{E} = -\frac{\mathrm{d}\phi^c}{\mathrm{d}x_c}\hat{e}_x$. La Figura 3.1.1 muestra una representación esquemática de un material de Dirac en presencia de un campo eléctrico y magnético externos.

Para campos eléctricos y magnéticos cuya dependencia es función únicamente de la coordenada $x$, se puede verificar que $[H, p_y^c] = 0$, por lo cual, las eigenfunciones pueden escribirse como:

$$\Psi(x, y) = \varphi(x_c)\mathrm{e}^{ik_y^c y_c}, \tag{3.1.11}$$

donde $k_y^c = \left(\sqrt{v_y/v_x}\right)k_y$ y $k_y$ es el número de onda en la dirección $y$, mientras que la función $\varphi(x_c)$ se define como

$$\varphi(x_c) = [\psi_1(x_c), i\,\psi_2(x_c)]^{\mathrm{T}}. \tag{3.1.12}$$

Expresando el operador de momento $p_y^c$ en la base de coordenadas y además usando (3.1.11) en (3.1.8), se tiene

$$\left\{-i\frac{\mathrm{d}}{\mathrm{d}x_c}\sigma_x + \left[k_y^c + A_y^c(x_c)\right]\sigma_y + \left[\frac{v_t}{v_y}A_y^c(x_c) - \frac{\nu}{v_y}\phi^c(x_c) - \bar{E}\right]\sigma_0\right\}\varphi(x_c) = 0, \tag{3.1.13}$$

con $\bar{E} = (\nu E - v_t k_y)/\sqrt{v_x v_y}$.

En las siguientes secciones se considerarán algunos perfiles específicos de campos externos eléctrico y magnético en (3.1.13) para poder hallar las funciones de onda y las energías que describen a los sistemas físicos correspondientes.

## 3.2 | Campos externos con perfil exponencial decreciente

Como primer caso, considérese un par de campos eléctrico y magnético con perfiles que decaen exponencialmente

$$\mathbf{E}(x) = \mathcal{E}_0 \, \mathrm{e}^{-\alpha x} \, \hat{e}_x, \tag{3.2.1}$$

$$\mathbf{B}(x) = \mathcal{B}_0 \, \mathrm{e}^{-\alpha x} \, \hat{e}_z, \tag{3.2.2}$$

donde $\mathcal{B}_0$ y $\mathcal{E}_0$ representan las intensidades de los campos magnético y eléctrico, respectivamente, mientras que $\alpha$ indica la longitud de penetración en la cual los campos externos son relevantes. El potencial vectorial y el potencial escalar para este caso son, respectivamente:

$$\mathbf{A}_y(x) = \frac{-\mathcal{B}_0}{\alpha} \left[ \mathrm{e}^{-\alpha x} - 1 \right] \hat{e}_y, \tag{3.2.3}$$

$$\phi(x) = \frac{\mathcal{E}_0}{\alpha} \left[ \mathrm{e}^{-\alpha x} - 1 \right]. \tag{3.2.4}$$

Aquí se puede ver que ambos campos tienen el mismo perfil de decaimiento exponencial. La forma de escribir los potenciales (3.2.3) y (3.2.4) permite recuperar el caso en el cual los campos magnético y eléctrico son constantes, tomando el límite $\alpha \to 0$ [Betancur-Ocampo et al., 2022, Díaz-Bautista, 2022], lo cual permite comparar con lo ya propuesto en la literatura como un caso limite. Este hecho no solo valida la consistencia del modelo propuesto con resultados bien establecidos en la literatura, sino que también resalta su carácter más general, al incluir como caso límite una situación ampliamente estudiada.

Sustituyendo los potenciales (3.2.3) y (3.2.4) en (3.1.13), y realizando el cambio de variable

$$z(x_c) = \frac{2D\sqrt{1-\beta_\nu^2}}{\alpha_c} \, \mathrm{e}^{-\alpha_c \, x_c}, \quad \text{con } z \in [0, \infty). \tag{3.2.5}$$

donde

$$\alpha_c = \sqrt{\frac{v_x}{v_y}}\alpha, \quad D = \frac{\mathcal{B}_0}{\alpha_c}, \quad \beta_\nu = \nu\frac{v_t}{v_y} + \frac{v_\mathrm{d}}{v_y}, \quad v_\mathrm{d} = \frac{\mathcal{E}_0}{\mathcal{B}_0}, \tag{3.2.6}$$

con $v_{\mathrm{d}}$ la velocidad de deriva, la ecuación (3.1.13) adopta la siguiente forma:

$$\left[ i\alpha_c z \frac{\mathrm{d}}{\mathrm{d}z}\sigma_x + \left( D + k_y^c - \frac{\alpha^c}{2\sqrt{1-\beta_\nu^2}}\,z \right)\sigma_y - \left( \bar{E} - \nu\beta_\nu D + \frac{\nu\beta_\nu\alpha^c}{2\sqrt{1-\beta_\nu^2}}z \right)\sigma_0 \right]\varphi(z) = 0.$$

$$(3.2.7)$$

Multiplicando a la izquierda por $-i\alpha_c\sigma_x$ y aplicando el operador diferencial $z\frac{\mathrm{d}}{\mathrm{d}z}$ a la expresión resultante, se obtiene:

$$z^2\frac{\mathrm{d}^2\varphi(z)}{\mathrm{d}z^2} + \left[ -\frac{z^2}{4} + z\mu + (\mathbb{K} - \lambda^2) \right]\varphi(z) = 0, \qquad (3.2.8)$$

donde $\mu$, $\lambda$ y $\mathbb{K}$ están dadas por

$$\mu = \frac{1}{\sqrt{1-\beta_\nu^2}}\left[ \left( \frac{\bar{E}}{\alpha_c} - \frac{\nu\beta_\nu D}{\alpha_c} \right)\nu\beta_\nu + \left( \frac{k_y^c}{\alpha_c} + \frac{D}{\alpha_c} \right) \right], \qquad (3.2.9)$$

$$\lambda = \frac{1}{\alpha_c}\sqrt{(k_y^c + D)^2 - (\bar{E} - \nu\beta_\nu D)^2}, \qquad (3.2.10)$$

$$\mathbb{K} = \frac{1}{\alpha_c}\begin{pmatrix} -(k_y^c + D) & -i(\bar{E} - \nu\beta_\nu D) \\ -i(\bar{E} - \nu\beta_\nu D) & k_y^c + D \end{pmatrix}. \qquad (3.2.11)$$

Para resolver (3.2.8), es necesario expresar la función de onda $\varphi(z)$ en la base donde la matriz $\mathbb{K}_{2\times 2}$ se diagonaliza, lo cual equivale a realizar un cambio de variables hacia los eigenvectores de la matriz $\mathbb{K}_{2\times 2}$. De esta manera, las componentes de $\varphi(z)$ dejan de estar acopladas y el problema se reduce a ecuaciones independientes asociadas a cada eigenvalor. Para llevar a cabo este proceso, resulta necesario determinar los eigenvectores $\chi_j$ y los correspondientes eigenvalores $\lambda_j$, con $j = 1, 2$, que caracterizan los modos propios del sistema.

Para este caso, los eigenvectores $\chi_1$ y $\chi_2$ están dados por

$$\chi_1 = N_1\begin{pmatrix} \frac{\bar{E}-\nu\beta\nu D}{\lambda+(k_y^c+D)} \\ i \end{pmatrix}, \qquad (3.2.12)$$

$$\chi_2 = N_2\begin{pmatrix} -i \\ \frac{\bar{E}-\nu\beta\nu D}{\lambda+(k_y^c+D)} \end{pmatrix}, \qquad (3.2.13)$$

donde $N_{1,2}$ son las constantes de normalización. En tanto, los eigenvalores toman la siguiente forma

$$\lambda_j = (-1)^{j+1}\lambda, \quad j = 1, 2. \qquad (3.2.14)$$

Por lo tanto, la ecuación diferencial (3.2.8) en la base $\chi_j$ puede escribirse como

$$\frac{\mathrm{d}^2\psi_j(z)}{\mathrm{d}z^2} + \left[-\frac{1}{4} + \frac{\mu}{z} + \frac{1/4 - m_j^2}{z^2}\right]\psi_j(z) = 0, \quad j = 1, 2. \tag{3.2.15}$$

Aquí $m_j$ se define como

$$m_j = \lambda + \frac{(-1)^j}{2}, \quad j = 1, 2. \tag{3.2.16}$$

La ecuación (3.2.15) es la ecuación de Whittaker [Whittaker, 1903], cuya solución es

$$\psi_j(z) = C_1 W_{\mu,m_j}(z) + C_2 M_{\mu,m_j}(z), \tag{3.2.17}$$

donde $C_{1,2}$ son constantes. En general, $M_{\mu,m_j}(z)$ y $W_{\mu,m_j}(z)$ son funciones multivaluadas en $z$ con puntos de ramificación en $z = 0$ y $z = \infty$. Cabe señalar que cuando $2m_j = -1, -2, -3, \ldots$, la función $M_{\mu,m_j}(z)$ no existe.

El comportamiento asintótico de las funciones $M_{\mu,m_j}(z)$ y $W_{\mu,m_j}(z)$ cuando $z \to 0$ es

$$M_{\mu,m_j}(z) = z^{m_j + \frac{1}{2}}(1 + O(z)), \quad 2m_j \neq -1, -2, -3, \ldots \tag{3.2.18}$$

$$W_{\mu,m_j}(z) = \begin{cases} (-1)^n (1 \pm 2m_j)_n z^{\frac{1}{2} \pm m_j} + O\left(z^{\frac{3}{2} \pm m_j}\right), & \frac{1}{2} - \mu \pm m_j = -n, \quad n + 1 \in \mathbb{N}, \\ \frac{\Gamma(2m_j)}{\Gamma(\frac{1}{2} + m_j - \mu)} z^{\frac{1}{2} - m_j} + O\left(z^{\frac{3}{2} - \mathrm{Re}(m_j)}\right), & \mathrm{Re}(m_j) \geq \frac{1}{2}, \quad m_j \neq \frac{1}{2}, \end{cases} \tag{3.2.19}$$

donde $(z)_n$ denota al símbolo de Pochhammer.

Ahora, cuando $z \to \infty$, las funciones $M_{\mu,m_j}(z)$ y $W_{\mu,m_j}(z)$ admiten el siguiente comportamiento

$$M_{\mu,m_j}(z) \sim \frac{\Gamma(1 + 2m_j)\mathrm{e}^{\frac{1}{2}z}z^{-\mu}}{\Gamma(\frac{1}{2} + m_j - \mu)}, \quad |\mathrm{ph}\, z| \leq \frac{1}{2}\pi - \delta', \quad m_j - \mu = -\frac{1}{2}, -\frac{3}{2}, \ldots \tag{3.2.20}$$

$$W_{\kappa,\mu}(z) \sim \mathrm{e}^{-\frac{1}{2}z}z^{\mu}, \quad |\mathrm{ph}\, z| \leq \frac{3}{2}\pi - \delta', \tag{3.2.21}$$

donde $\delta'$ es una constante positiva arbitrariamente pequeña. Para una mayor referencia de las funciones $M_{\mu,m_j}(z)$ y $W_{\mu,m_j}(z)$, véanse sus propiedades básicas en la Sección 13.14 de [Olver et al., 2010].

Para satisfacer la condición de frontera de Dirichlet, que establece que la función de onda (3.2.17) debe anularse en el límite $z \to \infty$, se impone que $C_2 = 0$. Asimismo, con el fin de asegurar un comportamiento físico aceptable, es decir, que la función de onda sea cuadrado–integrable y, por tanto, se excluyan automáticamente los términos divergentes de la función de onda cuando $z \to 0$, de (3.2.19) se impone la siguiente restricción.

$$\mu - m_j - \frac{1}{2} = n, \quad n+1 \in \mathbb{N} \tag{3.2.22}$$

y de la cual se deduce el espectro de energía:

$$E_n = v_F \beta_\nu \alpha_c n \sqrt{1-\beta_\nu^2} - v_d k_y + \kappa \nu v_F \sqrt{1-\beta_\nu^2} \sqrt{\left(k_y^c + D\right)^2 - \left(k_y^c + D - \alpha_c n \sqrt{1-\beta_\nu^2}\right)^2}, \tag{3.2.23}$$

donde $\kappa = +1$ ($\kappa = -1$) corresponde a la banda de conducción (valencia), mientras que $v_F = \sqrt{v_x v_y}$ (para más detalles véase [Mojica-Zárate et al., 2024]). La expresión (3.2.23) muestra una dependencia explícita de la anisotropía del material, lo cual implica que considera tanto la estructura de la red cristalina como las posibles deformaciones mecánicas aplicadas a la muestra.

Asimismo, bajo la condición (3.2.22), y de acuerdo con la ecuación 13.14.3 de [Olver et al., 2010], la función $W_{\mu,m_j}(z)$ se expresa como

$$W_{\mu=\frac{1}{2}+m_j+n,m_j}(z) \propto \mathrm{e}^{-z/2}\, z^{m_j+1/2} U(-n, 1+2m_j; z). \tag{3.2.24}$$

Aquí, la función de Kummer $U(-n, b; z)$, para $n$ entero no negativo, está relacionada con los polinomios asociados de Laguerre $L_n^k(z)$. En particular, a partir de la referencia [Olver et al., 2010], se tiene que

$$U(-n, 1+2m_j; z) \propto L_n^{2m_j}(z). \tag{3.2.25}$$

Así, las soluciones $\psi_{j,n}(z)$ en (3.2.17), se pueden escribir como

$$\psi_{j,n}(z) \propto \mathrm{e}^{-z/2}\, z^{m_j+1/2} L_n^{2m_j}(z). \tag{3.2.26}$$

Usando (3.2.16) en (3.2.26), las funciones de onda $\psi_{1,n}(z)$ y $\psi_{2,n}(z)$ del sistema están dadas por

$$\psi_{1,n}(z) = \mathcal{N}_1\, \mathrm{e}^{-z/2} z^{\lambda_n} L_n^{2\lambda_n-1}\left(z\right), \tag{3.2.27}$$

$$\psi_{2,n}(z) = \mathcal{N}_2\, \mathrm{e}^{-z/2} z^{\lambda_{n+1}+1} L_n^{2\lambda_{n+1}+1}\left(z\right), \; n = 0,1 ..., \tag{3.2.28}$$

donde

$$\lambda_n = \frac{1}{\alpha_c}\sqrt{(k_y^c + D)^2 - (\bar{E}_n - \nu\beta_\nu D)^2}, \tag{3.2.29}$$

$\bar{E}_n = (\nu E_n - v_t k_y)/\sqrt{v_x v_y}$, $\mathcal{N}_1$ y $\mathcal{N}_2$ son constantes de normalización:

$$\mathcal{N}_1 = \sqrt{\frac{\alpha\, n!}{\Gamma(2\lambda_n + n)}}, \tag{3.2.30}$$

$$\mathcal{N}_2 = \sqrt{\frac{\alpha\, n!}{\Gamma(2\lambda_{n+1} + n + 2)}}, \quad n = 0, 1 \dots. \tag{3.2.31}$$

Al sustituir la expresión de los pseudo-espinores dada en (3.1.12), junto con las relaciones (3.2.12) y (3.2.13), la solución en (3.1.11) se reescribe de la siguiente forma:

$$\Psi_n(x,y) = \mathcal{N}\exp\big(ik_y^c y\big)\left[C_1'\,\chi_1\psi_{1,n}(z) + iC_2'\,\chi_2\psi_{2,n-1}(z)\right],$$

$$= \mathcal{N}\exp\big(ik_y^c y\big)\begin{pmatrix} \frac{\bar{E}_n - \nu\beta\nu D}{\lambda + (k_y^c + D)}\psi_{1,n}(z) + (1 - \delta_{0n})\,\psi_{2,n-1}(z) \\ i\left[\psi_{1,n}(z) + \frac{\bar{E}_n - \nu\beta\nu D}{\lambda + (k_y^c + D)}(1 - \delta_{0n})\,\psi_{2,n-1}(z)\right] \end{pmatrix}, \tag{3.2.32}$$

$$= \mathcal{N}\exp\big(ik_y^c y\big)\begin{pmatrix} \frac{\bar{E}_n - \nu\beta\nu D}{\lambda + (k_y^c + D)} & -i \\ i & \frac{\bar{E}_n - \nu\beta\nu D}{\lambda + (k_y^c + D)} \end{pmatrix}\begin{pmatrix} \psi_{1,n}(z) \\ i(1 - \delta_{0n})\,\psi_{2,n-1}(z) \end{pmatrix}, \tag{3.2.33}$$

siendo $\mathcal{N}$ la constante de normalización del espinor dada por:

$$\mathcal{N} = \sqrt{\frac{\lambda + (k_y^c + D)}{2^{2-\delta_{0n}}\left[(D + k_y^c) + (\bar{E} - \nu\beta_\nu D)\,I_n\right]}}, \tag{3.2.34}$$

donde

$$I_n = 2^{\delta_{0n}}\sqrt{\frac{v_x}{v_y}}\,(1 - \delta_{0n})\int_{-\infty}^{\infty}\psi_{1,n}(x)\,\psi_{2,n-1}(x)\,\mathrm{d}x, \tag{3.2.35}$$

y se han elegido $C_1' = 1$ y $C_2' = 1 - \delta_0 n$.

Una cantidad física de gran relevancia en el estudio de estos materiales es la *densidad de probabilidad*, así como la *densidad de corriente de probabilidad*. Particularmente, la densidad de corriente proporciona información crucial sobre los mecanismos de transporte electrónico en el sistema. Para los Hamiltonianos descritos en las ecuaciones (3.1.1) y (3.1.2), ambas densidades se definen, respectivamente, como

$$\rho(x,y,t) = \Psi^\dagger\Psi, \quad \vec{\mathcal{J}}(x,y,t) = \Psi^\dagger\,\vec{j}\,\Psi, \tag{3.2.36}$$

donde el operador de densidad de corriente está dado por las siguientes componentes:

$$j_x = \nu v_x \sigma_x, \quad j_y = \nu v_y \sigma_y + v_t \sigma_0. \tag{3.2.37}$$

A partir de las expresiones anteriores, y considerando los pseudo-espinores asociados a los eigenvalores de energía del sistema, es posible obtener una formulación explícita para las componentes de la densidad de corriente de probabilidad. Para el $n$-ésimo estado, la densidad de probabilidad toma la forma

$$\rho_n(x) = \frac{\left[\left(D + k_y^c\right)|\Psi_n(z)|^2 + 2^{\delta_{0n}}(1 - \delta_{0n})\left(\bar{E} - \nu\beta_\nu D\right)\psi_{1,n}(z)\,\psi_{2,n-1}(z)\right]}{\left[\left(D + k_y^c\right) + \left(\bar{E} - \nu\beta_\nu D\right)I_n\right]}. \tag{3.2.38}$$

Además, la componente en la dirección $x$ se expresa como

$$\mathcal{J}_{x,n}(x,y,t) = -\frac{\nu\,v_x\lambda\;\Psi_n^\dagger(z)\sigma_x\Psi_n(z)}{\left[\left(D + k_y^c\right) + \left(\bar{E} - \nu\beta_\nu D\right)I_n\right]} = 0, \tag{3.2.39}$$

mientras que la componente en la dirección $y$ está dada por

$$\begin{aligned}
\mathcal{J}_{y,n}(x,y,t) &= \frac{\nu\left[\left(\bar{E} - \nu\beta_\nu D\right)v_y + \left(D + k_y^c\right)v_t\right]|\Psi_n(z)|^2}{\left[\left(D + k_y^c\right) + \left(\bar{E} - \nu\beta_\nu D\right)I_n\right]} \\
&+ \frac{2^{\delta_{0n}}(1 - \delta_{0n})\left[\left(\bar{E} - \nu\beta_\nu D\right)v_t + \left(D + k_y^c\right)v_y\right]\psi_{1,n}(z)\psi_{2,n-1}(z)}{\left[\left(D + k_y^c\right) + \left(\bar{E} - \nu\beta_\nu D\right)I_n\right]}.
\end{aligned} \tag{3.2.40}$$

Ambas densidades muestran cómo las contribuciones, ya sean provenientes del acoplamiento espinorial o del término de inclinación $v_t$, afectan la dinámica de los portadores de carga a tiempo fijo en cada dirección, dependiendo del eigenestado $n$, el parámetro de inclinación $\beta_\nu$, y las condiciones externas en los diferentes parámetros del material, como $\mathcal{B}_0$ y $\mathcal{E}_0$.

**Discusión**

Las Figuras 3.2.1 y 3.2.2 muestran el comportamiento del espectro de energía con respecto a la intensidad del campo eléctrico para los casos del grafeno prístino, el grafeno sometido a tensión uniaxial y el borofeno $8 - Pmmn$, respectivamente. Se elige estudiar estos materiales porque el Hamiltoniano que los describe corresponde al dado en la ecuación (3.1.1).

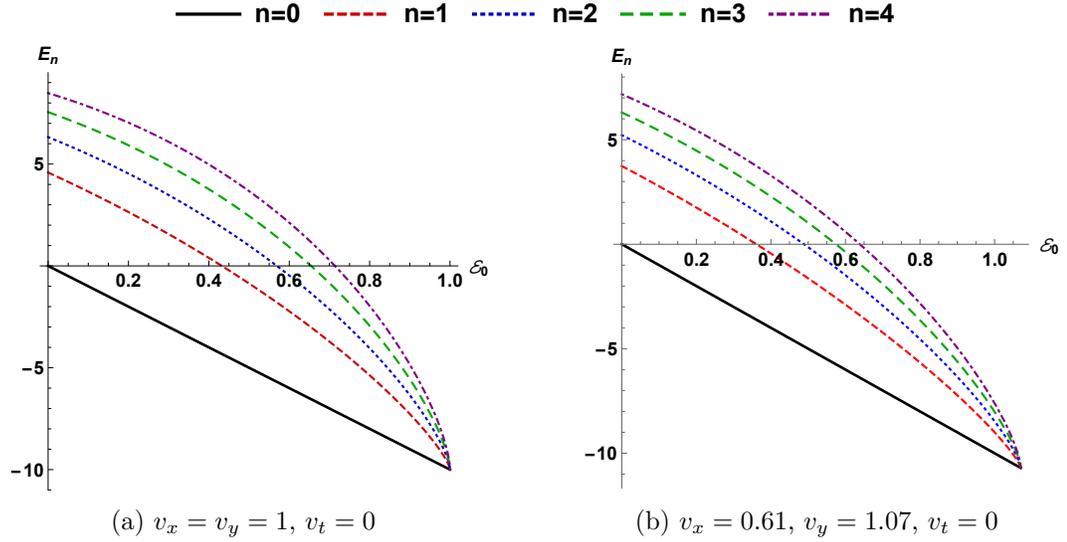

(a) $v_x = v_y = 1$, $v_t = 0$                 (b) $v_x = 0.61$, $v_y = 1.07$, $v_t = 0$

**Figura 3.2.1:** Espectro de energía $E_n$ en (3.2.23) caso exponencial con $\mathcal{B}_0 = 1$, $\alpha = 1$ y $k_y = 10$ en función del campo eléctrico $\mathcal{E}_0$ para (a) el grafeno prístino y (b) el grafeno bajo tensión a lo largo del eje $x$ con $\varepsilon = 0.15$.

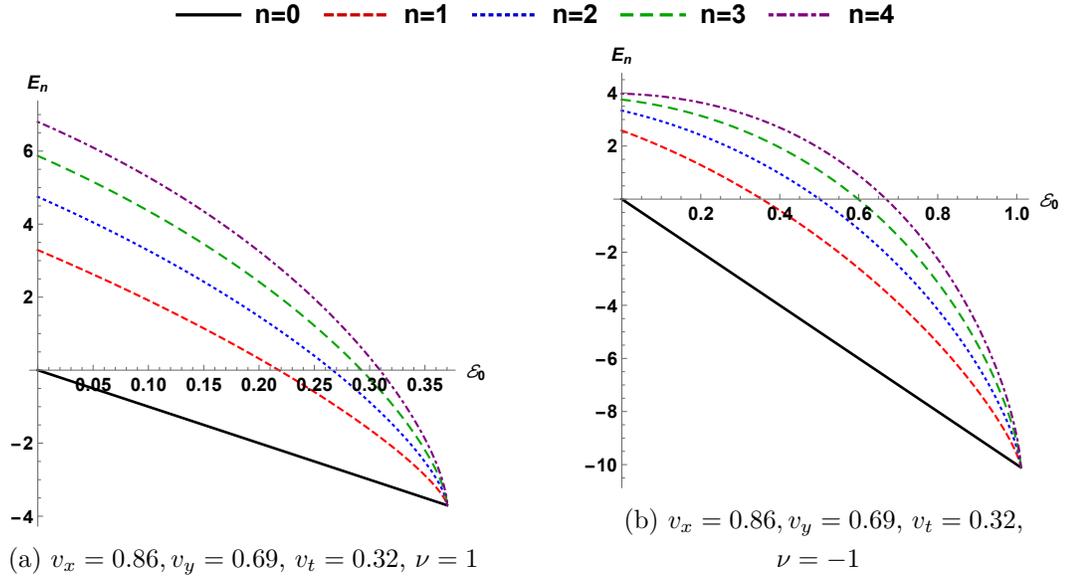

(a) $v_x = 0.86$, $v_y = 0.69$, $v_t = 0.32$, $\nu = 1$     (b) $v_x = 0.86$, $v_y = 0.69$, $v_t = 0.32$, $\nu = -1$

**Figura 3.2.2:** Espectro de energía $E_n$ en (3.2.23) caso exponencial con $\mathcal{B}_0 = 1$, $\alpha = 1$ y $k_y = 10$ en función del campo eléctrico $\mathcal{E}_0$ para el borofeno $8 - Pmmn$ en (a) el valle $K_+$ y (b) el valle $K_-$.

Este modelo permite analizar tres casos principales en relación con la anisotropía de los conos de Dirac, según los parámetros $\nu$, $v_x$, $v_y$ y $v_t$. En el caso del grafeno prístino, los conos de Dirac son idénticos en ambas bandas y no presentan inclinación, dado que $v_x = v_y$ y $v_t = 0$. Para el grafeno sometido a tensión uniaxial, los conos presentan una excentricidad distinta de cero; sin embargo, al igual que en el caso prístino, no existe inclinación en los conos ($v_x \neq v_y$ y $v_t = 0$).

Se puede observar que, en general, si la intensidad del campo eléctrico aumenta, habrá un valor crítico del campo eléctrico $\mathcal{E}_c$ dado por

$$\mathcal{E}_c = \mathcal{B}_0(v_y - \nu \, v_t), \tag{3.2.41}$$

para el cual los niveles de Landau en (3.2.23) colapsan. También de (3.2.23), se tiene que la energía del estado fundamental es $E_0 = -v_{\text{d},c}k_y$, y es el nivel de energía al cual colapsan el resto de niveles de Landau cuando la velocidad de deriva es igual a

$$v_{\text{d},c} = \frac{\mathcal{E}_c}{\mathcal{B}_0} = v_y - \nu \, v_t. \tag{3.2.42}$$

En términos físicos, el colapso de los niveles de Landau significa que al alcanzar el campo eléctrico crítico $\mathcal{E}_c$, la separación discreta entre niveles desaparece y todos se reducen a un único valor de energía, de modo que el espectro deja de estar cuantizado y pasa a ser continuo. En materiales de Dirac, el colapso de los niveles de Landau refleja la naturaleza relativista de sus portadores debido a su dispersión lineal y a la presencia de un nivel de Landau especial ($n = 0$) que no corresponde al estado fundamental en sentido convencional, sino que separa la parte de la banda de conducción y de valencia del espectro. Este fenómeno marca la transición hacia un régimen de transporte libre, lo que puede resultar relevante para la dinámica de los electrones y abrir posibles aplicaciones en nanoelectrónica 2D.

Los niveles de energía $E_n$ presentan una dependencia explícita con el parámetro $k_y$, y para ciertos valores de esta cantidad es posible la existencia de estados ligados. Específicamente, para cada valor de $k_y$, existe un número finito de niveles discretos permitidos, es decir, un número máximo de estados ligados. Este comportamiento se ilustra en las Figuras 3.2.3 y 3.2.4, en donde se muestra cómo varían los niveles de energía en función de $k_y$.

Además, se aprecia que los niveles discretos están contenidos dentro de una región delimitada por una curva envolvente, la cual actúa como una especie de límite superior. Esta línea envolvente indica el valor máximo de energía que pueden alcanzar los estados ligados; al superar este límite, los estados dejan de estar confinados y se

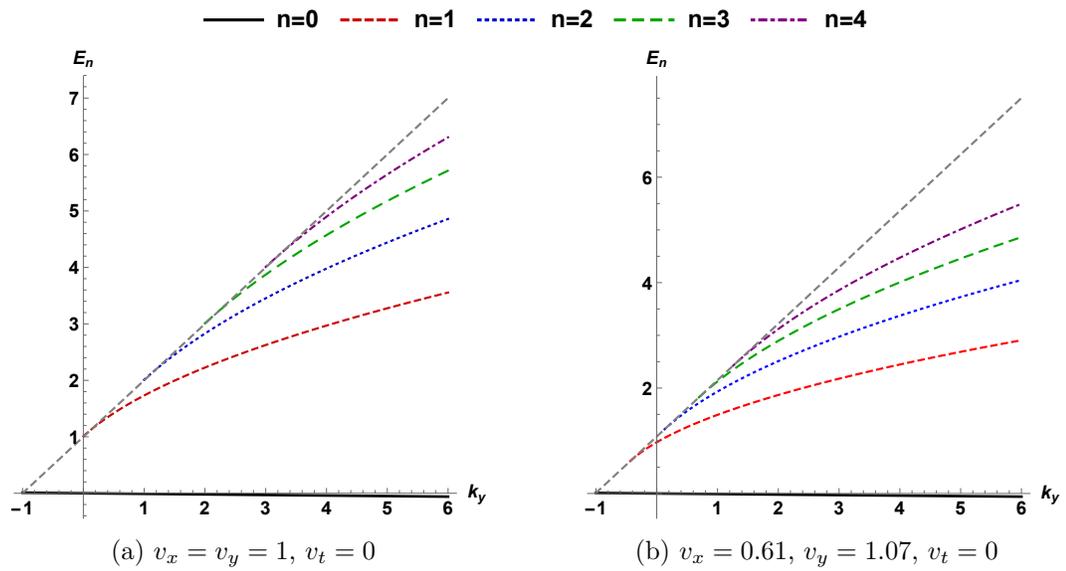

(a) $v_x = v_y = 1$, $v_t = 0$                    (b) $v_x = 0.61$, $v_y = 1.07$, $v_t = 0$

**Figura 3.2.3:** Espectro de energía $E_n$ en (3.2.23), caso exponencial, con $\mathcal{B}_0 = 1$, $\alpha = 1$ y $\mathcal{E}_0 = 0.01$ en función del momento $k_y$, para (a) el grafeno prístino y (b) el grafeno bajo tensión a lo largo del eje $x$ con $\varepsilon = 0.15$. La recta gris es la envolvente que indica el número de estados ligados.

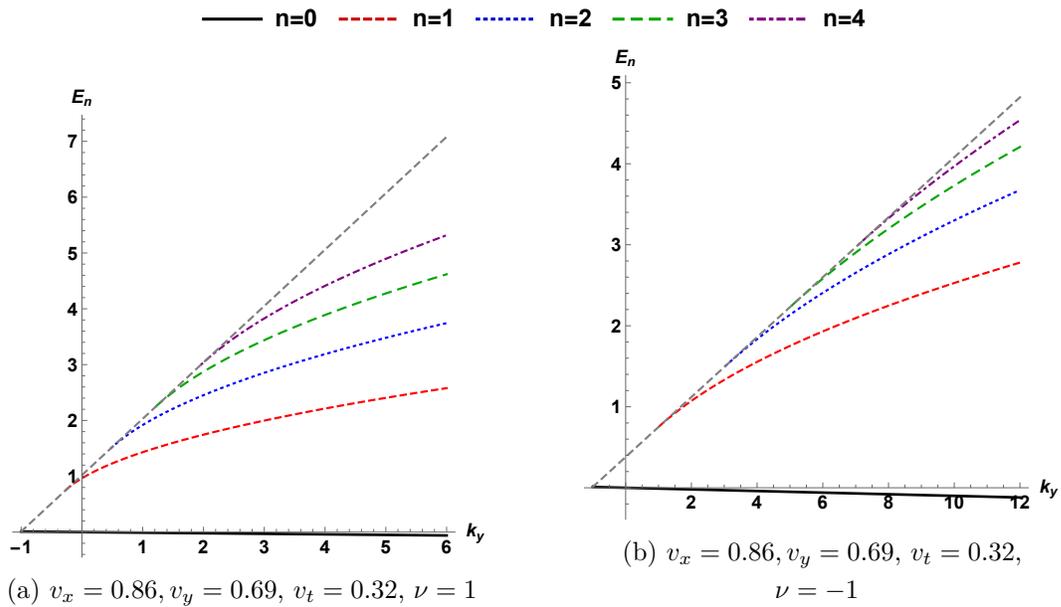

(a) $v_x = 0.86, v_y = 0.69, v_t = 0.32, \nu = 1$

(b) $v_x = 0.86, v_y = 0.69, v_t = 0.32,$
$\nu = -1$

**Figura 3.2.4:** Espectro de energía $E_n$ en (3.2.23), caso exponencial, con $\mathcal{B}_0 = 1$, $\alpha = 1$ y $\mathcal{E}_0 = 0.01$ en función del momento $k_y$, para el borofeno $8 - Pmmn$ en (a) al valle $K_+$ y (b) al valle $K_-$. La recta gris es la envolvente que indica el número de estados ligados.

comportan de manera dispersiva, propagándose libremente a través del material. La presencia de esta envolvente proporciona información clave sobre la estructura del espectro de $E_n$ y las condiciones necesarias para la existencia de estados ligados.

Por su parte, las densidades de probabilidad para los casos analizados se muestran en las Figuras 3.2.5-3.2.8. En los dos primeros escenarios (grafeno prístino y grafeno bajo tensión), la velocidad $v_x$ a lo largo del eje $x$ varía notablemente, lo que influye en la distribución espacial de los estados electrónicos. Para el caso del borofeno, tanto la densidad de probabilidad como la de corriente revelan diferencias marcadas entre los valles $\nu = \pm 1$, evidenciando una ruptura de simetría entre los puntos K$_+$ y K$_-$ en el espacio recíproco. Esta asimetría se traduce en una localización desigual de los estados en el espacio real, así como en direcciones opuestas del flujo de corriente, lo cual puede interpretarse como una forma de polarización de valle inducida por los campos externos.

Es importante señalar que aunque los puntos de Dirac K$_+$ y K$_-$ no aparecen como coordenadas directas en el espacio real, su influencia se manifiesta de forma clara en la estructura de las funciones de onda. En particular, los máximos en la densidad de probabilidad (véanse Figuras 3.2.5 y 3.2.6) indican regiones donde los estados electrónicos se localizan con mayor intensidad, lo cual está asociado a la proximidad de estos puntos críticos en el espacio recíproco y más evidente en el caso del borofeno. Más aún, los cambios de signo o las asimetrías observadas en la densidad de corriente (como se muestra en las Figuras 3.2.7 y 3.2.8) representan la respuesta del sistema a la estructura de valle, evidenciando indirectamente la distinción entre K$_+$ y K$_-$ en el espacio real. En este contexto, dichos puntos críticos funcionan como marcadores clave del comportamiento electrónico alrededor de cada valle, reflejándose en la distribución espacial de los estados cuánticos.

En síntesis, en materiales de Dirac como el grafeno, cuya red hexagonal impone una alta simetría, los estados en K$_+$ y K$_-$ son degenerados en ausencia de campos externos. No obstante, en materiales con mayor anisotropía estructural, como el borofeno, o en presencia de campos inhomogéneos, esta degeneración se rompe. Esto da lugar a una diferenciación efectiva entre los valles, lo que abre la posibilidad de controlar selectivamente estos grados de libertad. Esta propiedad resulta especialmente prometedora en el contexto de la electrónica de valles, donde se busca aprovechar dicha separación para el diseño de dispositivos electrónicos avanzados.

Ahora bien, desde un punto de vista dinámico, la velocidad $v_x$ a lo largo del eje

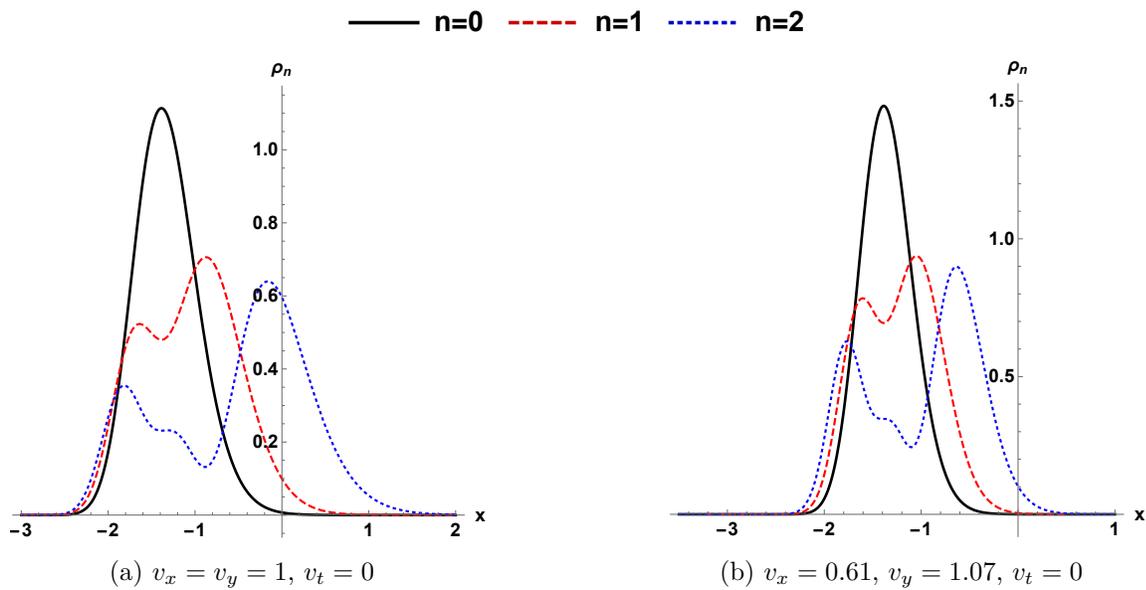

(a) $v_x = v_y = 1$, $v_t = 0$
(b) $v_x = 0.61$, $v_y = 1.07$, $v_t = 0$

**Figura 3.2.5:** Densidad de probabilidad $\rho_n(\mathbf{r})$ en el caso exponencial con $\mathcal{B}_0 = 1$, $\mathcal{E}_0 = 0.1$, $\alpha = 1$ y $k_y = 3$ a lo largo del eje $x$ para (a) el grafeno prístino y (b) el grafeno bajo tensión a lo largo del eje $x$ con $\varepsilon = 0.15$.

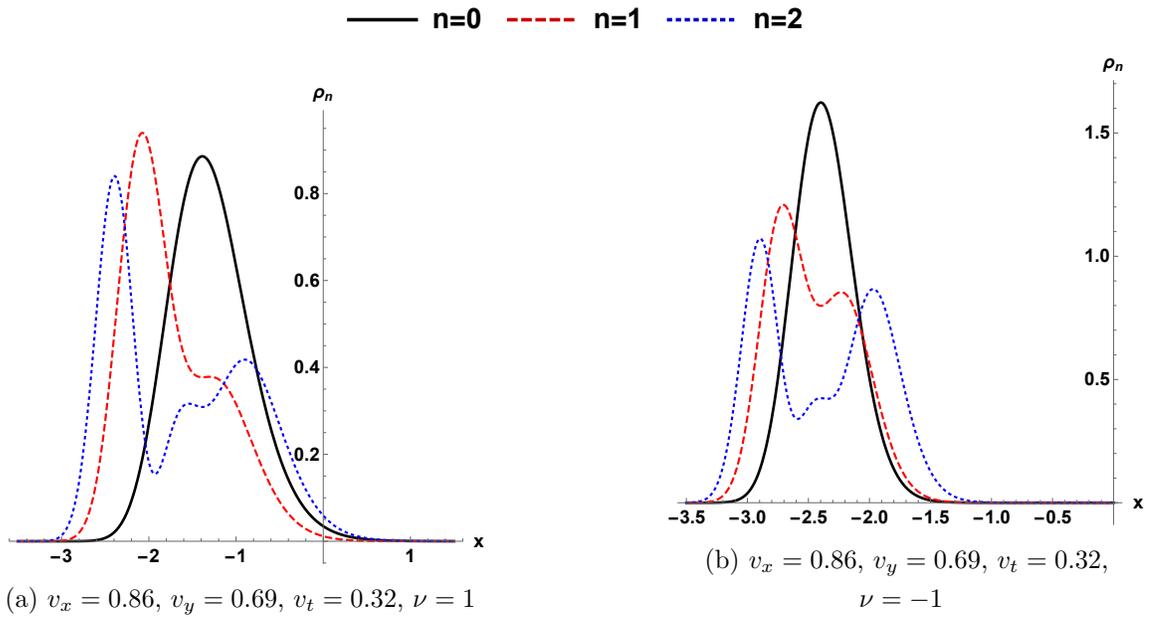

(a) $v_x = 0.86$, $v_y = 0.69$, $v_t = 0.32$, $\nu = 1$

(b) $v_x = 0.86$, $v_y = 0.69$, $v_t = 0.32$, $\nu = -1$

**Figura 3.2.6:** Densidad de probabilidad $\rho_n(\mathbf{r})$ en el caso exponencial con $\mathcal{B}_0 = 1$, $\mathcal{E}_0 = 0.1$, $\alpha = 1$ a lo largo del eje $x$ para el borofeno $8 - Pmmn$ con (a) $k_y = 3$ en el valle $K_+$, y con (b) $k_y = 8$ en el valle $K_-$.

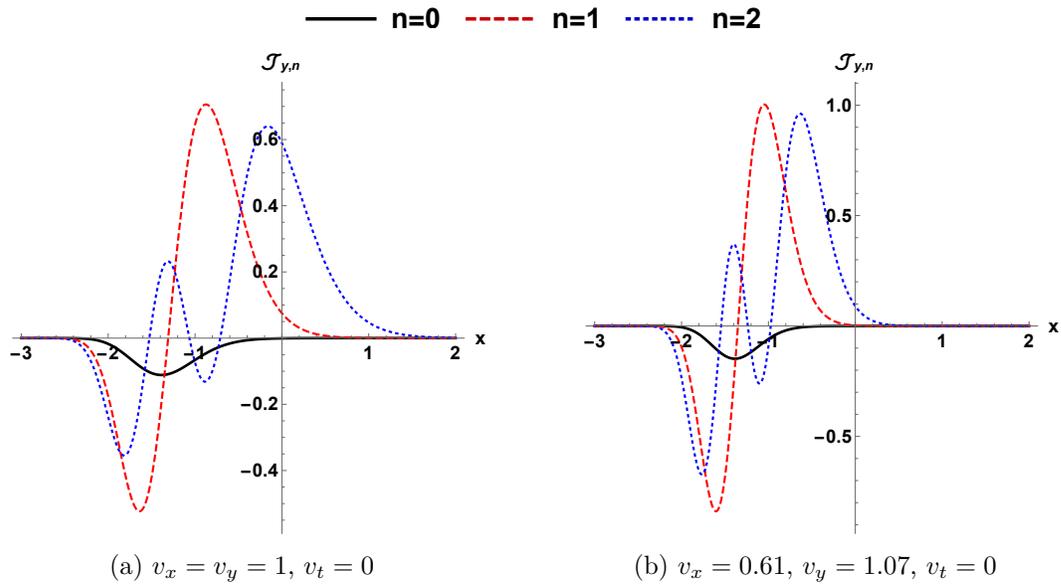

(a) $v_x = v_y = 1$, $v_t = 0$          (b) $v_x = 0.61$, $v_y = 1.07$, $v_t = 0$

**Figura 3.2.7:** Densidad de corriente $\mathcal{J}_{y,n}(x)$ en el caso exponencial con $\mathcal{B}_0 = 1$, $\mathcal{E}_0 = 0.1$, $\alpha = 1$ y $k_y = 3$ a lo largo del eje $x$ para: (a) grafeno prístino, (b) grafeno bajo tensión a lo largo del eje $x$ con $\varepsilon = 0.15$.

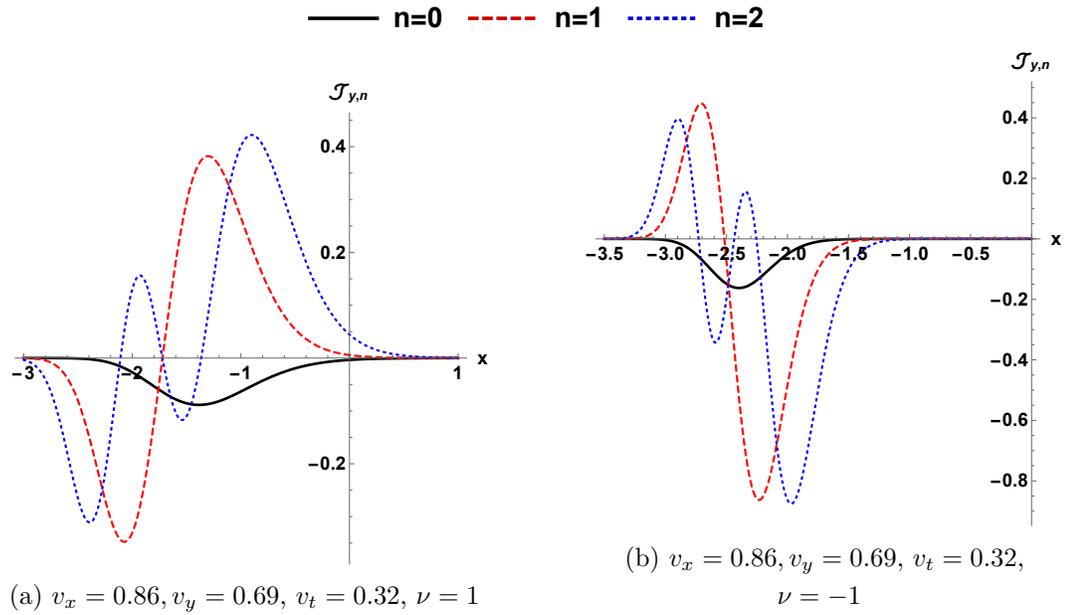

(a) $v_x = 0.86, v_y = 0.69, v_t = 0.32, \nu = 1$

(b) $v_x = 0.86, v_y = 0.69, v_t = 0.32,$
$\nu = -1$

**Figura 3.2.8:** Densidad de corriente $\mathcal{J}_{y,n}(x)$ en el caso exponencial con $\mathcal{B}_0 = 1$, $\mathcal{E}_0 = 0.1$, $\alpha = 1$ a lo largo del eje $x$ para el borofeno $8 - Pmmn$ con (a) $k_y = 3$ en el valle $K_+$, y con (b) $k_y = 8$ en el valle $K_-$.

longitudinal juega un papel importante en la localización de los estados electrónicos. Cuando $v_x$ es menor, los electrones permanecen más tiempo en las proximidades de los puntos críticos, lo que incrementa la probabilidad de encontrarlos en esas regiones. En cambio, si $v_x$ es mayor, los electrones atraviesan rápidamente estas zonas, reduciendo su densidad de probabilidad local. Este efecto se refleja en las gráficas mostradas en las Figuras 3.2.5 y 3.2.6 como un aumento en la densidad de probabilidad cerca de las posiciones espaciales asociadas, de forma indirecta, a los valles K$_+$ y K$_-$.

Por ejemplo, en el caso del borofeno, donde las velocidades $v_x$, $v_y$ y $v_t$ son constantes pero la simetría es menor, el comportamiento está gobernado por el índice de valle ($\nu = \pm 1$). Este índice asigna direcciones preferenciales al movimiento electrónico debido a la estructura cristalina del material. En las gráficas de las Figuras 3.2.5 y 3.2.6, esto se traduce en una distribución asimétrica de las densidades de probabilidad, que tienden a concentrarse en la región negativa del eje $x$. Este efecto es especialmente notable para el estado fundamental ($n = 0$), donde la función de onda muestra un único pico bien definido. En las gráficas de las Figuras 3.2.5-3.2.8, esto aparece como una curva estrecha y alta en la región $x < 0$. Para los estados excitados ($n > 0$), las funciones de onda presentan múltiples máximos, reflejando una mayor delocalización del electrón.

Por otro lado, el parámetro $k_y$ también juega un papel crucial en el aspecto dinámico de los estados electrónicos. En las gráficas de las Figuras 3.2.5-3.2.8, $k_y$ puede interpretarse como un parámetro que determina la dirección preferencial del movimiento electrónico. Por ejemplo, cuando $k_y$ es positivo, los electrones tienden a desplazarse hacia la región $x < 0$, como se evidencia en la concentración de la densidad de probabilidad en esa zona. Este comportamiento es análogo al de partículas clásicas en campos magnéticos, donde el momento transversal influye en la trayectoria.

Finalmente, al analizar las expresiones (3.2.23) y (3.2.33), es posible recuperar los resultados correspondientes a materiales isótropos sin conos inclinados al tomar en cuenta los siguientes valores

$$v_x = v_y = v_\mathrm{F} = 1, \quad v_t = 0. \tag{3.2.43}$$

Además, si se considera la ausencia de un campo eléctrico ($v_\mathrm{d} = 0$) en (3.2.23) y (3.2.33), el resultado coincide con el obtenido en el caso del grafeno prístino bajo la acción de un campo magnético exponencial [Ş. Kuru et al., 2009]. Para ese caso, las componentes del espinor de la ecuación (3.2.33) se reducen a:

$$\psi_{1,n}(z) = \frac{2\alpha\sqrt{n!}}{(2\lambda_n + n)} z^{\lambda_n - 1} e^{-z/2} L_n^{2\lambda_n - 2}(z),$$ (3.2.44)

$$\psi_{2,n}(z) = \frac{\alpha\sqrt{n!}}{(2\lambda_n + n)} \frac{2\lambda_n}{\sqrt{2\lambda_n + n}} z^{\lambda_n} e^{-z/2} L_n^{2\lambda_n}(z),$$ (3.2.45)

siendo ahora

$$z = \frac{2D}{\alpha} e^{-\alpha x},$$ (3.2.46)

$$\lambda_n = \frac{D + k_y - \alpha n}{\alpha}.$$ (3.2.47)

Por lo tanto, la solución completa es:

$$\psi_n(x,y) = N_n e^{ik_y y} \begin{pmatrix} (1 - \delta_{n0})\psi_{1,n-1}(z) \\ i\psi_{2,n}(z) \end{pmatrix},$$ (3.2.48)

donde la constante de normalización $N_n$ está dada por:

$$N_n = \frac{1}{\sqrt{2}} \frac{2\alpha}{\sqrt{\lambda_n (\lambda_n + n)}} \sqrt{n(2\lambda_n + n)}.$$ (3.2.49)

Por su parte, el espectro de energía toma la forma

$$E_n = v_{\mathrm{F}} \left[ (D + k_y)^2 - (D + k_y - \alpha n)^2 \right].$$ (3.2.50)

Lo descrito previamente se ilustra en la Figura 3.2.9, donde se reproduce fielmente el comportamiento reportado en [Ş. Kuru et al., 2009] utilizando los siguientes valores para los parámetros de interés: $k_y = 6$, $D = B = 1$, $\alpha = 1$. Es importante destacar que en la ecuación (3.2.50) se observa un desplazamiento en el parámetro $k_y \rightarrow k_y + D$, respecto a lo mostrado en [Ş. Kuru et al., 2009]. Este desfase, que surge de la elección de norma, establece un vínculo matemático que permite transitar entre las dos representaciones del problema.

Al introducir un campo eléctrico externo (véase la Figura 3.2.7), aparece una densidad de corriente finita para el estado fundamental ($n = 0$). Este fenómeno, que no se presenta en ausencia de campo eléctrico, pone en evidencia cómo la presencia de los campos externos considerados modifica sustancialmente las propiedades de transporte del sistema.

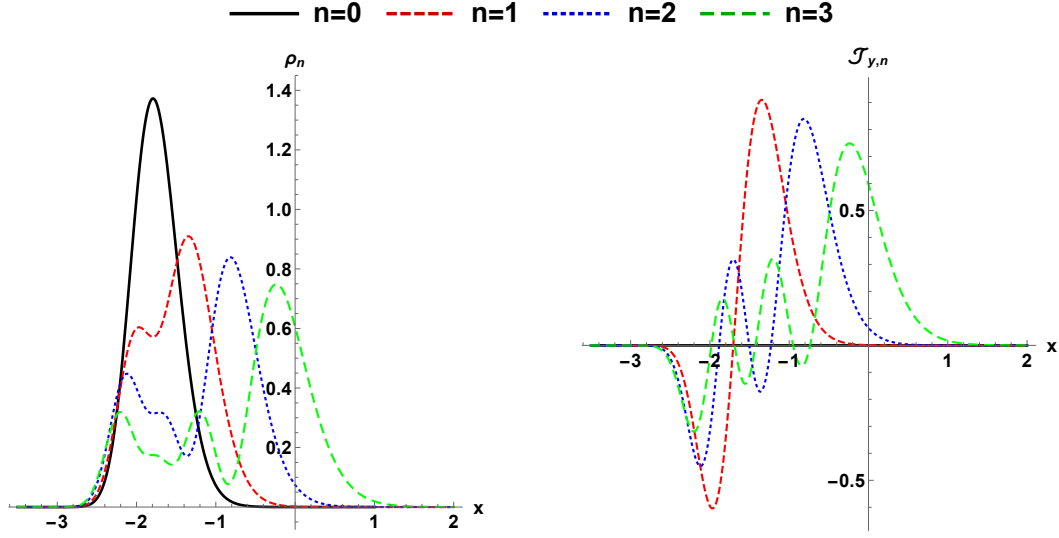

**Figura 3.2.9:** Densidad de probabilidad $\rho_n(\mathbf{r})$ y densidad de corriente $\mathcal{J}_{y,n}(x)$ en el caso exponencial con $\mathcal{B}_0 = D = 1$, $\mathcal{E} = 0$, $\alpha = 1$ y $k_y = 5$ a lo largo del eje $x$ para el grafeno prístino [S. Kuru et al., 2009].

## 3.3 | **Campos externos singulares**

Como segundo caso, considérese un campo magnético y eléctrico cuyos perfiles tienen la siguiente forma:

$$\mathbf{E}(x) = \frac{\mathcal{E}_0}{x^2}\,\hat{e}_x, \tag{3.3.1}$$

$$\mathbf{B}(x) = \frac{\mathcal{B}_0}{x^2}\,\hat{e}_z, \tag{3.3.2}$$

donde nuevamente $\mathcal{B}_0$ y $\mathcal{E}_0$ representan las intensidades del campo magnético y eléctrico, respectivamente.

Los correspondientes potenciales vectorial y escalar tienen la forma

$$\mathbf{A}_y(x) = \frac{-\mathcal{B}_0}{x}\hat{e}_y, \tag{3.3.3}$$

$$\phi(x) = \frac{\mathcal{E}_0}{x}. \tag{3.3.4}$$

Sustituyendo los potenciales (3.3.3) y (3.3.4) en (3.1.13), la ecuación de valores propios se expresa ahora como

$$\left[ -i\frac{\mathrm{d}}{\mathrm{d}x_c}\sigma_x + \left( k_y^c - \frac{\mathcal{B}_0}{x_c} \right)\sigma_y - \left( \bar{E} + \frac{\nu\beta_\nu\mathcal{B}_0}{x_c} \right)\sigma_0 \right]\varphi(x_c) = 0, \tag{3.3.5}$$

donde $\beta_\nu$ y $v_d$ están definidos de la misma forma que en (3.2.6). De manera análoga al caso anterior, al multiplicar por $-i\sigma_x$ la ecuación (3.3.6) por la izquierda y aplicar posteriormente el operador diferencial $\frac{\mathrm{d}}{\mathrm{d}x_c}$ a la expresión resultante, se obtiene lo siguiente:

$$\frac{\mathrm{d}^2\varphi(x_c)}{\mathrm{d}x_c^2} + \left\{ \left[ \bar{E}^2 - k_y^{c2} + \frac{2\mathcal{B}_0\left( \nu\beta_\nu\bar{E} + k_y^c \right)}{x_c} - \frac{\mathcal{B}_0^2\bar{\lambda}^2}{x_c^2} \right]\sigma_0 + \frac{\bar{\mathbb{K}}}{x_c^2} \right\}\varphi(x_c) = 0, \tag{3.3.6}$$

siendo $\bar{\lambda}$ y $\bar{\mathbb{K}}$ definidas como

$$\bar{\lambda} = \sqrt{1-\beta_\nu^2}, \tag{3.3.7}$$

$$\bar{\mathbb{K}} = \mathcal{B}_0 \begin{pmatrix} -1 & i\nu\beta_\nu \\ i\nu\beta_\nu & 1 \end{pmatrix}. \tag{3.3.8}$$

Para resolver la ecuación (3.3.6), es necesario expresar la función de onda $\varphi(z)$ en la base en la que la matriz $\bar{\mathbb{K}}$ es diagonal. Para ello, como en el caso anterior, se deben determinar los eigenvectores $\bar{\chi}_j$ y los eigenvalores $\bar{\lambda}_j$ de la matriz $\bar{\mathbb{K}}$.

En este caso, los eigenvectores $\bar{\chi}_1$ y $\bar{\chi}_2$ resultan ser

$$\bar{\chi}_1 = \bar{N}_1 \begin{pmatrix} \frac{-\nu\beta_\nu}{1+\sqrt{1-\beta_\nu^2}} \\ i \end{pmatrix}, \tag{3.3.9}$$

$$\bar{\chi}_2 = \bar{N}_2 \begin{pmatrix} -i \\ \frac{-\nu\beta_\nu}{1+\sqrt{1-\beta_\nu^2}} \end{pmatrix}. \tag{3.3.10}$$

donde $\bar{N}_{1,2}$ son las constantes de normalización y los valores propios se pueden expresar de la siguiente manera:

$$\bar{\lambda}_j = (-1)^{j+1}\bar{\lambda}, \quad j = 1, 2. \tag{3.3.11}$$

Así, la ecuación diferencial (3.3.6) se puede reescribir en la base de $\bar{\chi}_j$ de la siguiente forma

$$\frac{\mathrm{d}^2\varphi(x_c)}{\mathrm{d}x_c^2} + \left\{ \bar{E}^2 - k_c^2 + \frac{2\mathcal{B}_0\left( \nu\beta_\nu\bar{E} + k_y^c \right)}{x_c} - \frac{\mathcal{B}_0\left[ \mathcal{B}_0\bar{\lambda}^2 - \bar{\lambda}_j \right]}{x_c^2} \right\}\varphi(x_c) = 0. \tag{3.3.12}$$

Tras efectuar el siguiente cambio de variable

$$z(x_c) = 2x_c\sqrt{k_y^{c^2} - \bar{E}^2}, \quad \text{con } z \in [0, \infty), \tag{3.3.13}$$

la ecuación (3.3.12) se expresa como

$$\frac{\mathrm{d}^2\psi_j(z)}{\mathrm{d}z^2} + \left[-\frac{1}{4} + \frac{\bar{\mu}}{z} + \frac{1/4 - \bar{m}_j^2}{z^2}\right]\psi_j(z) = 0, \quad j = 1, 2, \tag{3.3.14}$$

donde $\bar{m}_j$ y $\bar{\mu}$ están dados por:

$$\bar{m}_j = \sqrt{\mathcal{B}_0\left(\mathcal{B}_0\bar{\lambda}^2 - \bar{\lambda}_j\right) + \frac{1}{4}}, \quad j = 1, 2, \tag{3.3.15}$$

$$\bar{\mu} = \frac{\mathcal{B}_0\left(k_y^c + \nu\beta_\nu\bar{E}\right)}{\sqrt{k_y^{c^2} - \bar{E}^2}}. \tag{3.3.16}$$

Introduciendo la cantidad $F = \mathcal{B}_0\sqrt{1 - \beta_\nu^2}$, y sustituyendo el valor de $\bar{\lambda}^2$ y $\bar{\lambda}_j$, es posible expresar a $\bar{m}_j$ como

$$\bar{m}_j = \begin{cases} F - \frac{1}{2} & j = 1, \\ F + \frac{1}{2} & j = 2. \end{cases} \tag{3.3.17}$$

Como puede verse en la ecuación (3.3.14), la ecuación de Whittaker resultante puede resolverse de la misma manera que en la sección anterior, por lo que la solución está dada por la función mostrada en (3.2.17). Al analizar el comportamiento asintótico de las funciones $M_{\bar{\mu}, \bar{m}_j}(z)$ y $W_{\bar{\mu}, \bar{m}_j}(z)$ en los límites $z \to 0$ y $z \to \infty$, es posible imponer las mismas condiciones de frontera que en el caso anterior ($C_2 = 0$), obteniéndose ahora la solución

$$\psi_j(z) = C_1 W_{\bar{\mu}, \bar{m}_j}, \tag{3.3.18}$$

junto con la condición de cuadrado-integrabilidad

$$\bar{\mu} - \bar{m}_j - \frac{1}{2} = n, \quad n + 1 \in \mathbb{N}, \tag{3.3.19}$$

a partir de la cual se obtiene el espectro de energía del sistema:

$$E_n = k_y\nu\,v_t - \frac{\mathcal{B}_0^2\,\beta_\nu\,k_y^c v_F + \kappa v_F\,k_y^c\,(F + n)\,\sqrt{\mathcal{B}_0^2\left(\beta_\nu^2 - 1\right) + (F + n)^2}}{(F + n)^2 + \mathcal{B}_0^2\beta_\nu^2}. \tag{3.3.20}$$

Similar al caso anterior, la condición cuadrado-integrabilidad (3.3.19) permite expresar a la función $W_{\bar{\mu},\bar{m}_j}(z)$ en términos de los polinomios asociados de Laguerre:

$$W_{\bar{\mu},\bar{m}_j}(z) \propto L_n^{2\bar{m}_j}(z). \tag{3.3.21}$$

Así, las soluciones $\psi_{j,n}(z)$ en (3.3.18) toman la siguiente forma:

$$\psi_{j,n}(z) \propto \mathrm{e}^{-z/2}\, z^{\bar{m}_j+1/2} L_n^{2\bar{m}_j}(z). \tag{3.3.22}$$

Usando (3.3.17) en (3.3.22), las funciones de onda $\psi_{1,n}(z)$ y $\psi_{2,n}(z)$ del sistema toman la forma

$$\psi_{1,n}(z) = \bar{\mathcal{N}}_1\, \mathrm{e}^{-z_n/2} z_n^F L_n^{2F-1}(z_n)\,, \tag{3.3.23}$$

$$\psi_{2,n}(z) = \bar{\mathcal{N}}_2\, \mathrm{e}^{-z_{n+1}/2} z_{n+1}^{F+1} L_n^{2F+1}(z_{n+1})\,, \tag{3.3.24}$$

donde $\bar{\mathcal{N}}_1$ y $\bar{\mathcal{N}}_2$ son constantes de normalización dadas por

$$\bar{\mathcal{N}}_1 = \sqrt{\frac{\sqrt{\frac{v_y}{v_x}(k_c^2 - \bar{E}_n^2)}\, n!}{(n+F)\,\Gamma(2F+n)}}, \tag{3.3.25}$$

$$\bar{\mathcal{N}}_2 = \sqrt{\frac{\sqrt{\frac{v_y}{v_x}(k_c^2 - \bar{E}_{n+1}^2)}\,(n+1)!}{(n+F+1)\,\Gamma(2F+n+2)}}, \quad n = 0,1\ldots, \tag{3.3.26}$$

con $\bar{E}_n = (\nu E_n - v_t k_y)/\sqrt{v_x v_y}$.

Considerando la expresión dada en (3.1.12) para los pseudo-espinores, y junto con las relaciones (3.2.12) y (3.2.13), la solución adquiere la siguiente forma:

$$
\begin{aligned}
\Psi_n(x,y) &= \bar{\mathcal{N}} \exp\!\big(ik_y^c y\big) \left[ \bar{C}_1\, \bar{\chi}_1 \psi_{1,n}(z) + i\bar{C}_2\, \bar{\chi}_2 \psi_{2,n-1}(z) \right], \\
&= \bar{\mathcal{N}} \exp\!\big(ik_y^c y\big) \begin{pmatrix} \mathcal{B}_0\, \nu\, \beta_\nu\, \psi_{1,n}(z) + \mathcal{B}_0\,(1-\delta_{0n})\, \psi_{2,n-1}(z) \\ i\left[ \mathcal{B}_0\, \psi_{1,n}(z) + \mathcal{B}_0\, \nu\, \beta_\nu\, \psi_{2,n-1}(z) \right] \end{pmatrix}, \\
&= \bar{\mathcal{N}} \exp\!\big(ik_y^c y\big) \begin{pmatrix} \mathcal{B}_0\, \nu\, \beta_\nu & -i\, \mathcal{B}_0 \\ i\, \mathcal{B}_0 & \mathcal{B}_0\, \nu\, \beta_\nu \end{pmatrix} \begin{pmatrix} \psi_{1,n}(z) \\ i(1-\delta_{0n})\, \psi_{2,n-1}(z) \end{pmatrix}, \tag{3.3.27}
\end{aligned}
$$

siendo la constante de normalización $\bar{\mathcal{N}}$ expresada como:

$$\bar{\mathcal{N}} = \sqrt{\frac{\bar{\lambda} + \mathcal{B}_0}{2^{2-\delta_{0n}} \mathcal{B}_0\, \nu \beta_\nu\, I_n}}, \quad \text{con } \bar{I}_n = 2^{\delta_{0n}} \sqrt{\frac{v_x}{v_y}}\,(1-\delta_{0n}) \int_0^\infty \psi_{1,n}(x)\, \psi_{2,n-1}(x)\, \mathrm{d}x, \tag{3.3.28}$$

y donde se han elegido $\bar{C}_1 = 1$ y $\bar{C}_2 = 1 - \delta_{0n}$.

Sustituyendo la solución dada en (3.3.27) en las expresiones (3.2.36), es posible obtener la *densidad de probabilidad* y la *densidad de corriente de probabilidad*:

$$\rho_n(x) = \frac{|\Psi_n(z)|^2 - 2^{\delta_{0n}}(1 - \delta_{0n})\,\nu\beta_\nu\,\psi_{1,n}(z)\,\psi_{2,n-1}(z)}{1 - \nu\beta_\nu\,I_n}, \tag{3.3.29}$$

$$\mathcal{J}_{x,n}(x,y,t) = \frac{\nu\,v_x\bar{\lambda}\,\Psi_n^\dagger(z)\sigma_x\Psi_n(z)}{\mathcal{B}_0\,[1 - \nu\beta_\nu\,I_n]} = 0, \tag{3.3.30}$$

$$\mathcal{J}_{y,n}(x,y,t) = \frac{\nu\,[v_t - \nu\beta_\nu\,v_y]\,|\Psi_n(z)|^2}{1 - \nu\beta_\nu\,I_n} + \frac{2^{\delta_{0n}}(1 - \delta_{0n})\,[v_y - \nu\beta_\nu\,v_t]\,\psi_{1,n}(z)\psi_{2,n-1}(z)}{1 - \nu\beta_\nu\,I_n}. \tag{3.3.31}$$

### Discusión

Para entender el comportamiento del espectro de energía en (3.3.20), es importante analizar su dependencia respecto a cada parámetro de interés que aparece en ella. Las Figuras 3.3.1 y 3.3.2 muestran el comportamiento del espectro de energía con respecto a la intensidad del campo eléctrico para el grafeno prístino, el grafeno sometido a tensión uniaxial y el borofeno $8 - Pmmn$, respectivamente. Tal como se observó en el caso de los campos eléctrico y magnético con perfil exponencial, en los perfiles de campo singulares también se presenta un valor crítico de la intensidad del campo eléctrico, $\mathcal{E}_c$, para el cual los niveles de Landau mostrados en la ecuación (3.3.20) colapsan, como se aprecia en las Figuras 3.3.1 y 3.3.2. Este valor crítico $\mathcal{E}_c$ se define de la misma manera como el que se obtuvo en la ecuación (3.2.41).

La dependencia de la energía $E_n$ respecto del parámetro $k_y$ se muestra en las Figuras 3.3.3 y 3.3.4. Esto permite observar la existencia de un número máximo de estados ligados para cada valor de $k_y$, el cual está delimitado por una línea envolvente.

En el caso de las densidades de probabilidad mostradas en las Figuras 3.3.5-3.3.8, se observa un comportamiento marcadamente asimétrico, localizado exclusivamente en la región $x > 0$, en contraste con el perfil exponencial decreciente. La densidad de probabilidad asociada al estado fundamental ($n = 0$) exhibe un pico agudo y estrecho en $x > 0$, lo que indica una alta localización del electrón cerca de la singularidad. Para los estados excitados ($n > 0$), la densidad de probabilidad presenta oscilaciones cuya amplitud decrece conforme $x$ aumenta.

En particular, si se considera $v_x = v_y$, $v_t = 0$, y la ausencia del campo eléctrico

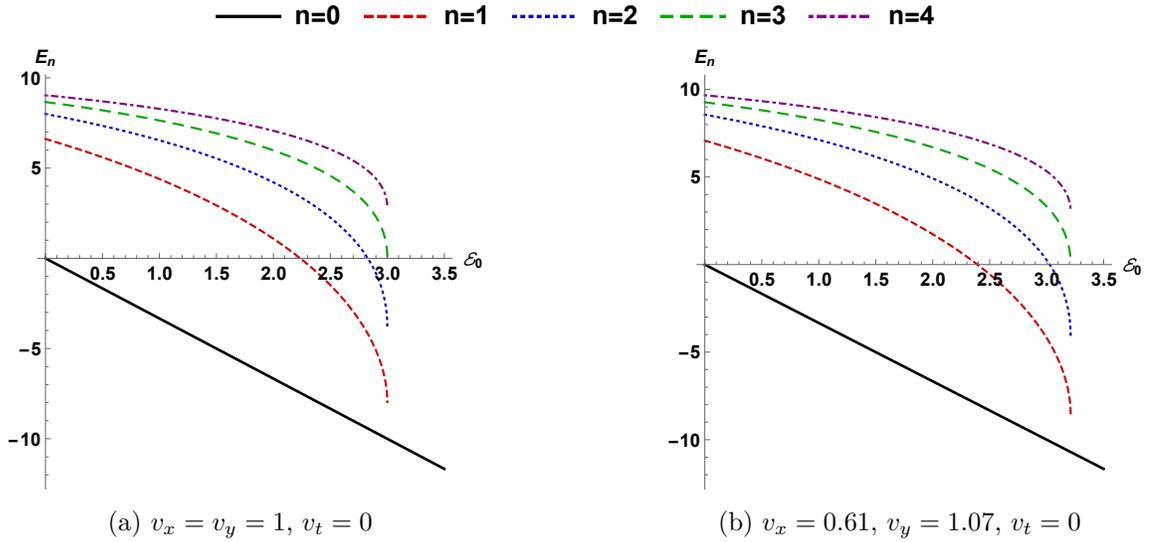

(a) $v_x = v_y = 1$, $v_t = 0$        (b) $v_x = 0.61$, $v_y = 1.07$, $v_t = 0$

**Figura 3.3.1:** Espectro de energía $E_n$ en (3.3.20), caso singular, con $\mathcal{B}_0 = 3$ y $k_y = 10$ en función del campo eléctrico $\mathcal{E}_0$ para (a) el grafeno prístino y (b) el grafeno bajo tensión a lo largo del eje $x$ con $\varepsilon = 0.15$.

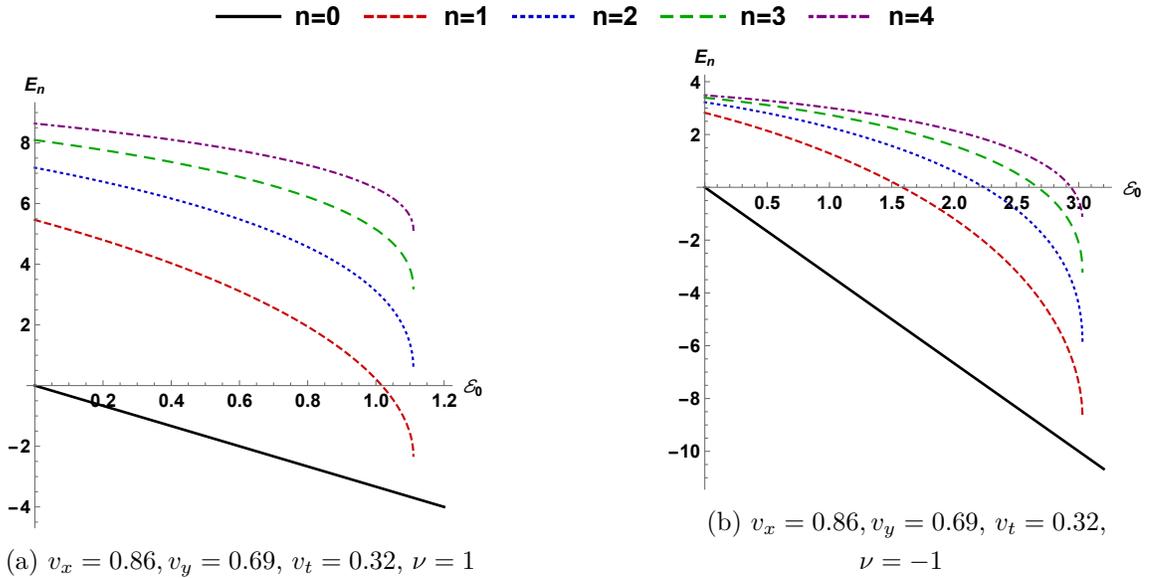

(a) $v_x = 0.86, v_y = 0.69, v_t = 0.32, \nu = 1$

(b) $v_x = 0.86, v_y = 0.69, v_t = 0.32,$
$\nu = -1$

**Figura 3.3.2:** Espectro de energía $E_n$ en (3.3.20), caso singular, con $\mathcal{B}_0 = 3$ y $k_y = 10$ en función del campo eléctrico $\mathcal{E}_0$ para el borofeno $8 - Pmmn$ en (a) el valle $K_+$ y (b) el valle $K_-$.

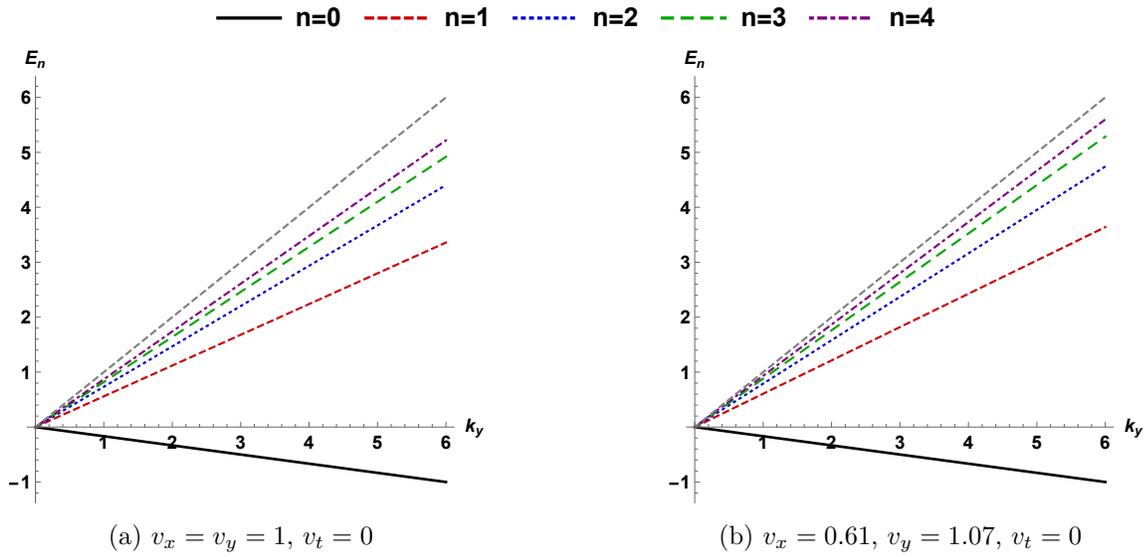

(a) $v_x = v_y = 1$, $v_t = 0$

(b) $v_x = 0.61$, $v_y = 1.07$, $v_t = 0$

**Figura 3.3.3:** Espectro de energía $E_n$ en (3.3.20), caso singular, con $\mathcal{B}_0 = 3$ y $\mathcal{E}_0 = 0.5$ en función del momento $k_y$, para (a) el grafeno prístino y (b) el grafeno bajo tensión a lo largo del eje $x$ con $\varepsilon = 0.15$. La recta gris es la envolvente que indica el número de estados ligados.

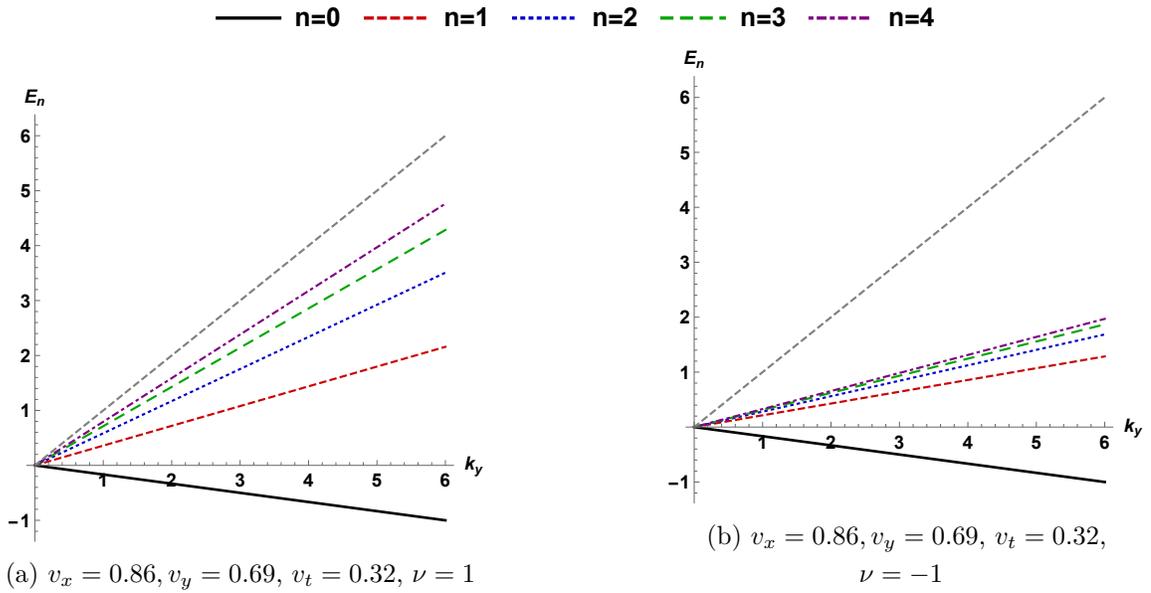

(a) $v_x = 0.86, v_y = 0.69, v_t = 0.32, \nu = 1$

(b) $v_x = 0.86, v_y = 0.69, v_t = 0.32,$
$\nu = -1$

**Figura 3.3.4:** Espectro de energía $E_n$ en (3.3.20), caso singular, con $\mathcal{B}_0 = 3$ y $\mathcal{E}_0 = 0.5$ en función del momento $k_y$, para el borofeno $8 - Pmmn$ en (a) el valle $K_+$ y (b) el valle $K_-$. La recta gris es la envolvente que indica el número de estados ligados.

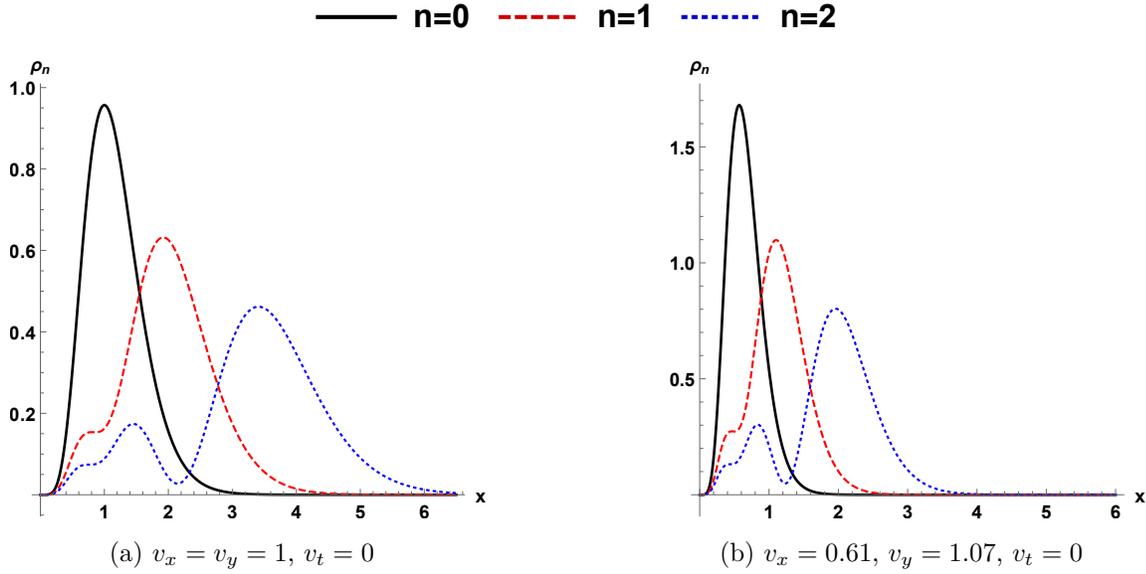

(a) $v_x = v_y = 1$, $v_t = 0$

(b) $v_x = 0.61$, $v_y = 1.07$, $v_t = 0$

**Figura 3.3.5:** Densidad de probabilidad $\rho_n(\mathbf{r})$ en el caso singular con $\mathcal{B}_0 = 3$, $\mathcal{E} = 0.5$ y $k_y = 3$ a lo largo del eje $x$ para (a) el grafeno prístino y (b) el grafeno bajo tensión a lo largo del eje $x$ con $\varepsilon = 0.15$.

($v_\mathrm{d} = 0$), las funciones (3.3.27) toman la siguiente forma:

$$\psi_{1,n}(z) = z_{n+1}^{F+1} \mathrm{e}^{-z_{n+1}/2} L_n^{2F+1}(z_{n+1}), \qquad (3.3.32)$$

$$\psi_{2,n}(z) = z_n^{F+1} \mathrm{e}^{-z_n/2} L_n^{2F+1}(z_n). \qquad (3.3.33)$$

Así, la función de onda espinorial resulta ser

$$\psi_n(x,y) = N_n \mathrm{e}^{ik_y y} \begin{pmatrix} (1 - \delta_{n0}) \psi_{1,n-1}(z) \\ i\psi_{2,n}(z) \end{pmatrix}, \qquad (3.3.34)$$

siendo $z_n$ y $z_{n+1}$ las variables definidas como

$$z_n = \frac{2k\,D}{n+D}\,x, \quad z_{n+1} = \frac{2k\,D}{n+D+1}\,x, \qquad (3.3.35)$$

con $D \equiv \mathcal{B}_0$ un parámetro característico del sistema.
Por otro lado, el espectro de energía dado en (3.3.20) adopta la siguiente forma:

$$E_n = \kappa\,k_y\,F \sqrt{\frac{1}{D^2} - \frac{1}{(n+D)^2}}. \qquad (3.3.36)$$

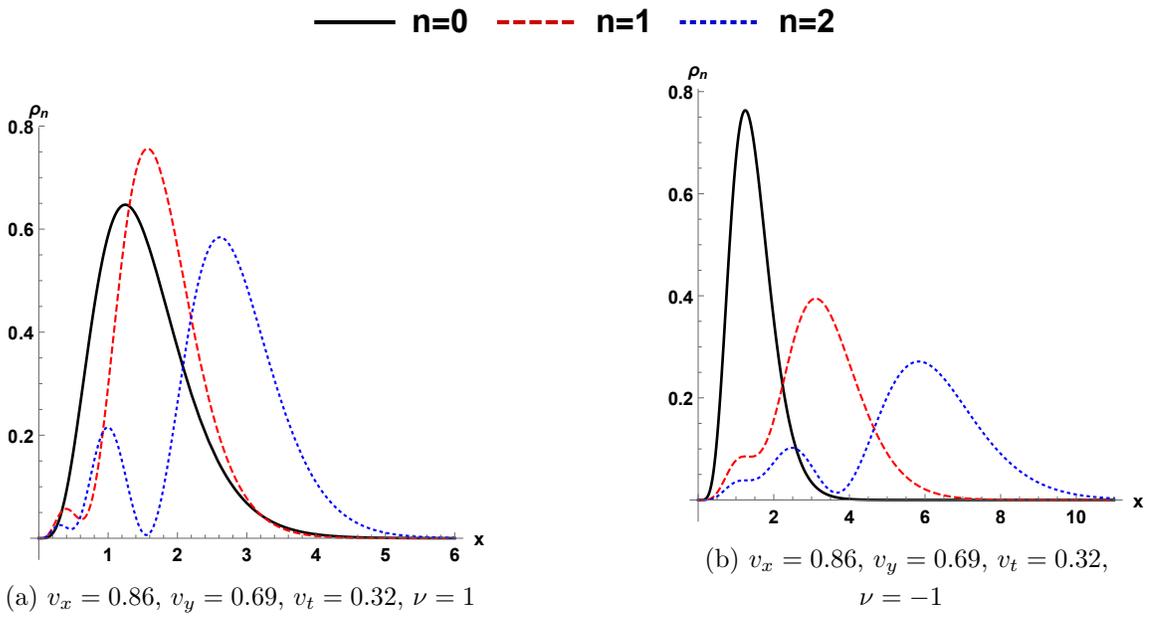

(a) $v_x = 0.86$, $v_y = 0.69$, $v_t = 0.32$, $\nu = 1$

(b) $v_x = 0.86$, $v_y = 0.69$, $v_t = 0.32$, $\nu = -1$

**Figura 3.3.6:** Densidad de probabilidad $\rho_n(\mathbf{r})$ en el caso singular con $\mathcal{B}_0 = 3$, $\mathcal{E} = 0.5$, $\alpha = 1$ y $k_y = 3$ a lo largo del eje $x$ para el borofeno $8 - Pmmn$ en (a) el valle $K_+$ y (b) el valle $K_-$.

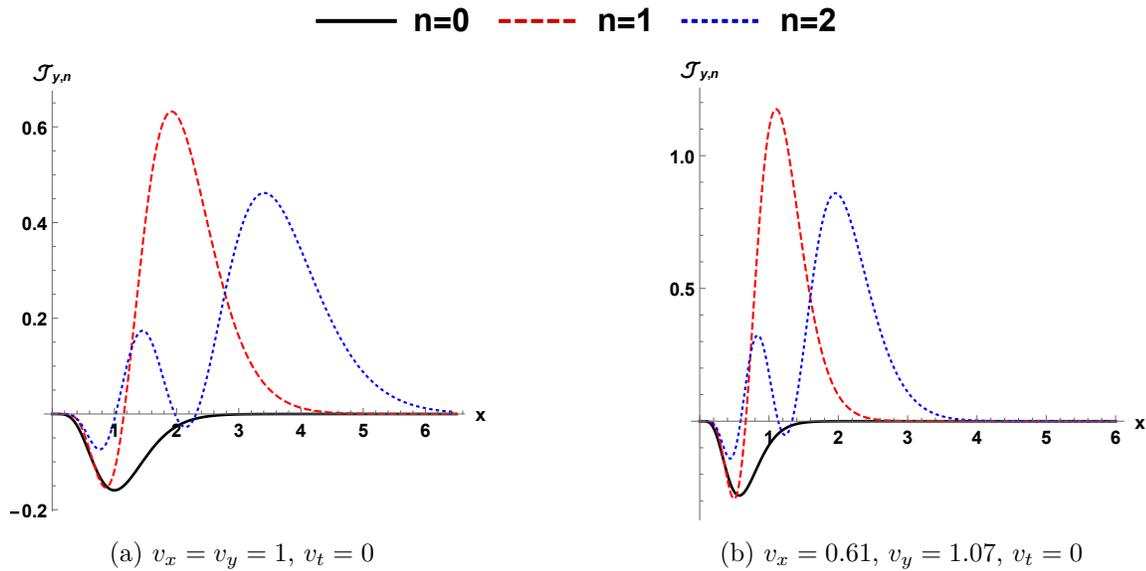

(a) $v_x = v_y = 1$, $v_t = 0$          (b) $v_x = 0.61$, $v_y = 1.07$, $v_t = 0$

**Figura 3.3.7:** Densidad de corriente $\mathcal{J}_{y,n}(x)$ en el caso singular con $\mathcal{B}_0 = 3$, $\mathcal{E} = 0.5$, $\alpha = 1$ y $k_y = 3$ a lo largo del eje $x$ para (a) el grafeno prístino y (b) el grafeno bajo tensión a lo largo del eje $x$ con $\varepsilon = 0.15$.

La Figura 3.3.9 presenta los resultados para grafeno prístino, donde se reproduce el comportamiento reportado en [Ş. Kuru et al., 2009] al utilizar los siguientes valores para los parámetros de interés: $k_y = 10$, $B = D = 3$.

A diferencia del perfil mostrado en la Sección 3.2, en el caso actual no se observa un desfase en el número de onda $k_y$. Además, contrario a lo reportado en [Ş. Kuru et al., 2009], en el sistema bajo estudio aparece una densidad de corriente finita para el estado fundamental ($n = 0$), lo cual puede atribuirse directamente a la presencia del campo eléctrico externo. Nuevamente, esto último refleja cómo la presencia y el perfil de los campos externos aplicados tienen un impacto en el transporte electrónico del sistema.

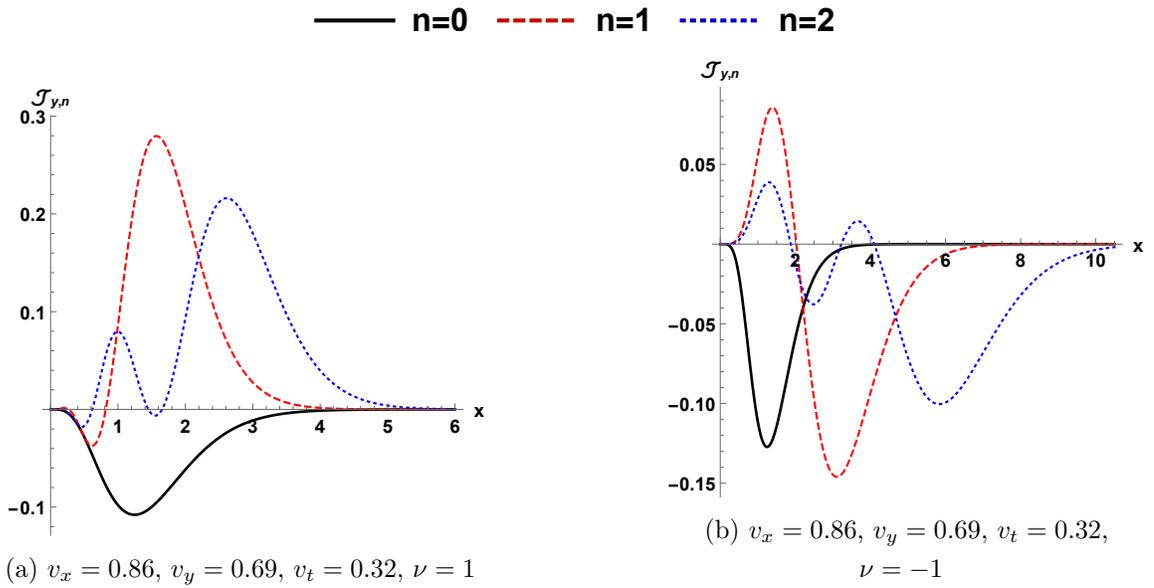

(a) $v_x = 0.86$, $v_y = 0.69$, $v_t = 0.32$, $\nu = 1$

(b) $v_x = 0.86$, $v_y = 0.69$, $v_t = 0.32$, $\nu = -1$

**Figura 3.3.8:** Densidad de corriente $\mathcal{J}_{y,n}(x)$ en el caso singular con $\mathcal{B}_0 = 3$, $\mathcal{E} = 0.5$, $\alpha = 1$ y $k_y = 3$ a lo largo del eje $x$ para el borofeno $8 - Pmmn$ en (a) el valle $K_+$ y (b) el valle $K_-$.

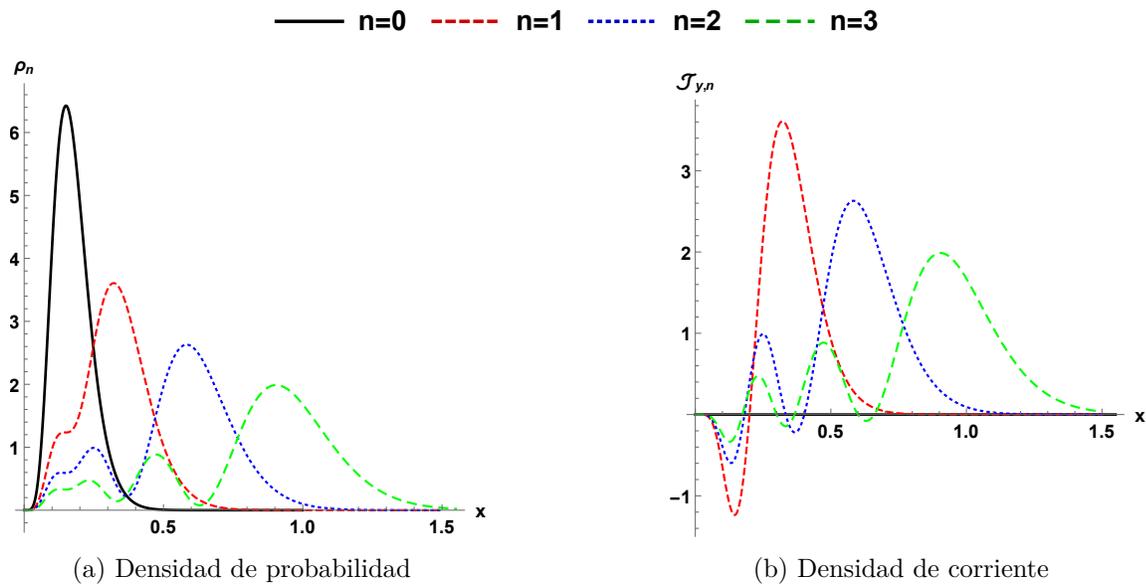

(a) Densidad de probabilidad                     (b) Densidad de corriente

**Figura 3.3.9:** Densidad de probabilidad $\rho_n(\mathbf{r})$ y densidad de corriente $\mathcal{J}_{y,n}(x)$ en el caso singular con $\mathcal{B}_0 = D = 3$, $\mathcal{E} = 0$, $\alpha = 1$ y $k_y = 20$ a lo largo del eje $x$ para el grafeno prístino [S. Kuru et al., 2009].

# 4 | Materiales de Dirac con campos eléctricos y magnéticos externos en una región finita

El método de iteración asintótica (AIM, por sus siglas en inglés), propuesto inicialmente por Ciftci et al. [Ciftci et al., 2003], destaca por su capacidad para calcular eigenvalores y eigenfunciones de ecuaciones diferenciales homogéneas de segundo orden. Es aplicable tanto de manera analítica como numérica, y ha demostrado ser útil en diversos casos específicos.

Por ejemplo, en el ámbito de la relatividad general, para el estudio de los modos cuasinormales de agujeros negros se requiere resolver ecuaciones diferenciales de segundo orden, cuyos cálculos suelen realizarse numéricamente. Por ello, Cho [Cho et al., 2012, López et al., 2024] empleó el AIM para calcular las frecuencias de estos modos en diversos tipos de agujeros negros, evidenciando su aplicabilidad en problemas complejos.

Por otro lado, en la física atómica, el análisis de la ecuación de Schrödinger con potenciales arbitrarios representa un desafío constante. Mientras que algunos potenciales, como el de Coulomb, de Kratzer, de Pöschl–Teller o el oscilador armónico, admiten soluciones exactas, otros (como los potenciales de Yukawa, Morse y Hulthén [Flügge, 1994, Bayrak and Boztosun, 2006]) presentan dificultades para obtener eigenvalores de energía en forma analítica. En este escenario, el AIM surge como una estrategia viable, como lo demuestran los trabajos de O. Bayrak [Bayrak and Boztosun, 2007], ofreciendo una vía alternativa para abordar problemas no triviales.

En este capítulo se retoman los perfiles de campo presentados en el Capítulo 3, pero ahora dichos perfiles se considerarán dentro de una región finita, acotada por una frontera situada en $x = x_0$. Aquí, la ecuación de Whittaker se resolverá aplicando el método de iteración asintótica de forma perturbativa. También se presentará el espectro de energía, así como las densidades de probabilidad y de corriente, destacando las diferencias entre los casos confinado y no confinado. La consideración de regiones acotadas resulta de gran importancia, ya que introduce condiciones de frontera que reflejan situaciones físicas más realistas, modifican la estructura del espectro y permiten analizar el impacto del confinamiento en las propiedades de transporte de los electrones en materiales bidimensionales.

## 4.1 | Método de iteración asintótica

Para implementar el método de iteración asintótica, se parte de la siguiente ecuación diferencial homogénea de segundo orden:

$$y''(x) - \lambda_0 y'(x) - s_0 y(x) = 0, \tag{4.1.1}$$

donde $\lambda_0$ y $s_0$ son funciones de clase $C^\infty(a, b)$ (es decir, son continuas e indefinidamente diferenciables en el intervalo $(a, b)$), y $f'(x)$ denota la derivada respecto a $x$.

El primer paso para iniciar con este método consiste en derivar la ecuación (4.1.1) con respecto a $x$, obteniendo:

$$y'''(x) - \lambda_0 y''(x) - \lambda_0' y'(x) - s_0 y'(x) - s_0' y(x) = 0. \tag{4.1.2}$$

Sustituyendo la ecuación (4.1.1) en (4.1.2), se obtiene

$$y'''(x) - \lambda_1 y'(x) - s_1 y(x) = 0. \tag{4.1.3}$$

A continuación, se repite el proceso de derivación respecto a $x$ para obtener ecuaciones diferenciales recursivas de orden superior. De esta manera, al sustituir nuevamente la derivada de segundo orden mostrada en la ecuación (4.1.1) en las expresiones resultantes, y siguiendo un procedimiento iterativo, se llega a

$$y''''(x) - \lambda_2 y'(x) - s_2 y(x) = 0, \tag{4.1.4}$$

en donde se han definido las siguientes expresiones

$$\lambda_2 = \lambda_1' + \lambda_1 \lambda_0 + s_1, \tag{4.1.5}$$

$$s_2 = s_1' + s_0 \lambda_1. \tag{4.1.6}$$

Este proceso puede repetirse $n$ veces para obtener una ecuación diferencial de orden $n+2$. De esta manera, se obtiene la siguiente expresión

$$y^{(n+2)} - \lambda_n y'(x) - s_n y(x) = 0, \tag{4.1.7}$$

donde, similarmente, $\lambda_n$ y $s_n$ se definen de forma recursiva como:

$$\lambda_n = \lambda'_{n-1} + \lambda_{n-1}\lambda_0 + s_{n-1}, \tag{4.1.8}$$
$$s_n = s'_{n-1} + s_0\lambda_{n-1}. \tag{4.1.9}$$

De la ecuación previa a (4.1.7), es decir, la ecuación diferencial de orden $n+1$, se tiene que

$$y^{(n+1)} - \lambda_{n-1} y'(x) - s_{n-1} y(x) = 0. \tag{4.1.10}$$

A continuación, el método requiere calcular el cociente entre las ecuaciones (4.1.7) y (4.1.10), esto es:

$$\frac{y^{(n+2)}}{y^{(n+1)}} = \frac{\mathrm{d}}{\mathrm{d}x} \ln\left(y^{(n+1)}\right) = \frac{\lambda_n(y' + \frac{s_n}{\lambda_n}y)}{\lambda_{n-1}(y' + \frac{s_{n-1}}{\lambda_{n-1}}y)}. \tag{4.1.11}$$

El carácter asintótico del método indica que, para un $n$ adecuada, se debe satisfacer la relación (4.1.12), donde las razones se igualan en un mismo valor $\eta_n(x)$, condición que asegura la convergencia y la obtención de los eigenvalores.

$$\frac{s_n}{\lambda_n} = \frac{s_{n-1}}{\lambda_{n-1}} = \eta_n(x). \tag{4.1.12}$$

Al sustituir la expresión (4.1.12) en (4.1.11), se obtiene la siguiente igualdad

$$\frac{\mathrm{d}}{\mathrm{d}x} \ln\left(y^{(n+1)}\right) = \frac{\lambda_n}{\lambda_{n-1}}, \tag{4.1.13}$$

cuya solución es:

$$y(x)^{(n+1)} = C_1 \exp\left[\int^x \frac{\lambda_n(t)}{\lambda_{n-1}(t)}\,\mathrm{d}t\right], \tag{4.1.14}$$

donde $C_1$ una constante a determinar.

Usando ahora la ecuación (4.1.12) en (4.1.14), resulta:

$$\frac{\lambda_n}{\lambda_{n-1}} = \frac{\lambda'_{n-1} + \lambda_{n-1}\lambda_0 + s_{n-1}}{\lambda_{n-1}} = \frac{\mathrm{d}}{\mathrm{d}x}\ln(\lambda_{n-1}) + \eta_n(x) + \lambda_0. \tag{4.1.15}$$

Sustituyendo la ecuación anterior en (4.1.14), es posible conseguir el siguiente resultado:

$$y(x)^{(n+1)} = C_1 \lambda_{n-1} \exp\left[\int^x (\eta_n(t) + \lambda_0(t))\ dt\right]. \tag{4.1.16}$$

Si se desea obtener una solución para $y$, se debe sustituir (4.1.16) en (4.1.10), de donde se obtiene

$$y' + \eta_n(t)y = C_2 \exp\left[\int^x (\eta_n(t) + \lambda_0(t))\ dt\right]. \tag{4.1.17}$$

Para resolver la ecuación (4.1.17), se puede aplicar el método del factor integrante, con lo cual se obtiene la siguiente solución:

$$y(x) = \exp\left(-\int^x \eta_n(t)\ dt\right)\left[C_2 + C_1 \int^x \exp\left(\int^t (\lambda_0(\tau) + 2\eta_n(\tau))\ d\tau\right)\ dt\right], \tag{4.1.18}$$

donde $C_2$ es una nueva constante de integración.

Es importante enfatizar que el segundo termino de (4.1.18) usualmente no posee un buen comportamiento, por lo que en muchos casos se establece $C_1 = 0$ (para una mayor referencia de este punto, véase [Bayrak and Boztosun, 2006]). Bajo esta condición, la solución de (4.1.1) es

$$y(x) = C_2 \exp\left(-\int^x \eta_n(t)\ dt\right). \tag{4.1.19}$$

Ahora bien, una forma alternativa de (4.1.12) es

$$\delta_k(x) \equiv s_k(x)\lambda_{k-1}(x) - \lambda_k(x)s_{k-1}(x) = 0, \quad k = 1, 2, 3\dots, \tag{4.1.20}$$

en donde $k$ denota el numero de iteraciones a realizar y $\delta_k(x) = 0$ se le suele llamar la *condición de cuantización*. Las raíces de la condición cuantización proporcionan los eigenvalores de (4.1.1).

Cuando el sistema es exactamente soluble, los eigenvalores de energía se obtienen directamente al resolver la ecuación diferencial correspondiente. En casos donde esto no es posible, se emplea un enfoque aproximado mediante un proceso iterativo. Para aplicar este método, se selecciona un punto $z_0$ que sirva como referencia para construir una solución confiable. Dicho punto suele elegirse en función de las características del sistema, por ejemplo, en el entorno donde la función de onda alcanza su valor máximo o cerca del mínimo del potencial efectivo. La elección de $z_0$ no es arbitraria, ya que influye en la convergencia y precisión del método, un valor adecuado asegura que la aproximación reproduzca de manera fiel el comportamiento físico del sistema, mientras que una mala elección puede llevar a resultados inestables o poco representativos. De hecho, en muchos casos prácticos $z_0$ se sitúa en un punto estacionario del potencial efectivo o en una región donde la función de onda cambia lentamente, lo cual optimiza la estabilidad de la expansión.

Los eigenvalores de energía se determinan como las raíces de la ecuación resultante al considerar valores suficientemente grandes del parámetro de iteración $k$. A medida que $k$ aumenta, las soluciones numéricas tienden a estabilizarse, lo que permite obtener una estimación precisa de los niveles de energía.

### 4.1.1 |Método de iteración asintótica: teoría de perturbaciones

En esta sección se presenta, de manera breve, la aplicación del método de iteración asintótica en su formulación perturbativa, propuesto por Ciftci, Hall y Saad [Ciftci et al., 2005]. Este enfoque resulta particularmente útil, dado que en numerosos problemas de física es necesario recurrir a técnicas perturbativas para obtener soluciones aproximadas. La idea fundamental del método guarda una estrecha relación con los procedimientos utilizados en la teoría de perturbaciones de la mecánica cuántica.

Para potenciales que describen un sistema físico y que pueden expresarse como

$$V(x) = V_1(x) + \lambda V_2(x), \tag{4.1.21}$$

donde $V_1(x)$ representa un potencial exactamente soluble, mientras que $V_2(x)$ actúa como una perturbación del sistema y $\lambda$ es un parámetro de control, el método asume que los eigenvalores pueden expresarse como una serie de potencias en función del parámetro $\lambda$:

$$E = E^{(0)} + \lambda E^{(1)} + \lambda^2 E^{(2)} + \lambda^3 E^{(3)} + \cdots. \tag{4.1.22}$$

Aquí, el propósito de la teoría de perturbaciones es calcular los coeficientes $E^{(j)}$, con $j = 0, 1, 2, \ldots$. Para este fin, se parte de la ecuación de Schrödinger independiente del tiempo:

$$\left[ -\frac{\mathrm{d}^2}{\mathrm{d}x^2} + V_1(x) + \lambda V_2(x) \right] \psi(x) = E\psi(x). \tag{4.1.23}$$

El siguiente paso en el método de iteración asintótica perturbativo es proponer a la función de onda $\psi(x)$ como $\psi(x) = y_0(x)y(x)$, donde usualmente $y_0(x)$ se obtiene analizando las soluciones de la ecuación (4.1.23) en los puntos asintoticos. Sustituyendo la redefinición de $\psi(x)$ en (4.1.23), se tiene la siguiente ecuación diferencial para $y(x)$:

$$y''(x) = \lambda_0(x, \lambda) y'(x) + s_0(x, \lambda) y(x), \tag{4.1.24}$$

donde $\lambda_0(x, \lambda)$ y $s_0(x, \lambda)$ se definen, respectivamente, como

$$\lambda_0(x, \lambda) = -\frac{2\, y_0'(x)}{y_0(x)}, \tag{4.1.25}$$

$$s_0(x, \lambda) = V_1(x) + \lambda V_2(x) - \frac{y_0''(x)}{y_0(x)} - E. \tag{4.1.26}$$

Aplicando el método de iteración asintótica descrito anteriormente, es posible calcular las funciones $\lambda_n(x, \lambda)$ y $s_n(x, \lambda)$ a partir de $\lambda_0(x, \lambda)$ y $s_0(x, \lambda)$. De la misma forma, se construyen las funciones $\delta(x, \lambda)$. Entonces, se tiene

$$\delta(x, \lambda) = s_n(x, \lambda)\lambda_{n+1}(x, \lambda) - s_{n+1}(x, \lambda)\lambda_n(x, \lambda). \tag{4.1.27}$$

En este caso, el método de iteración asintótica establece que la ecuación (4.1.27) debe desarrollarse en una serie de Taylor alrededor del parámetro perturbativo $\lambda = 0$, el valor $\lambda = 0$ corresponde al caso no perturbado, es decir, al sistema base, mientras que los términos con potencias de $\lambda$ representan las correcciones sucesivas. Para que la expansión sea válida, $\lambda$ debe ser un parámetro pequeño, de manera que los términos de orden superior resulten cada vez menos significativos, es decir:

$$\delta(x, \lambda) = \sum_{k=0}^{\infty} \lambda^k \delta^{(k)}(x), \tag{4.1.28}$$

donde los coeficientes $\delta^{(k)}(x)$ están dados por:

$$\delta^{(k)}(x) = \frac{1}{k!}\left.\frac{\partial^k \delta(x, \lambda)}{\partial \lambda^k}\right|_{\lambda=0}. \tag{4.1.29}$$

Para garantizar que $\delta(x, \lambda) = 0$ se cumpla para cualquier valor de $\lambda$, es necesario que cada uno de los coeficientes $\delta^{(k)}(x)$ sea igual a cero, es decir:

$$\delta^{(0)}(x) = 0, \quad \delta^{(1)}(x) = 0, \quad \delta^{(2)}(x) = 0, \quad \text{y así sucesivamente.} \tag{4.1.30}$$

La primera de estas condiciones, $\delta^{(0)}(x) = 0$, proporciona los eigenvalores para el caso no perturbado $E^{(0)}$. La siguiente condición, $\delta^{(1)}(x) = 0$, permite encontrar la corrección a la energía a primer orden $E^{(1)}$, y así sucesivamente para $k = 2, 3 \ldots$.

Este procedimiento permite calcular los eigenvalores de (4.1.23) al orden deseado.

Cabe señalar que en varios casos, al resolver la ecuación $\delta^{(k)}(x) = 0$ no será posible encontrar una solución exacta en todos los puntos $x$. En lugar de eso, se selecciona un punto específico $x_0$ alrededor del cual se resolverá la ecuación para hallar los valores de $y(x)$. Este punto $x_0$ se elige porque corresponde a una región representativa: un máximo o mínimo del potencial $V_1(x)$ refleja el comportamiento dominante de la función de onda, mientras que una raíz de $\lambda_0(x) = 0$ asegura estabilidad en la expansión. De esta manera, se obtiene una región adecuada para encontrar la solución de manera más eficiente.

Para hallar las eigenfunciones de (4.1.23), es necesario primero desarrollar $\eta(x, \lambda)$ en una serie de potencias respecto de $\lambda$, obteniendo así:

$$\eta(x, \lambda) = \frac{s_k(x, \lambda)}{\lambda_k(x, \lambda)} = \sum_{j=0}^{\infty} \lambda^j \eta^{(j)}(x), \tag{4.1.31}$$

$$\eta^{(j)}(x) = \frac{1}{j!} \left. \frac{\partial^j}{\partial \lambda^j} \left( \frac{s_k(x, \lambda)}{\lambda_k(x, \lambda)} \right) \right|_{\lambda=0}. \tag{4.1.32}$$

Para continuar con el método se debe de sustituir (4.1.32) en (4.1.19) para obtener

$$y(x) = C_2 \exp\left( -\int \left[ \eta^{(0)}(t) + \lambda \eta^{(1)}(t) + \lambda^2 \eta^{(2)}(t) + \cdots \right] \mathrm{d}t \right). \tag{4.1.33}$$

Esta expresión se puede descomponer en productos de exponenciales. Entonces, (4.1.33) se puede expresar de la siguiente forma

$$y(x) = C_2 \exp\left( -\int \eta^{(0)}(t) \, \mathrm{d}t \right) \exp\left( -\lambda \int \eta^{(1)}(t) \, \mathrm{d}t \right) \exp\left( -\lambda^2 \int \eta^{(2)}(t) \, \mathrm{d}t \right) \cdots. \tag{4.1.34}$$

Reescribiendo (4.1.34), se tiene

$$y(x) = C_2 \prod_{k=0}^{\infty} y^{(k)}(x), \tag{4.1.35}$$

donde $y^{(k)}(x)$ se define como

$$y^{(k)}(x) = \exp\left( -\lambda^k \int \eta^{(k)}(t) \, \mathrm{d}t \right). \tag{4.1.36}$$

## 4.2 |Modificación de las condiciones de frontera

En esta sección se resolverá la ecuación de Whittaker, la cual surge del análisis de perfiles de campos eléctricos y magnéticos con decaimiento exponencial y singulares, como se mostró en el capítulo anterior. No obstante, se introduce aquí una variación importante: en lugar de considerar que dichos perfiles están definidos en todo el intervalo $x \in (-\infty, \infty)$, se introduce una variable $x_0$ que restringe el dominio en el que están definidos los campos eléctricos y magnéticos, como se muestra en la Figura 4.2.1. Este cambio permite analizar cómo la solución y los niveles de energía se ven afectados por la reducción del intervalo en el que se definen los perfiles de los campos eléctrico y magnético $x \in (-\infty, x_0]$.

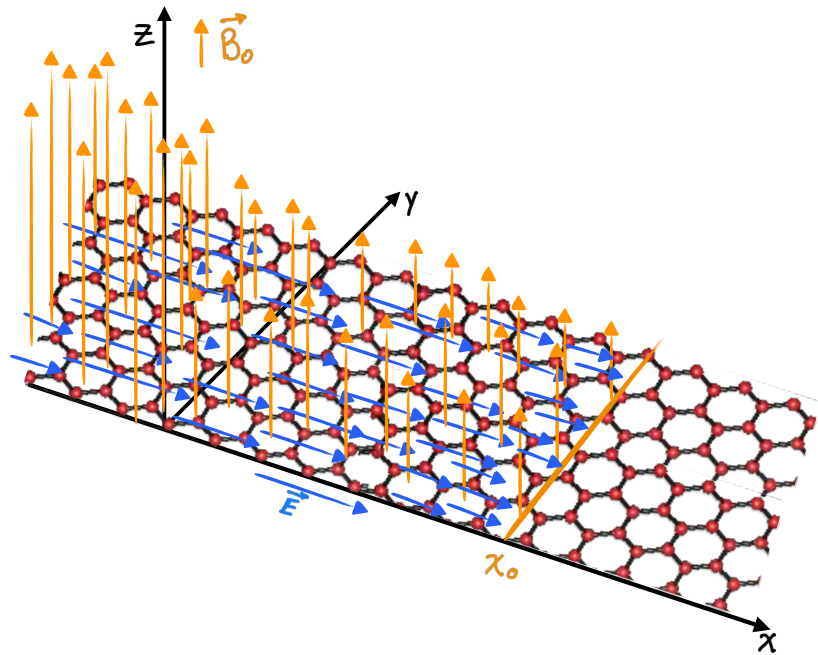

**Figura 4.2.1:** Monocapa de un material de Dirac (red hexagonal) interactuando con un campo eléctrico **E** y un campo magnético **B** externos existentes en el intervalo $(-\infty, x_0]$, ambos con una simetría de traslación en la coordenada $y$.

La modificación en la región de definición de los campos eléctrico y magnético conlleva a la necesidad de aplicar una nueva condición de frontera en el punto $x_0$ (cabe mencionar que en la variable $z$, se tiene $z \in [0, z_0(x_0)]$). Esta modificación también permite considerar una representación más realista de muchos sistemas físicos, ya que en los experimentos de laboratorio es común trabajar con sistemas definidos en una región finita, mientras que los modelos ideales que se extienden en todo el espacio no reflejan las restricciones físicas de un espacio limitado. De este modo, al imponer la condición en $z_0$, se obtiene una descripción más adecuada y relevante para sistemas confinados en dominios finitos.

Partiendo de la ecuación de Whittaker genérica (similar a las ecuaciones (3.2.15) y (3.3.14)),

$$\frac{\mathrm{d}^2\psi_j(z)}{\mathrm{d}z^2} + \left[-\frac{1}{4} + \frac{k}{z} + \frac{1/4 - \mu_j^2}{z^2}\right]\psi_j(z) = 0, \quad j = 1, 2, \quad \text{con } z \in [0, \infty) \quad (4.2.1)$$

Para valores de $z$ alrededor de cero ($z \to 0$), la ecuación (4.2.1) toma la forma

$$\frac{\mathrm{d}^2\psi_j(z)}{\mathrm{d}z^2} + \frac{\frac{1}{4} - \mu_j^2}{z^2}\psi_j(z) = 0, \quad (4.2.2)$$

cuyas soluciones generales son

$$\psi_j(z) = \bar{C}\, z^{\frac{1}{2}+\mu_j} + \bar{D}\, z^{\frac{1}{2}-\mu_j}. \quad (4.2.3)$$

Al imponer la condición de frontera tipo Dirichlet en $z = 0$, es decir, $\psi_j(0) = 0$, se descarta el término divergente y se fija $\bar{D} = 0$. Por lo tanto, la solución queda dada por

$$\psi_j(z) = \bar{C}\, z^{\frac{1}{2}+\mu_j}. \quad (4.2.4)$$

en donde $\bar{C}$ es una constante

Ahora bien, para el caso en el que $z \to z_0$, la ecuación de Whittaker toma la forma de

$$\frac{\mathrm{d}^2\psi_j(z)}{\mathrm{d}z^2} - \xi_j^2\psi_j(z) = 0, \quad j = 1, 2, \quad (4.2.5)$$

que es una ecuación lineal de segundo orden con coeficientes constantes. Su solución general corresponde a una combinación de exponenciales crecientes y decrecientes:

$$\psi_j(z) = \bar{A}\mathrm{e}^{\xi_j z} + \bar{B}\mathrm{e}^{-\xi_j z}, \quad (4.2.6)$$

donde $\bar{A}$ y $\bar{B}$ son contantes, mientras que $\xi_j$ esta definido como

$$\xi_j = \xi_j(z_0) = \sqrt{\frac{1}{4} - \frac{k}{z_0} - \frac{1/4 - \mu_j^2}{z_0^2}}\,. \quad (4.2.7)$$

la solución (4.2.6) puede escribirse como

$$\psi_j(z) = \bar{A}\,e^{\xi_j z} + \bar{B}\,e^{-\xi_j z} = C\,\sinh\big[\xi_j(z - z_0)\big] + D\,\cosh\big[\xi_j(z - z_0)\big], \qquad (4.2.8)$$

con $C = \bar{A} - \bar{B}$ y $D = \bar{A} + \bar{B}$. Si se impone la condición de frontera tipo Dirichlet, $\psi_j(z_0) = 0$, se obtiene $D = 0$ y es posible obtener la siguiente expresión

$$\psi_j(z) = C\,\sinh\big[\xi_j(z - z_0)\big] = \tilde{B}\Big(e^{-\xi_j z} - e^{\xi_j(z - 2z_0)}\Big), \qquad (4.2.9)$$

Sin embargo, al analizar el comportamiento de la solución (4.2.9) en el límite $z \to z_0$ y simultáneamente $z_0 \to \infty$, se observa un comportamiento anómalo, ya que la función puede presentar términos no normalizables en dicho límite. Para resolver este inconveniente, se propone la forma modificada

$$\psi_j(z) = c_1\left(e^{-\xi_j z} - e^{-\xi_j z_0}\right), \qquad (4.2.10)$$

la cuál satisface de manera exacta la condición de frontera de Dirichlet en $z \to z_0$ y, adicionalmente, en el límite $z_0 \to \infty$ recupera la solución presentada en (ver apendice A). De esta manera, (4.2.10) asegura un comportamiento bien definido tanto en la frontera finita (caso confinado) como en el caso extendido (caso no confinado).

Para implementar el método de iteración asintótica, se propone que la función $\psi_j$ sea de la forma:

$$\psi_j(z) = z^{\frac{1}{2} + \mu_j}\left(e^{-\xi_j z} - e^{-\xi_j z_0}\right) y_j(z). \qquad (4.2.11)$$

Introduciendo $y_j(z)$ en (4.2.1), se obtiene

$$\frac{\mathrm{d}^2 y_j(z)}{\mathrm{d}z^2} + \left(\frac{1 + 2\mu_j}{z} - \frac{\xi_j e^{-\xi_j z}}{e^{-\xi_j z} - e^{-\xi_j z_0}}\right)\frac{\mathrm{d}y_j(z)}{\mathrm{d}z}$$
$$- \left(\frac{z - 4k}{2z} + \frac{\xi_j e^{-\xi_j z}\left(2 + 4\mu_j - z\right)}{2z\left(e^{-\xi_j z} - e^{-\xi_j z_0}\right)}\right) y_j(z) = 0. \qquad (4.2.12)$$

De acuerdo con el método de iteración asintótica, las funciones $\lambda_0(z)$ y $s_0(z)$ son

$$\lambda_0(z, z_0) = -\frac{1/2 + \mu_j}{z} + \gamma_1, \qquad (4.2.13)$$

$$s_0(z, z_0) = \frac{z - 4k}{4z} + \gamma_2, \qquad (4.2.14)$$

donde

$$\gamma_1(z, z_0) \equiv \gamma_1 = \frac{2\,\xi_j \mathrm{e}^{-\xi_j z}}{\mathrm{e}^{-\xi_j z} - \mathrm{e}^{-\xi_j z_0}}, \tag{4.2.15}$$

$$\gamma_2(z, z_0) \equiv \gamma_2 = \frac{\xi_j \mathrm{e}^{-\xi_j z}\left(1 + 2\,\mu_j - z\,\xi\right)}{z\left(\mathrm{e}^{-\xi_j z} - \mathrm{e}^{-\xi_j z_0}\right)}. \tag{4.2.16}$$

Para seguir con el método, es necesario calcular las funciones $\lambda_n(z, z_0)$ y $s_n(z, z_0)$ definidas en las ecuaciones (4.1.8) y (4.1.9) de la subsección anterior. Se tiene entonces:

$$\lambda_1(z, z_0) = \frac{\partial \gamma_1}{\partial z} + \frac{\gamma_2 z - 4k + z}{4z} + \frac{\left[\gamma_1 z - (2\mu_j + 1)\right]^2 + (2\mu_j + 1)}{z^2}, \tag{4.2.17}$$

$$s_1(z, z_0) = \frac{\partial \gamma_2}{\partial z} + \frac{\left(4\gamma_2 z - 4k + z\right)\left[\gamma_1 z - (2\mu_j + 1)\right] + k}{4z^2}, \tag{4.2.18}$$

$$\lambda_2(z, z_0) = \frac{\partial^2 \gamma_1}{\partial z^2} + \frac{\left[2z^3\gamma_1 - 4(4\mu_j + 2)z^2\right]\frac{\partial \gamma_1}{\partial z} + \left[2\gamma_1 z - (4\mu_j + 4)\right](4\mu_j + 2)}{z^3} + \frac{2\left(z^2 \frac{\partial \gamma_2}{\partial z} + k\right)}{z^2}$$
$$+ \frac{\left[\gamma_1 z - (2\mu_j + 1)^2\right]\left[z^2\frac{\partial \gamma_1}{\partial z} + z\gamma_1\left[z\gamma_1 - (4\mu_j + 2)\right] + z\left(2z\gamma_2 - 2k + \frac{z}{2}\right) + (2\mu_j + 2)(2\mu_j + 1)\right]}{z^3}, \tag{4.2.19}$$

$$s_2(z, z_0) = \frac{\partial^2 \gamma_2}{\partial z^2} - \frac{2k}{z^3} + \frac{\left(4z\frac{\partial \gamma_1}{\partial z} + 4\gamma_2 + 1\right)\left[z\gamma_1 - (2\mu_j + 1)\right] + (4z\gamma_2 - 4k + z)\left(z\frac{\partial \gamma_1}{\partial z} + \gamma_1\right)}{4z^2}$$
$$+ \frac{(4z\gamma_2 + 4k + z)\left[4z^2\frac{\partial \gamma_1}{\partial z} + 4z\gamma_1\left[z\gamma_1 - (4\mu_j + 4)\right] + z\left(4z\gamma_2 - 4k + z\right) + (4\mu_j + 8)(8\mu_j + 4)\right]}{16z^3}. \tag{4.2.20}$$

Como las expresiones de $\lambda_n(z, z_0)$ y $s_n(z, z_0)$ con $n = 3, 4$ resultan bastante extensas, no se incluyen de manera explícita en el presente documento. Sin embargo, su forma explícita puede obtenerse directamente aplicando el método descrito anteriormente. Cabe señalar que cuando $z_0$ tiende a infinito se recupera las soluciones obtenidas de $\lambda_n(z)$ y $s_n(z)$, con $n = 1, 2, 3$ del Apéndice A.

El próximo paso del método consiste en usar la condición de cuantización en (4.1.29) de la Subsección 4.1.1. Para ello, se utilizan las expresiones en (4.2.13), (4.2.14) y (4.2.17) - (4.2.19), lo cual permite encontrar los eigenvalores del sistema, calculando

$\delta_1(z, z_0)$, $\delta_2(z, z_0)$, $\delta_3(z, z_0)$ y $\delta_4(z, z_0)$:

$$\delta_1(z) = \left(\frac{\partial \gamma_2}{\partial z} + \frac{[4\gamma_2 z - 4k + z]\left(\gamma_1 z - (2\mu_j + 1)\right) + k}{4z^2}\right)\left(-\frac{1/2 + \mu_j}{z} + \gamma_1\right)$$

$$- \left(\frac{\partial \gamma_1}{\partial z} + \frac{\gamma_2 z - 4k + z}{4z} + \frac{[\gamma_1 z - (2\mu_j + 1)]^2 + (2\mu_j + 1)}{z^2}\right)\left(\frac{z - 4k}{4z} + \gamma_2\right),$$
$$(4.2.21)$$

$$\delta_2(z) = \left(\frac{\partial^2 \gamma_2}{\partial z^2} - \frac{2k}{z^3} + \frac{\left(4z\frac{\partial \gamma_1}{\partial z} + 4\gamma_2 + 1\right)(z\gamma_1 - (2\mu_j + 1)) + (4z\gamma_2 - 4k + z)\left(z\frac{\partial \gamma_1}{\partial z} + \gamma_1\right)}{4z^2}\right.$$

$$\left. + \frac{(4z\gamma_2 + 4k + z)\left[4z^2\frac{\partial \gamma_1}{\partial z} + 4z\gamma_1(z\gamma_1 - (4\mu_j + 4)) + z(4z\gamma_2 - 4k + z) + (4\mu_j + 8)(8\mu_j + 4)\right]}{16z^3}\right)$$

$$\times \left(\frac{\partial \gamma_1}{\partial z} + \frac{\gamma_2 z - 4k + z}{4z} + \frac{(\gamma_1 z - (2\mu_j + 1))^2 + (2\mu_j + 1)}{z^2}\right)$$

$$- \left(\frac{\partial^2 \gamma_1}{\partial z^2} + \frac{(2z^3\gamma_1 - 4(4\mu_j + 2)z^2)\frac{\partial \gamma_1}{\partial z} + (2\gamma_1 z - (4\mu_j + 4))(4\mu_j + 2)}{z^3}\right.$$

$$\left. + \frac{2\left(z^2\frac{\partial \gamma_2}{\partial z} + k\right)}{z^2} + \frac{(\gamma_1 z - (2\mu_j + 1)^2)}{z^3}\left[\begin{array}{l}z^2\frac{\partial \gamma_1}{\partial z} + z\gamma_1(z\gamma_1 - (4\mu_j + 2)) \\ + z\left(2z\gamma_2 - 2k + \frac{z}{2}\right) + (2\mu_j + 2)(2\mu_j + 1)\end{array}\right]\right)$$

$$\times \left(\frac{\partial \gamma_2}{\partial z} + \frac{(4\gamma_2 z - 4k + z)\left(\gamma_1 z - (2\mu_j + 1)\right) + k}{4z^2}\right).$$
$$(4.2.22)$$

Debido a que las expresiones para $\delta_3(z, z_0)$ y $\delta_4(z, z_0)$ son particularmente extensas, se ha optado por no incluirlas explícitamente en el desarrollo principal. No obstante, su obtención se sigue de manera directa a partir del método previamente descrito.

La atención se centrará ahora en el enfoque perturbativo descrito en la Subsección 4.1.1. En particular, se parte de la ecuación (4.1.29), donde se realiza un desarrollo en serie de Taylor alrededor del límite $z_0 \to \infty$.

Retomando lo expuesto en la Subsección 4.1.1, el cálculo de los eigenvalores a partir del término $\delta_n(z, z_0)$ presenta dificultades debido a su dependencia explícita con respecto a $z$. Para superar esta complicación, se evalúa $\delta_n(z, z_0)$ en el punto $z = z_1$, donde $z_1$ es la raíz de la ecuación $\lambda_0(z = z_1, z_0) = 0$.

Así, el valor de $z_1$ queda determinado por:

$$z_1 = -\frac{1}{2}\frac{-2\mu_j + 2\,\mathrm{W}\left(\left(\frac{1}{2} + \mu_j\right)\mathrm{e}^{-\xi_j z_0 + \frac{1}{2} + \mu_j}\right) - 1}{\xi_j}, \tag{4.2.23}$$

donde $\mathrm{W}(x)$ es la función de Lambert [Olver et al., 2010].

A continuación, se emplean las funciones $\delta_n(z_1, z_0)$ realizando un desarrollo en serie de Taylor en torno $z_0 \to \infty$, considerando únicamente el término de primer orden.

Para $n = 1$, se tiene

$$\delta_1(z_1, z_0) = \delta_1^{(0)} + \frac{1}{z_0}\delta_1^{(1)}, \tag{4.2.24}$$

con

$$\delta_1^{(0)} = -\frac{\left(\mu_j - k + \frac{3}{2}\right)\left(\mu_j - k + \frac{1}{2}\right)}{(1 + 2\mu_j)^2}, \tag{4.2.25}$$

$$\delta_1^{(1)} = \frac{4k\,\left(\mu_j - k + 2\right)\left(\mu_j - k + \frac{1}{2}\right)}{\left(1 + 2\mu_j\right)^2 z_0}. \tag{4.2.26}$$

Mientras que para $n = 2$, se tiene

$$\delta_2(z_1, z_0) = \delta_2^{(0)} + \frac{1}{z_0}\delta_1^{(2)}, \tag{4.2.27}$$

donde $\delta_2^{(0)}$ y $\delta_2^{(1)}$ se definen como

$$\delta_2^{(0)} = \frac{\left(\mu_j - k + \frac{5}{2}\right)\left(\mu_j - k + \frac{3}{2}\right)\left(\mu_j - k + \frac{1}{2}\right)}{\left(1 + 2\mu_j\right)^3}, \tag{4.2.28}$$

$$\delta_2^{(1)} = -\frac{6k\left(\mu_j - k + \frac{7}{2}\right)\left(\mu_j - k + \frac{3}{2}\right)\left(\mu_j - k + \frac{1}{2}\right)}{4\left(1 + 2\mu_j\right)^3 z_0}. \tag{4.2.29}$$

Si $n = 3$, se puede escribir

$$\delta_3(z_1, z_0) = \delta_1^{(0)} + \frac{1}{z_0}\delta_3^{(1)}, \tag{4.2.30}$$

donde se tiene

$$\delta_3^{(0)} = -\frac{\left(\mu_j - k + \frac{7}{2}\right)\left(\mu_j - k + \frac{5}{2}\right)\left(\mu_j - k + \frac{3}{2}\right)\left(\mu_j - k + \frac{1}{2}\right)}{(1 + 2\mu_j)^4}, \tag{4.2.31}$$

$$\delta_3^{(1)} = \frac{8k\left(\mu_j - k + 5\right)\left(\mu_j - k + \frac{5}{2}\right)\left(\mu_j - k + \frac{3}{2}\right)\left(\mu_j - k + \frac{1}{2}\right)}{4\left(1 + 2\mu_j\right)^4 z_0}. \tag{4.2.32}$$

Se observa que, en el límite $z_0 \to \infty$, las expresiones (4.2.24), (4.2.27) y (4.2.30) se reducen a las ecuaciones (A.0.14), (A.0.15) y (A.0.16), las cuales proporcionan, de manera evidente, los eigenvalores del sistema no perturbado, (A.0.21).

Siguiendo con el método de iteración asintótica en su forma perturbativa, el cálculo de los eigenvalores al orden dominante se realiza a partir de las ecuaciones (4.2.25), (4.2.28) y (4.2.31). Al resolver $\delta_n^{(0)} = 0$ para $n = 1, 2, 3, \ldots$, es posible obtener la siguiente condición:

$$k - \mu_j - \frac{1}{2} = n, \quad \text{con} \quad n = 0, 1, 2, \ldots. \tag{4.2.33}$$

Para mayores detalles sobre el cálculo, se puede consultar el Apéndice A, donde la ecuación (A.0.33) proporciona los niveles de energía $E_n^{(0)}$ para perfiles de campo exponencial, mientras que la ecuación (A.0.36) describe el caso de perfiles de campo singulares.

En el cálculo de los niveles de energía de primer orden, $E_n^{(1)}$, es necesario determinar los eigenvalores a partir de la condición $\delta_n^{(1)} = 0$, con $n = 1, 2, 3, \ldots$. En particular, se requiere calcular los ceros de dichas ecuaciones, obteniéndose así las siguientes raíces:

$$\delta_1^{(1)}(z) = 0 \quad \Rightarrow \quad \boxed{\mu_j - k + 2 = 0}, \quad \mu_j - k + \frac{1}{2} = 0, \tag{4.2.34}$$

$$\delta_2^{(1)}(z) = 0 \quad \Rightarrow \quad \boxed{\mu_j - k + \frac{7}{2} = 0}, \quad \mu_j - k + \frac{3}{2} = 0, \quad \mu_j - k + \frac{1}{2} = 0, \tag{4.2.35}$$

$$\delta_3^{(1)}(z) = 0 \quad \Rightarrow \quad \boxed{\mu_j - k + 5 = 0}, \quad \mu_j - k + \frac{5}{2} = 0, \quad \mu_j - k + \frac{3}{2} = 0, \tag{4.2.36}$$

$$\mu_j - k + \frac{1}{2} = 0, \tag{4.2.37}$$

$$\delta_4^{(1)}(z) = 0 \quad \Rightarrow \quad \boxed{\mu_j - k + \frac{13}{2} = 0}, \quad \mu_j - k + \frac{7}{2} = 0, \quad \mu_j - k + \frac{5}{2} = 0, \tag{4.2.38}$$

$$\mu_j - k + \frac{3}{2} = 0, \quad \mu_j - k + \frac{1}{2} = 0.$$

De las ecuaciones anteriores, las más relevantes son aquellas enmarcadas en un recuadro. Estas pueden expresarse, de manera general, de la siguiente forma:

$$\boxed{k - \mu_j - \frac{1}{2} = \frac{3}{2}(n+1)} \quad \text{con} \quad n = 0, 1, 2, \ldots, \tag{4.2.39}$$

A partir de (4.2.39) es posible obtener la expresión de la energía de primer orden, $E_n^{(1)}$, la cual dependerá de cada caso y será retomada más adelante.

A continuación, se discute la obtención de las funciones de onda $\psi_j(z)$, el primer paso consiste en calcular las funciones auxiliares $\eta_n(z, z_0)$ dadas por la ecuación (4.1.31),

$$\eta_n(z, z_0) = \frac{s_n(z, z_0)}{\lambda_n(z, z_0)}, \tag{4.2.40}$$

y posteriormente realizar un desarrollo en serie de Taylor alrededor de $z_0 \to \infty$ (o, de manera equivalente, $1/z_0 \to 0$).

De este modo, las eigenfunciones a primer orden de la ecuación (4.2.12) toman la forma

$$y_n(z) = \mathcal{C}_n y_n^{(0)}(z) y_n^{(1)}(z), \quad n = 0, 1, 2, \ldots, \tag{4.2.41}$$

donde $\mathcal{C}_n$ es una constante de normalización y las funciones $y_n^{(0)}(z)$ y $y_n^{(1)}(z)$ estan definidas como

$$y_n^{(0)}(z) = \exp\left[-\int \eta_n^{(0)}(z)\Big|_{k=\mu_j+\frac{1}{2}+n} \mathrm{d}z\right], \tag{4.2.42}$$

$$y_n^{(1)}(z) = \exp\left[-\frac{1}{z_0}\int \eta_n^{(1)}(z)\Big|_{k=\mu_j+\frac{1}{2}+\frac{3}{2}(n+1)} \mathrm{d}z\right]. \tag{4.2.43}$$

Para proseguir con el cálculo es necesario determinar $\eta_n(z, z_0)$ para el estado base, es decir para $n = 0$. De (4.1.31) resulta:

$$\eta_0(z, z_0) = \frac{(-4\xi_j^2 z + (8\mu_j + 4)\xi_j - 4k + z)\mathrm{e}^{-\xi_j z} + (4k - z)\mathrm{e}^{-\xi_j z_0}}{(8\xi_j z - 8\mu_j - 4)\mathrm{e}^{-\xi_j z} + (8\mu_j + 4)\mathrm{e}^{-\xi_j z_0}}, \tag{4.2.44}$$

y luego, realizando el desarrollo en serie de Taylor en la variable $z_0$ en (4.2.44), se tiene

$$\eta_0(z, z_0) = \eta_0^{(0)}(z) + \frac{1}{z_0}\eta_0^{(1)}(z) + \ldots. \tag{4.2.45}$$

Aquí, el primer término, $\eta_0^{(0)}(z)$, coincide con el caso en el cual las condiciones de frontera no se modifican, mientras que el término $\eta_0^{(1)}(z)$ describe la corrección a primer orden cuando las condiciones de frontera son modificadas. Ambos se expresan de la siguiente forma:

$$\eta_0^{(0)}(z) = \frac{k - \mu_j - \frac{1}{2}}{-z + 2\mu_j + 1}, \tag{4.2.46}$$

$$\eta_0^{(1)}(z) = \frac{2(k - \mu_j - \frac{1}{2})zk}{(z - 2\mu_j - 1)^2} + k. \tag{4.2.47}$$

Evaluando estas expresiones en el eigenvalor correspondiente, se tiene:

$$\eta_0^{(0)}(z)\Big|_{k=\mu_j+\frac{1}{2}} = \frac{k-\mu_j-\frac{1}{2}}{-z+2\mu_j+1}\Big|_{k-\mu_j-\frac{1}{2}=0} = 0, \tag{4.2.48}$$

$$\eta_0^{(1)}(z)\Big|_{k=\mu_j+\frac{1}{2}+\frac{3}{2}} = \frac{2zk\left(k-\mu_j-\frac{1}{2}\right)}{\left(z-2\mu_j-1\right)^2}+k\Big|_{k-\mu_j-2=0} = \frac{3z\left(\mu_j+2\right)}{\left(z-2\mu_j-1\right)^2}+\mu_j+2. \tag{4.2.49}$$

Entonces, introduciendo (4.2.48) en (4.2.42) y (4.2.49) en (4.2.43), es posible calcular las funciones $y_0^{(0,1)}(z)$ del estado base, obteniendo

$$y_{0j}^{(0)}(z) = \exp\left[-\int \eta_0^{(0)}(z)\,\mathrm{d}z\right] = 1, \tag{4.2.50}$$

$$y_{0j}^{(1)}(z) = \exp\left[-\frac{1}{z_0}\int \eta_0^{(1)}(z)\,\mathrm{d}z\right] = \mathrm{e}^{-\frac{z(\mu_j+2)}{z_0}}\mathrm{e}^{-\frac{3(\mu_j+2)(2\mu_j+1)}{z_0(z-2\mu_j-1)}}\left(z-2\mu_j-1\right)^{\frac{-3(\mu_j+2)}{z_0}}. \tag{4.2.51}$$

Combinando estas expresiones con (4.2.41) se obtiene la función $y_0(z)$,

$$y_0(z) = \mathcal{C}_{0j}\,y_0^{(0)}(z)y_0^{(1)}(z) = \mathcal{C}_{0j}\mathrm{e}^{-\frac{z(\mu_j+2)}{z_0}}\mathrm{e}^{-\frac{3(\mu_j+2)(2\mu_j+1)}{z_0(z-2\mu_j-1)}}\left(z-2\mu_j-1\right)^{\frac{-3(\mu_j+2)}{z_0}}, \tag{4.2.52}$$

y, en consecuencia, al sustituir $y_0(z)$ en (4.2.11), la eigenfunción de primer orden para el estado fundamental toma la siguiente forma:

$$\psi_{0,j}^{(1)}(z) = \mathcal{C}_{0j}z^{\frac{1}{2}+\mu_j}\left(\mathrm{e}^{-\xi_j z}-\mathrm{e}^{-\xi_j z_0}\right)\mathrm{e}^{-\frac{z(\mu_j+2)}{z_0}}\mathrm{e}^{-\frac{3(\mu_j+2)(2\mu_j+1)}{z_0(z-2\mu_j-1)}}\left(z-2\mu_j-1\right)^{\frac{-3(\mu_j+2)}{z_0}}. \tag{4.2.53}$$

Cabe destacar que (4.2.53) presenta una discontinuidad en el punto

$$z_0 = 2\mu_j+1. \tag{4.2.54}$$

Este aspecto será abordado más adelante.

Continuando con el primer estado excitado ($n=1$), se calcula $\eta_1(z,z_0)$ y, posteriormente, se realiza el desarrollo de Taylor en la variable $z_0$, obteniéndose:

$$\eta_1(z,z_0) = \eta_1^{(0)}(z) + \frac{1}{z_0}\eta_1^{(1)}(z) + \dots, \tag{4.2.55}$$

con $\eta_1^{(0)}(z)$ y $\eta_1^{(1)}(z)$ definidas como

$$\eta_1^{(0)}(z) = \frac{2(z - 2\mu_j - 2)(k - \mu_j - \frac{1}{2})}{-2z^2 + (6\mu_j + 2k + 3)z - 8\mu_j^2 - 12\mu_j - 4}, \tag{4.2.56}$$

$$\eta_1^{(1)}(z) = \frac{12zk(z - 2\mu_j - 2)\left(k - \mu_j - \frac{1}{2}\right)(2\mu_j + 1 - z)}{(2kz - 8\mu_j^2 + 6z\mu_j - 2z^2 - 12\mu_j + 3z - 4)^2}$$
$$- \frac{2k(2kz + 4\mu_j^2 - 6z\mu_j + z^2 + 6\mu_j - 3z + 2)}{(2kz - 8\mu_j^2 + 6z\mu_j - 2z^2 - 12\mu_j + 3z - 4)^2}. \tag{4.2.57}$$

Siguiendo un procedimiento análogo al del estado fundamental, es posible calcular las funciones $y_1^{(0,1)}(z)$ correspondientes al primer estado excitado, obteniéndose:

$$y_1^{(0)}(z) = \exp\left[-\int \left.\eta_1^{(0)}(z)\right|_{k-\mu_j-\frac{3}{2}=0} \mathrm{d}z\right] = z - 2\mu_j - 1\,, \tag{4.2.58}$$

$$y_1^{(1)}(z) = \exp\left[\int \left.\eta_1^{(1)}(z)\right|_{k-\mu_j-\frac{7}{2}=0} \mathrm{d}z\right]$$
$$= \mathrm{e}^{-\frac{z(\mu_j+\frac{7}{2})}{z_0}} \exp\left(\frac{9(2\mu_j + 7)\left[2(4\mu_j + 5)(2\mu_j + 1)(\mu_j + 1) - (8\mu_j^2 + 28\mu_j + 21)z\right]}{(16\mu_j + 17)[2(\mu_j + 1)(2\mu_j + 1) - (4\mu_j + 5)z + z^2]z_0}\right)$$
$$\times \left(\frac{\sqrt{16\mu_j + 17} + 2z - 4\mu_j - 5}{\sqrt{16\mu_j + 17} - 2z + 4\mu_j + 5}\right)^{\frac{36(2\mu_j+7)(2\mu_j+1)(\mu_j+1)}{z_0(16\mu_j+17)^{3/2}}}. \tag{4.2.59}$$

Finalmente, al sustituir $y_1^{(0)}(z)$ y $y_1^{(1)}(z)$ en (4.2.11), se obtiene la eigenfunción del primer estado excitado a primer orden:

$$\psi_{1,j}^{(1)}(z) = \mathcal{C}_{1j}\, z^{\frac{1}{2}+\mu_j}(z - 2\mu_j - 1)\left(\frac{\sqrt{16\mu_j + 17} + 2z - 4\mu_j - 5}{\sqrt{16\mu_j + 17} - 2z + 4\mu_j + 5}\right)^{\frac{36(2\mu_j+7)(2\mu_j+1)(\mu_j+1)}{z_0(16\mu_j+17)^{3/2}}} \mathrm{e}^{-\frac{z(\mu_j+\frac{7}{2})}{z_0}}$$
$$\times \exp\left(\frac{9(2\mu_j + 7)\left[2(4\mu_j + 5)(2\mu_j + 1)(\mu_j + 1) - (8\mu_j^2 + 28\mu_j + 21)z\right]}{(16\mu_j + 17)[2(\mu_j + 1)(2\mu_j + 1) - (4\mu_j + 5)z + z^2]z_0}\right)\left(\mathrm{e}^{-\xi_j z} - \mathrm{e}^{-\xi_j z_0}\right). \tag{4.2.60}$$

Del análisis de (4.2.60) se observa que persisten discontinuidades en los puntos

$$z_1 = \frac{4\mu_j + 5}{2} - \frac{\sqrt{16\mu_j + 17}}{2}, \tag{4.2.61}$$

$$z_2 = \frac{4\mu_j + 5}{2} + \frac{\sqrt{16\mu_j + 17}}{2}. \tag{4.2.62}$$

Estas singularidades serán abordadas más adelante.

En las siguientes secciones se analizarán por separado los dos casos previamente discutidos, con el fin de examinar a detalle la modificación de las condiciones de frontera.

# 4.3 | Campos externos con perfil exponencial decreciente.

En este caso, se retoman los valores de la Sección 3.2, considerando los $\mu_j$ dados por (3.2.16) y $k$ (denominado $\mu$ en dicha sección) definido en (3.2.9). Estos pueden escribirse como:

$$k = \frac{1}{\sqrt{1-\beta_\nu^2}}\left[\left(\frac{\bar{E}}{\alpha_c} - \frac{\nu\beta_\nu D}{\alpha_c}\right)\nu\beta_\nu + \left(\frac{k_y^c}{\alpha_c} + \frac{D}{\alpha_c}\right)\right], \qquad (4.3.1)$$

$$\mu_j = \lambda + \frac{(-1)^j}{2}, \quad j = 1, 2. \qquad (4.3.2)$$

Al sustituir estas cantidades en la ecuación (4.2.33), se obtiene la expresión para la corrección a la energía al orden dominante para el caso en que se modifican las condiciones de frontera. Esta expresión coincide con la ecuación (3.2.23) presentada en la Sección 3.2. De esta forma se tiene:

$$E_n^{(0)} = v_F\beta_\nu\alpha_c n\sqrt{1-\beta_\nu^2} - v_d k_y + \kappa\nu v_F\sqrt{1-\beta_\nu^2}\sqrt{\left(k_y^c + D\right)^2 - \left(k_y^c + D - \alpha_c n\sqrt{1-\beta_\nu^2}\right)^2}. \qquad (4.3.3)$$

Introduciendo nuevamente (4.3.1) y (4.3.2) en (4.2.39), se obtiene una expresión para la corrección de la energía a primer orden cuando se modifican las condiciones de frontera, como se muestra a continuación:

$$E_n^{(1)} = \frac{3}{2}v_F\beta_\nu\alpha_c\left(n+1\right)\sqrt{1-\beta_\nu^2} - v_d k_y$$
$$+ \kappa\nu v_F\sqrt{1-\beta_\nu^2}\sqrt{\left(k_y^c + D\right)^2 - \left(k_y^c + D - \frac{3}{2}\alpha_c\left(n+1\right)\sqrt{1-\beta_\nu^2}\right)^2}. \qquad (4.3.4)$$

Utilizando las ecuaciones (4.3.3) y (4.3.4) en la expresión (4.1.22), se obtiene la energía a primer orden para un perfil exponencial de los campos eléctrico y magnético, considerando la modificación en las condiciones de frontera:

$$E_n = E_n^{(0)} + \frac{1}{z_0}E_n^{(1)}. \tag{4.3.5}$$

A partir del resultado anterior, se sigue que

$$\Delta E_n = E_n - E_n^{(0)} = \frac{1}{z_0}E_n^{(1)}. \tag{4.3.6}$$

Por lo tanto, es claro que $\Delta E_n \to 0$ cuando $z_0 \to \infty$. En las Figuras 4.3.1-4.3.3 se muestra el comportamiento de $\Delta E_n$ como función de la intensidad del campo eléctrico $\mathcal{E}_0$ y del momento $k_y$, para un valor de $z_0 = 600$.

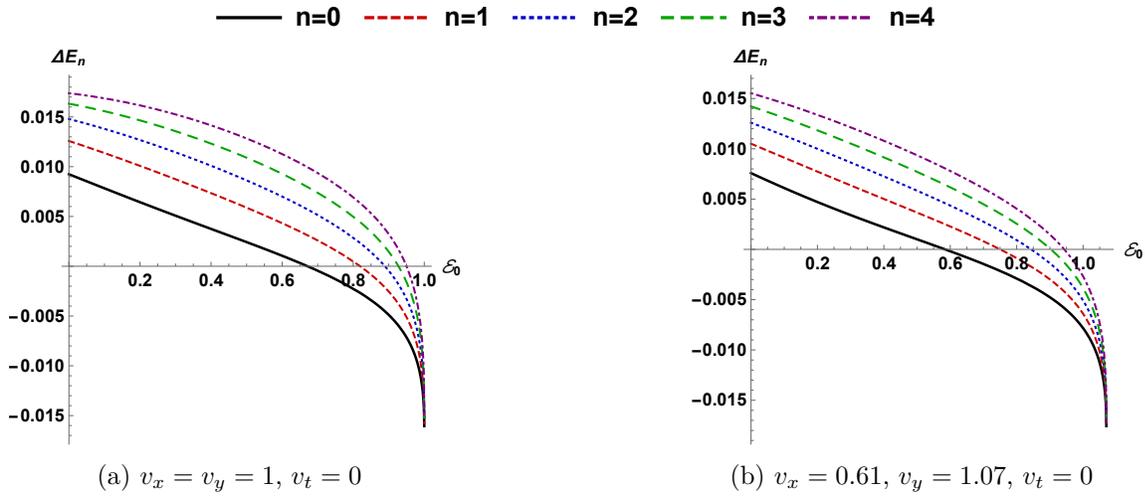

(a) $v_x = v_y = 1$, $v_t = 0$

(b) $v_x = 0.61$, $v_y = 1.07$, $v_t = 0$

**Figura 4.3.1:** Espectro de energía $\Delta E_n$ en (4.3.6) para el caso exponencial, con $\mathcal{B}_0 = 1$, $\alpha = 1$ y $k_y = 10$ en función del campo eléctrico $\mathcal{E}_0$ para (a) el grafeno prístino y (b) el grafeno bajo tensión a lo largo del eje $x$ con $\varepsilon = 0.15$.

A continuación, en las Figuras 4.3.5-4.3.8, se presentan las densidades de probabilidad $\rho_0(\mathbf{r})$ y las densidades de corriente $\mathcal{J}_{y,0}(x)$ correspondientes al estado fundamental ($n = 0$) para el caso de campos externos con perfil exponencial decreciente. Se consideran distintos materiales –grafeno prístino, grafeno bajo tensión uniaxial y borofeno $8 - Pmmn$– con parámetros característicos y distintos valores del momento transversal $k_y$. Las gráficas muestran cómo se distribuyen espacialmente las probabilidades y las corrientes a lo largo del eje $x$ bajo las condiciones del campo eléctrico aplicado.

En las Figuras 4.3.9-4.3.12, se presentan las densidades de probabilidad $\rho_1(\mathbf{r})$ y las densidades de corriente $\mathcal{J}_{y,1}(x)$ correspondientes al primer estado excitado ($n = 1$) en el caso exponencial. Se analizan diferentes materiales, incluyendo el grafeno prístino,

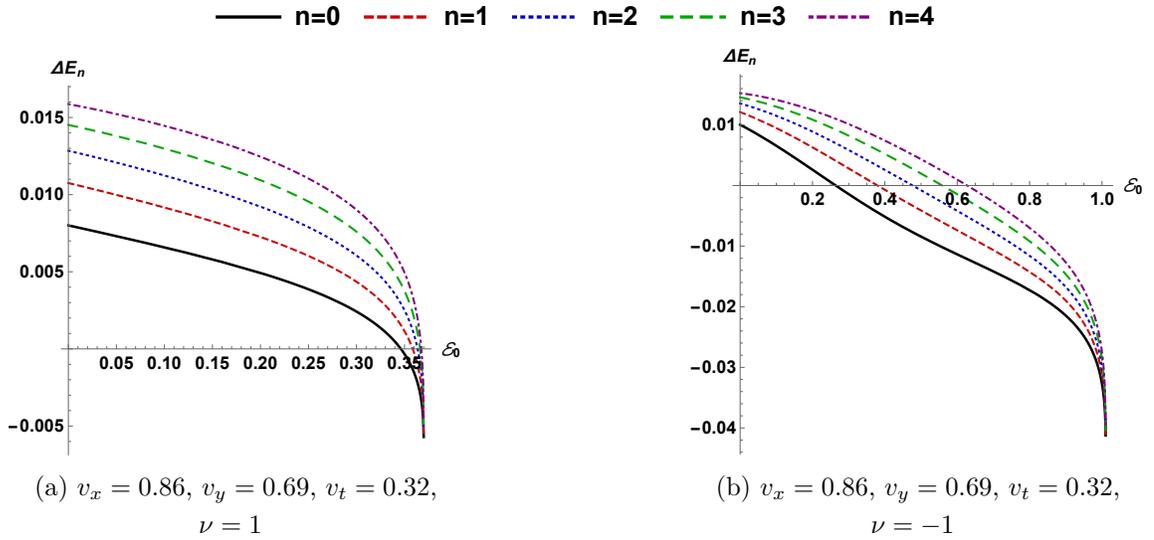

(a) $v_x = 0.86$, $v_y = 0.69$, $v_t = 0.32$, $\nu = 1$

(b) $v_x = 0.86$, $v_y = 0.69$, $v_t = 0.32$, $\nu = -1$

**Figura 4.3.2:** Espectro de energía $\Delta E_n$ en (4.3.6) para el caso exponencial, con $\mathcal{B}_0 = 1$, $\alpha = 1$ en función del campo eléctrico $\mathcal{E}_0$ para el borofeno $8 - Pmmn$ en (a) el valle $\mathrm{K}_+$ con $k_y = 10$ y (b) el valle $\mathrm{K}_-$ con $k_y = 25$.

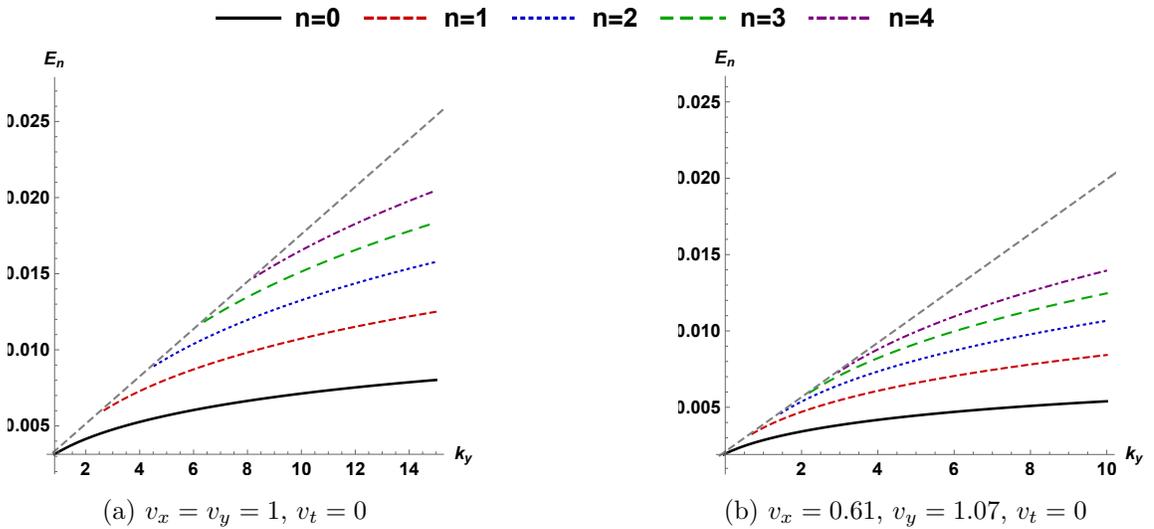

(a) $v_x = v_y = 1$, $v_t = 0$

(b) $v_x = 0.61$, $v_y = 1.07$, $v_t = 0$

**Figura 4.3.3:** Espectro de energía $\Delta E_n$ en (4.3.6) para el caso exponencial, con $\mathcal{B}_0 = 1$, $\alpha = 1$ y $\mathcal{E}_0 = 0.15$ en función del momento $k_y$ para (a) el grafeno prístino y (b) el grafeno bajo tensión a lo largo del eje $x$ con $\varepsilon = 0.15$.

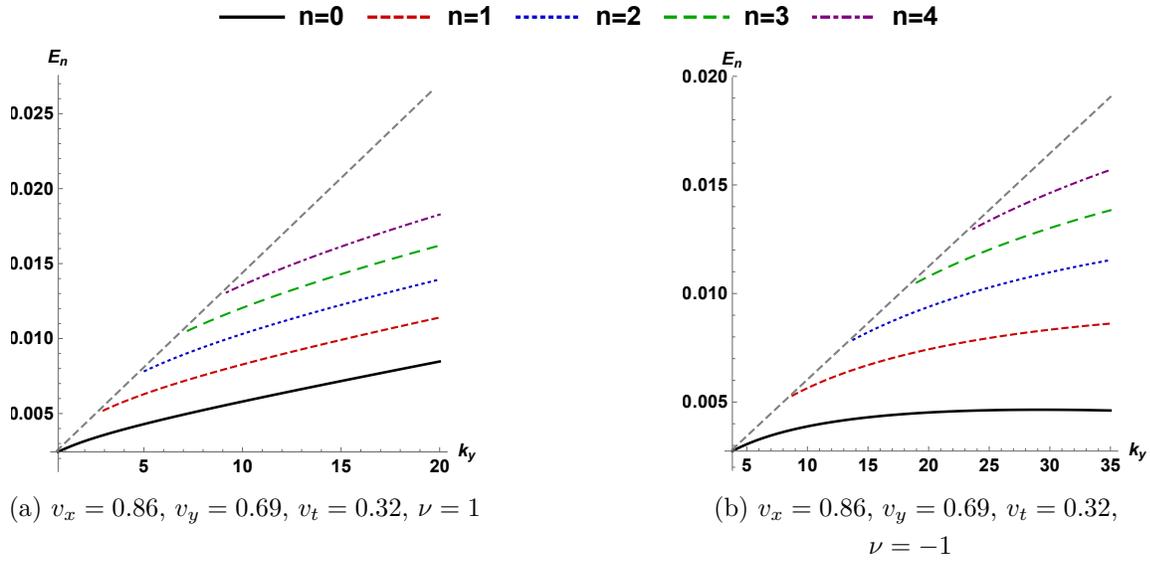

(a) $v_x = 0.86$, $v_y = 0.69$, $v_t = 0.32$, $\nu = 1$

(b) $v_x = 0.86$, $v_y = 0.69$, $v_t = 0.32$, $\nu = -1$

**Figura 4.3.4:** Espectro de energía $\Delta E_n$ en (4.3.6) para el caso exponencial, con $\mathcal{B}_0 = 1$, $\alpha = 1$ y $\mathcal{E}_0 = 0.15$ en función del momento $k_y$ para el borofeno $8 - Pmmn$ en (a) el valle $K_+$ y (b) el valle $K_-$.

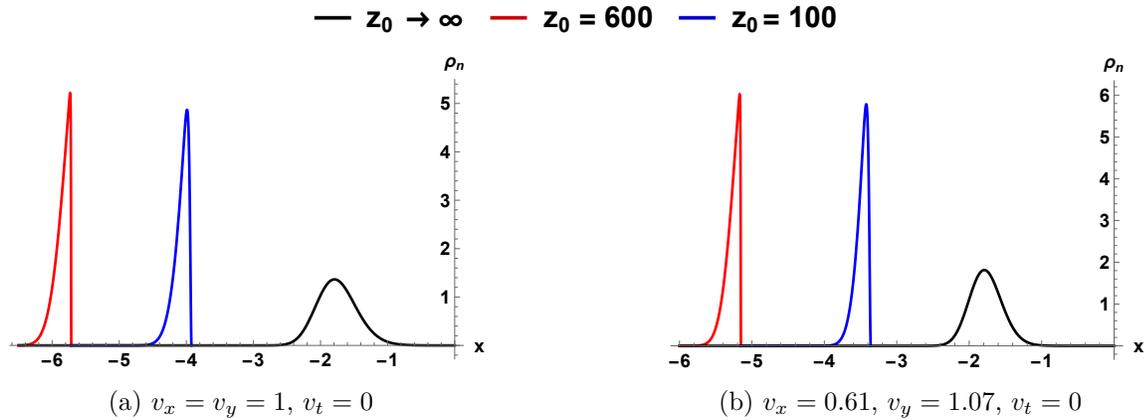

(a) $v_x = v_y = 1$, $v_t = 0$

(b) $v_x = 0.61$, $v_y = 1.07$, $v_t = 0$

**Figura 4.3.5:** Densidad de probabilidad $\rho_0(\mathbf{r})$ para el caso exponencial, con $\mathcal{B}_0 = 1$, $\mathcal{E}_0 = 0.15$, $\alpha = 1$ y $k_y = 5$ a lo largo del eje $x$ para (a) el grafeno prístino y (b) el grafeno bajo tensión a lo largo del eje $x$ con $\varepsilon = 0.15$.

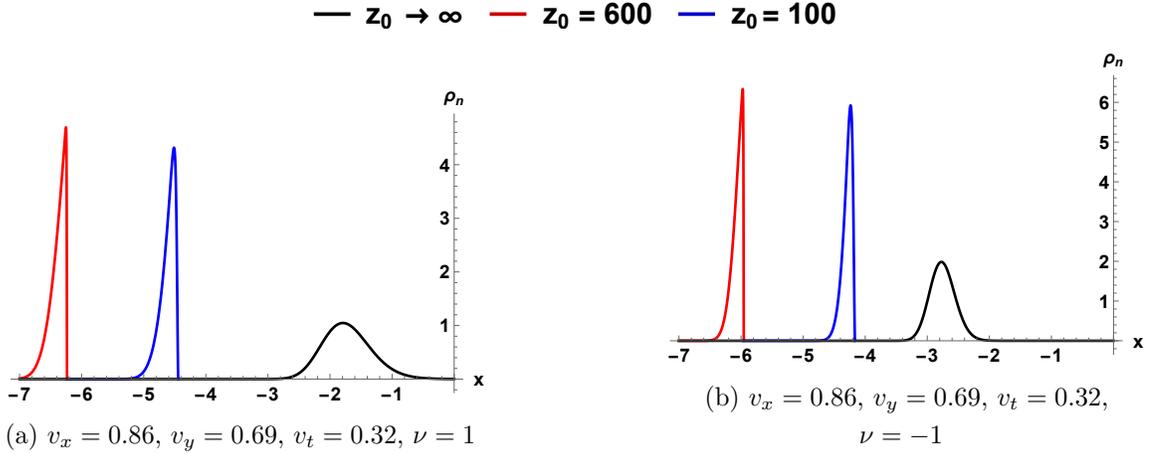

(a) $v_x = 0.86$, $v_y = 0.69$, $v_t = 0.32$, $\nu = 1$

(b) $v_x = 0.86$, $v_y = 0.69$, $v_t = 0.32$,
$\nu = -1$

**Figura 4.3.6:** Densidad de probabilidad $\rho_0(\mathbf{r})$ para el caso exponencial, con $\mathcal{B}_0 = 1$,
$\mathcal{E}_0 = 0.15$, $\alpha = 1$ a lo largo del eje $x$ para el borofeno $8 - Pmmn$ con (a) $k_y = 5$ en el
valle $\mathrm{K}_+$, y con (b) $k_y = 15$ en el valle $\mathrm{K}_-$.

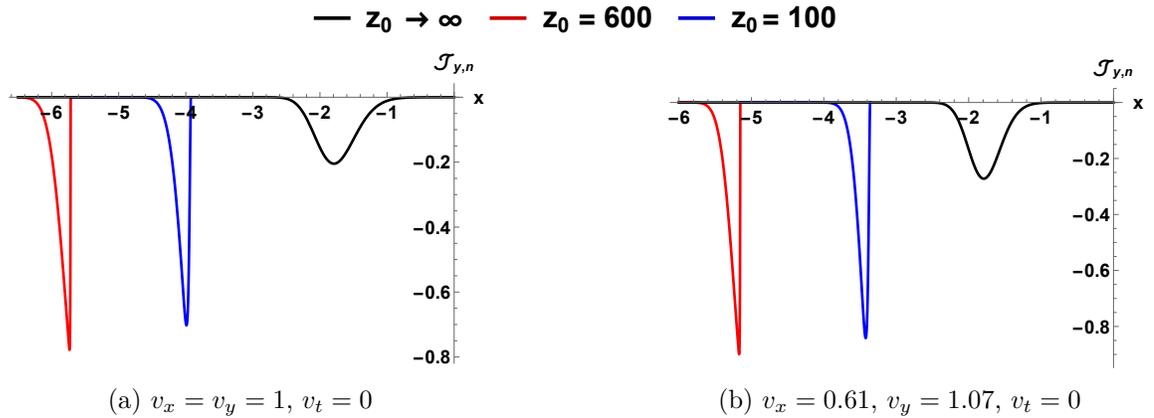

(a) $v_x = v_y = 1$, $v_t = 0$

(b) $v_x = 0.61$, $v_y = 1.07$, $v_t = 0$

**Figura 4.3.7:** Densidad de corriente $\mathcal{J}_{y,0}(x)$ para el caso exponencial, con $\mathcal{B}_0 = 1$,
$\mathcal{E}_0 = 0.15$, $\alpha = 1$ y $k_y = 5$ a lo largo del eje $x$ para (a) el grafeno prístino y (b) el
grafeno bajo tensión a lo largo del eje $x$ con $\varepsilon = 0.15$.

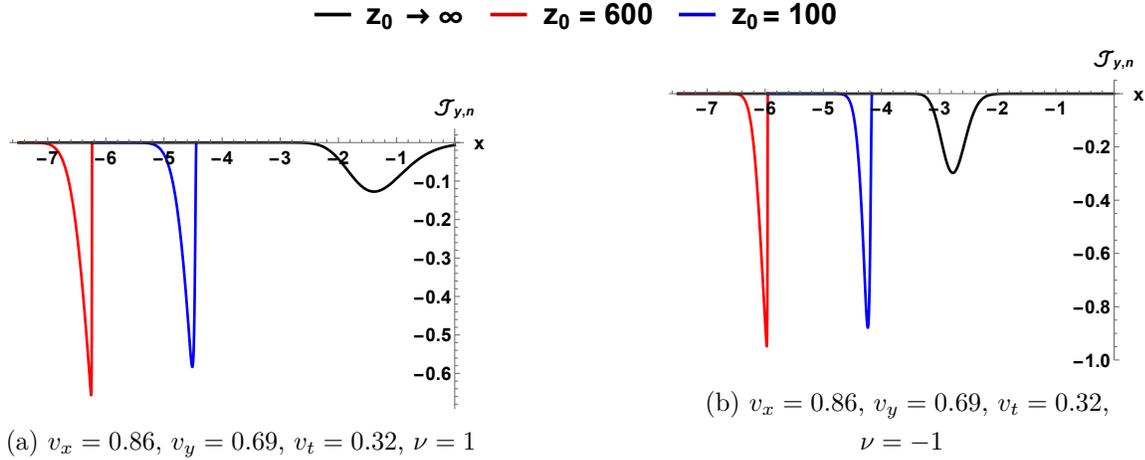

(a) $v_x = 0.86$, $v_y = 0.69$, $v_t = 0.32$, $\nu = 1$

(b) $v_x = 0.86$, $v_y = 0.69$, $v_t = 0.32$, $\nu = -1$

**Figura 4.3.8:** Densidad de corriente $\mathcal{J}_{y,0}(x)$ para el caso exponencial, con $\mathcal{B}_0 = 1$, $\mathcal{E}_0 = 0.15$, $\alpha = 1$ a lo largo del eje $x$ para el borofeno $8 - Pmmn$ con (a) $k_y = 5$ en el valle $K_+$, y con (b) $k_y = 15$ en el valle $K_-$.

el grafeno sometido a tensión uniaxial y el borofeno $8 - Pmmn$, bajo la acción de un campo eléctrico con parámetros fijos. Las gráficas muestran la distribución espacial de las probabilidades y corrientes para distintos valles y valores del momento $k_y$.

## 4.4 | Campos externos singulares

En el caso de campos singulares, los valores de $k$ y $\mu_j$ están dados por las expresiones (3.3.16) y (3.3.17), respectivamente, de la Sección 3.3. De esta forma se tiene:

$$k = \frac{\mathcal{B}_0 \left( k_y^c + \nu \beta_\nu \bar{E} \right)}{\sqrt{k_y^{c2} - \bar{E}^2}}, \tag{4.4.1}$$

$$\mu_j = \begin{cases} F - \frac{1}{2} & j = 1, \\ F + \frac{1}{2} & j = 2. \end{cases} \tag{4.4.2}$$

De forma análoga a lo mostrado para los perfiles exponenciales de campos eléctrico y magnético, se sustituyen (4.4.1) y (4.4.2) en las ecuaciones (4.2.33) y (4.2.39), lo que

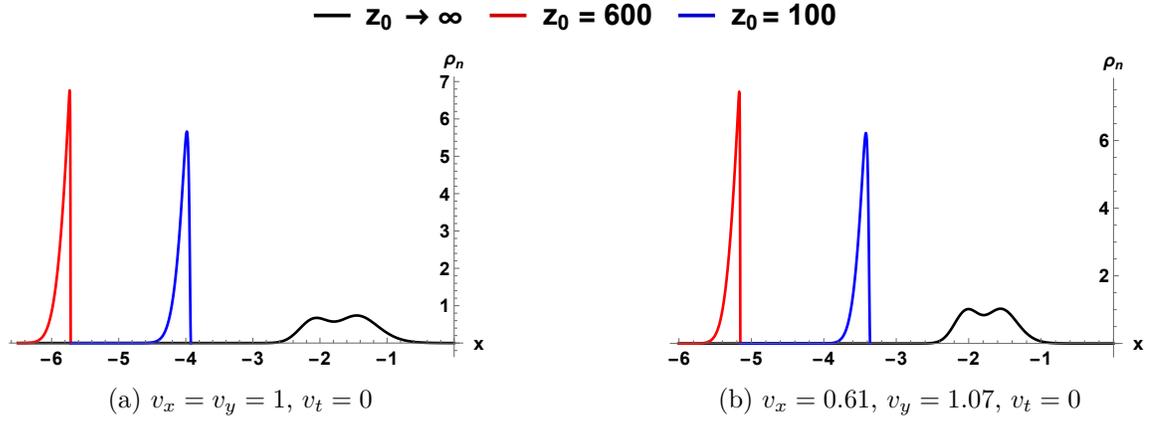

(a) $v_x = v_y = 1$, $v_t = 0$

(b) $v_x = 0.61$, $v_y = 1.07$, $v_t = 0$

**Figura 4.3.9:** Densidad de probabilidad $\rho_1(\mathbf{r})$ para el caso exponencial, con $\mathcal{B}_0 = 1$, $\mathcal{E}_0 = 0.15$, $\alpha = 1$ y $k_y = 5$ a lo largo del eje $x$ para (a) el grafeno prístino y (b) el grafeno bajo tensión a lo largo del eje $x$ con $\varepsilon = 0.15$.

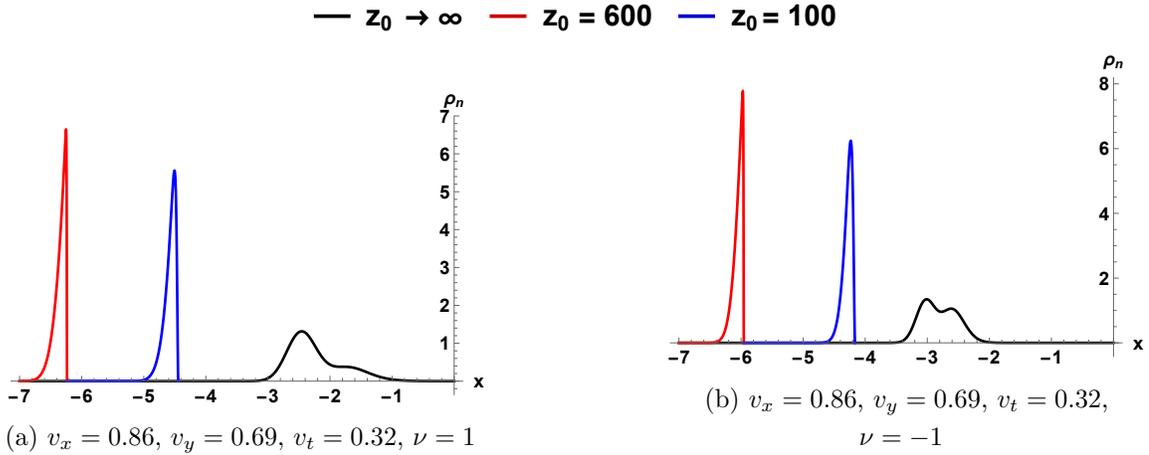

(a) $v_x = 0.86$, $v_y = 0.69$, $v_t = 0.32$, $\nu = 1$

(b) $v_x = 0.86$, $v_y = 0.69$, $v_t = 0.32$, $\nu = -1$

**Figura 4.3.10:** Densidad de probabilidad $\rho_1(\mathbf{r})$ para el caso exponencial, con $\mathcal{B}_0 = 1$, $\mathcal{E}_0 = 0.15$, $\alpha = 1$ a lo largo del eje $x$ para el borofeno $8 - Pmmn$ con (a) $k_y = 5$ en el valle $K_+$, y con (b) $k_y = 15$ en el valle $K_-$.

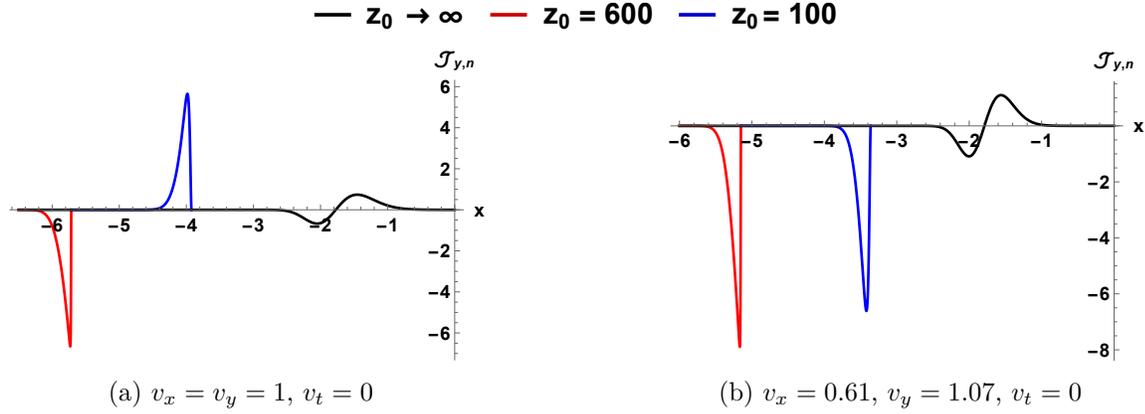

(a) $v_x = v_y = 1$, $v_t = 0$

(b) $v_x = 0.61$, $v_y = 1.07$, $v_t = 0$

**Figura 4.3.11:** Densidad de corriente $\mathcal{J}_{y,1}(x)$ para el caso exponencial, con $\mathcal{B}_0 = 1$, $\mathcal{E}_0 = 0.15$, $\alpha = 1$ y $k_y = 5$ a lo largo del eje $x$ para (a) el grafeno prístino y (b) el grafeno bajo tensión a lo largo del eje $x$ con $\varepsilon = 0.15$.

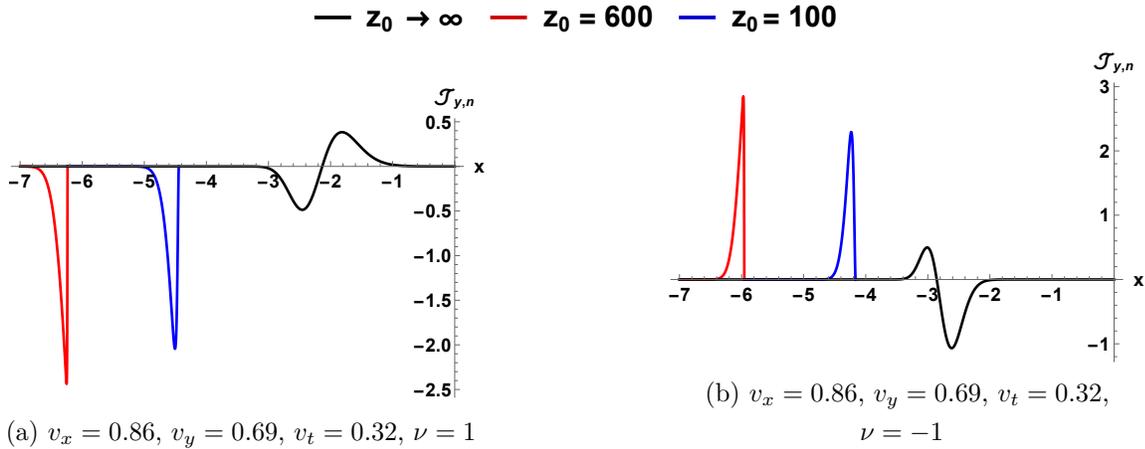

(a) $v_x = 0.86$, $v_y = 0.69$, $v_t = 0.32$, $\nu = 1$

(b) $v_x = 0.86$, $v_y = 0.69$, $v_t = 0.32$, $\nu = -1$

**Figura 4.3.12:** Densidad de corriente $\mathcal{J}_{y,1}(x)$ para el caso exponencial, con $\mathcal{B}_0 = 1$, $\mathcal{E}_0 = 0.15$, $\alpha = 1$ a lo largo del eje $x$ para el borofeno $8 - Pmmn$ con (a) $k_y = 5$ en el valle $\mathrm{K}_+$, y con (b) $k_y = 15$ en el valle $\mathrm{K}_-$.

permite obtener:

$$
E_n^{(0)} = k_y \nu\, v_t - \frac{\mathcal{B}_0^2\,\beta_\nu\,k_y^c v_\mathrm{F} + \kappa v_\mathrm{F}\,k_y^c\,(F+n)\,\sqrt{\mathcal{B}_0^2\,(\beta_\nu^2 - 1) + (F+n)^2}}{(F+n)^2 + \mathcal{B}_0^2\beta_\nu^2},
\tag{4.4.3}
$$

$$
E_n^{(1)} = k_y \nu\, v_t + \frac{\kappa\left[k_y^c v_\mathrm{F}\,(3 + 2F + 3n)\,\sqrt{(3 + 2F + 3n)^2 + 4\mathcal{B}_0^2\,(\beta_\nu^2 - 1)}\right] - 4\mathcal{B}_0^2 k_y^c v_\mathrm{F}\beta_\nu}{(3 + 2F + 3n)^2 + 4\mathcal{B}_0\beta_\nu^2}.
\tag{4.4.4}
$$

De acuerdo con (4.1.22), la energía para perfiles singulares de los campos eléctrico y magnético, considerando una modificación en las condiciones de frontera a primer orden, toma la siguiente forma:

$$
E_n = E_n^{(0)} + \frac{1}{z_0} E_n^{(1)}.
\tag{4.4.5}
$$

En las Figuras 4.4.1-4.4.4, muestra $\Delta E_n = E_n - E_n^{(0)}$ para $z_0 = 600$ y para bajo distintos conjuntos de parámetros característicos del sistema. Mientras que en las Figuras 4.4.5-4.4.8 y 4.4.9-4.4.12, se muestran las densidades de probabilidad y de corriente para el estado fundamental ($n = 0$) y el primer estado excitado ($n = 1$), respectivamente.

## 4.5 | Discusión

De las gráficas en las Figuras 4.3.3–4.3.4 y 4.4.3–4.4.4 se observa que, de forma similar al caso en que las condiciones de frontera no han sido modificadas, en los perfiles exponencial y singular discutidos en el capítulo anterior, aquí también persiste la existencia de estados ligados. Además, como muestran las Figuras 4.3.1–4.3.2 y 4.4.1–4.4.2, se produce un colapso en los niveles de Landau para un valor crítico del campo eléctrico dado por la expresión (3.2.41).

Un parámetro fundamental en el análisis, cuando se modifican las condiciones de frontera, es $z_0$ (o $x_0$), que actúa como regulador de la región donde se definen los campos eléctricos y magnéticos. En el cálculo de las energías y funciones de onda, se observa que en el límite $z_0 \to \infty$ (o $x_0 \to \infty$) dichas soluciones convergen al caso en que las condiciones de frontera no están modificadas. Este comportamiento se refleja en las gráficas de densidad de probabilidad y corriente (representadas en azul y rojo), donde al aumentar el valor de $z_0$ el sistema se aproxima progresivamente a la situación mostrada en el capítulo tres. Por construcción, $z_0$ debe ser suficientemente

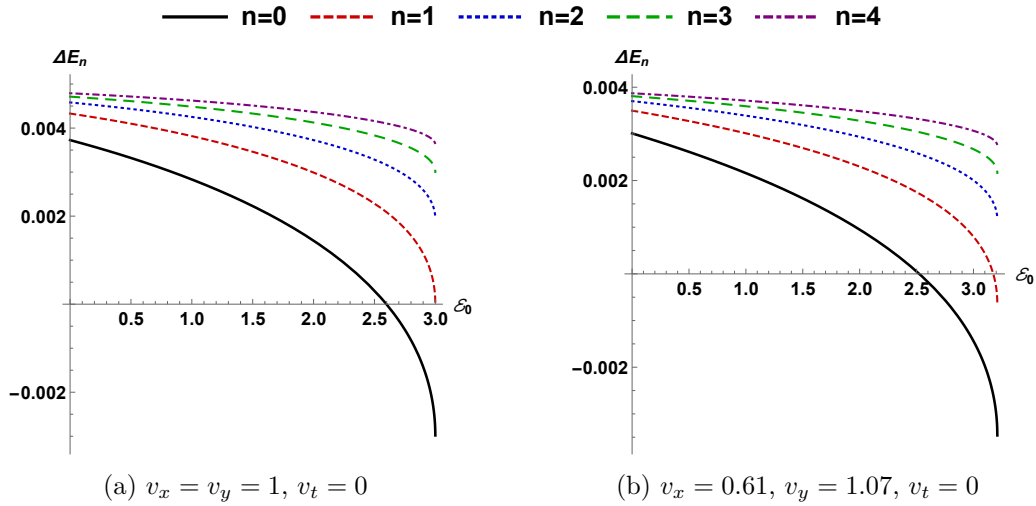

(a) $v_x = v_y = 1$, $v_t = 0$          (b) $v_x = 0.61$, $v_y = 1.07$, $v_t = 0$

**Figura 4.4.1:** Espectro de energía $\Delta E_n$ en (4.3.6) para el caso singular, con $\mathcal{B}_0 = 3$ y $k_y = 3$ en función del campo eléctrico $\mathcal{E}_0$ para (a) el grafeno prístino y (b) el grafeno bajo tensión a lo largo del eje $x$ con $\varepsilon = 0.15$.

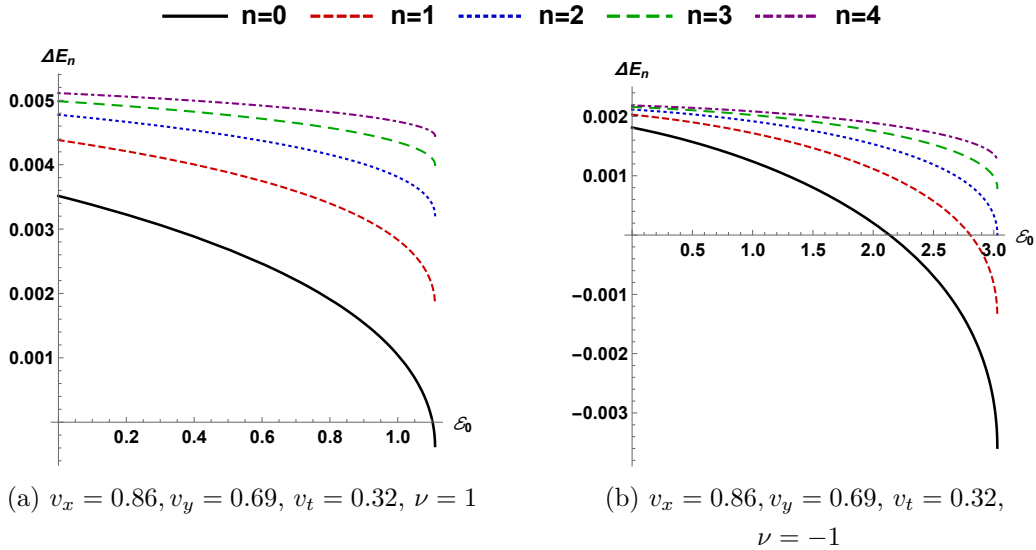

(a) $v_x = 0.86$, $v_y = 0.69$, $v_t = 0.32$, $\nu = 1$    (b) $v_x = 0.86$, $v_y = 0.69$, $v_t = 0.32$, $\nu = -1$

**Figura 4.4.2:** Espectro de energía $\Delta E_n$ en (4.3.6) para el caso singular, con $\mathcal{B}_0 = 3$, y $k_y = 3$ en función del campo eléctrico $\mathcal{E}_0$ para el borofeno $8 - Pmmn$ en (a) el valle $K_+$ y (b) el valle $K_-$.

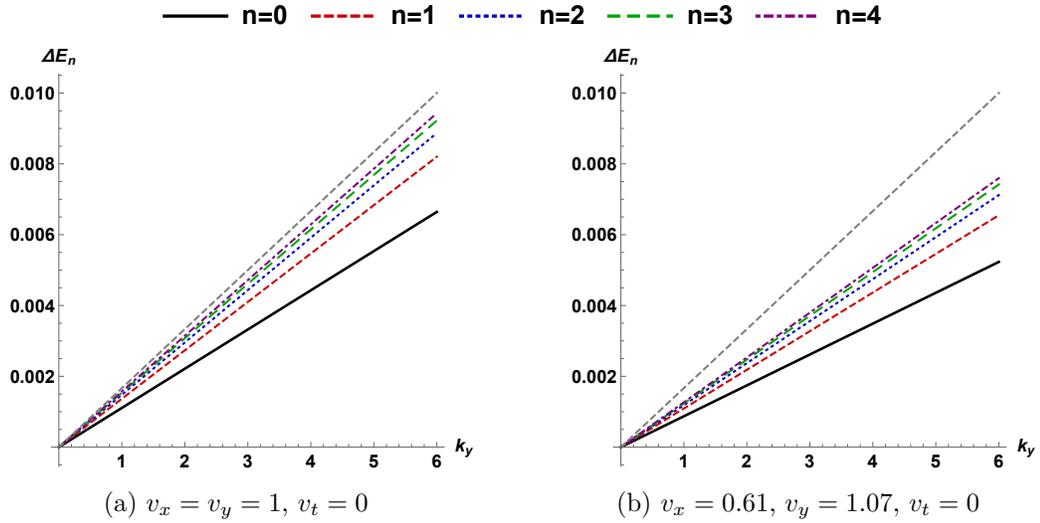

(a) $v_x = v_y = 1$, $v_t = 0$

(b) $v_x = 0.61$, $v_y = 1.07$, $v_t = 0$

**Figura 4.4.3:** Espectro de energía $\Delta E_n$ en (4.3.6) para el caso singular, con $\mathcal{B}_0 = 3$ y $\mathcal{E}_0 = 0.5$ en función del momento $k_y$ para (a) el grafeno prístino y (b) el grafeno bajo tensión a lo largo del eje $x$ con $\varepsilon = 0.15$.

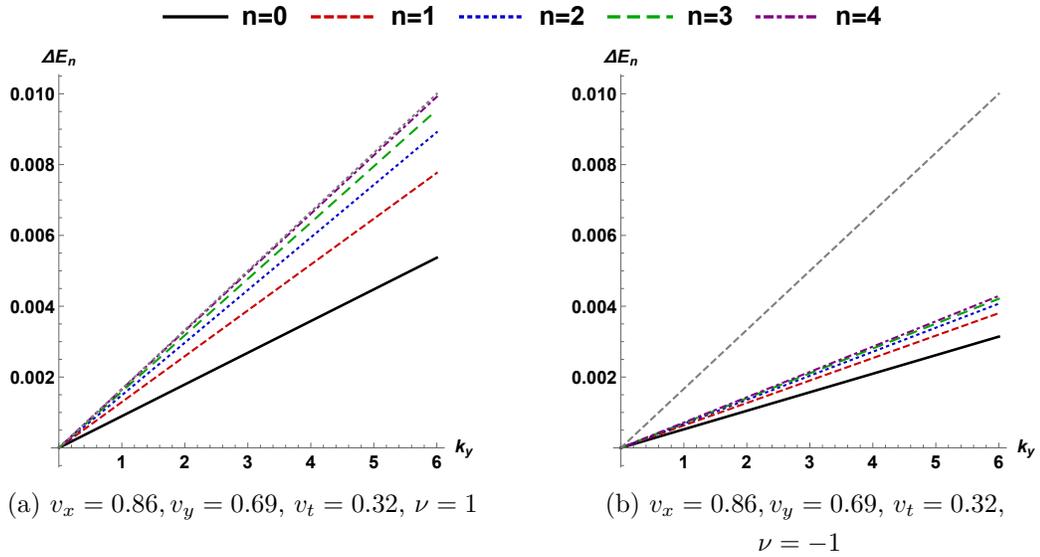

(a) $v_x = 0.86$, $v_y = 0.69$, $v_t = 0.32$, $\nu = 1$

(b) $v_x = 0.86$, $v_y = 0.69$, $v_t = 0.32$, $\nu = -1$

**Figura 4.4.4:** Espectro de energía $\Delta E_n$ (4.3.6) para el caso singular, con $\mathcal{B}_0 = 3$ y $\mathcal{E}_0 = 0.5$ en función del momento $k_y$ para el borofeno $8 - Pmmn$ en (a) el valle $K_+$ y (b) el valle $K_-$.

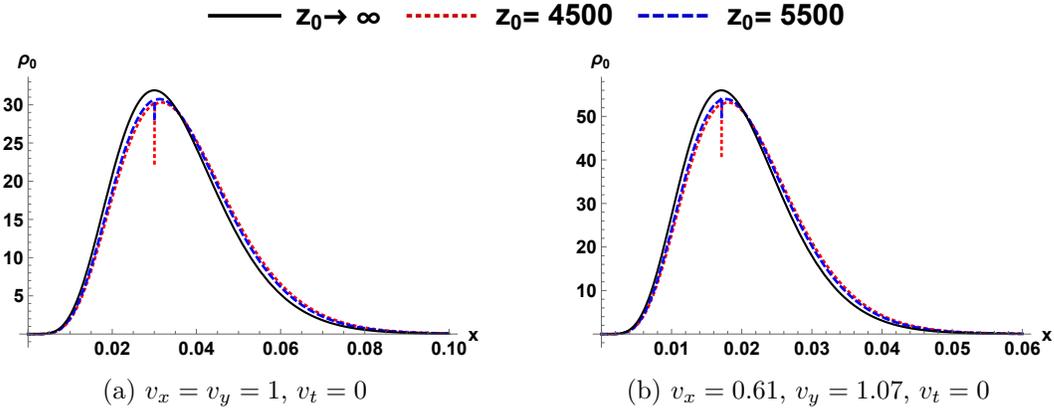

(a) $v_x = v_y = 1$, $v_t = 0$

(b) $v_x = 0.61$, $v_y = 1.07$, $v_t = 0$

**Figura 4.4.5:** Densidad de probabilidad $\rho_0(\mathbf{r})$ para el caso singular, con $\mathcal{B}_0 = 3$, $\mathcal{E} = 0.5$ y $k_y = 100$ a lo largo del eje $x$ para el (a) grafeno prístino y (b) el grafeno bajo tensión a lo largo del eje $x$ con $\varepsilon = 0.15$.

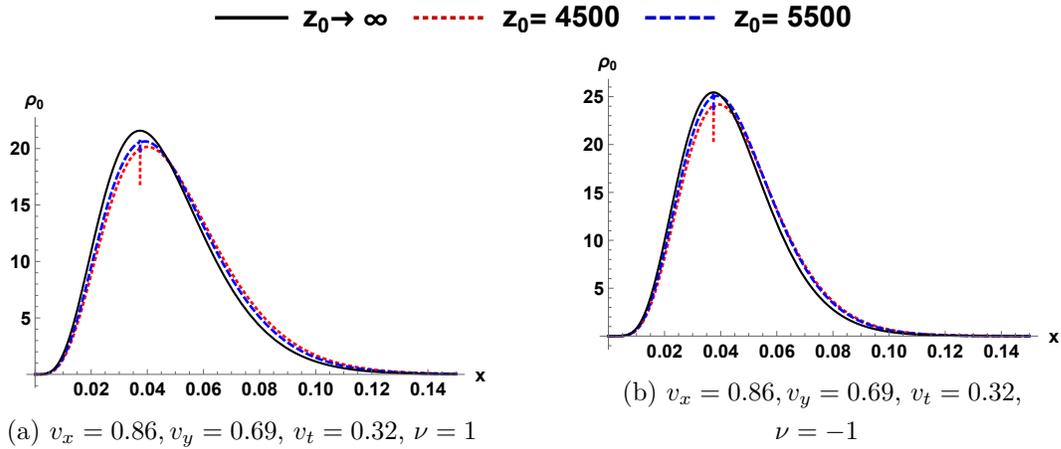

(a) $v_x = 0.86$, $v_y = 0.69$, $v_t = 0.32$, $\nu = 1$

(b) $v_x = 0.86$, $v_y = 0.69$, $v_t = 0.32$, $\nu = -1$

**Figura 4.4.6:** Densidad de probabilidad $\rho_0(\mathbf{r})$ para el caso singular, con $B_0 = 3$, $\mathcal{E} = 0.5$, $\alpha = 1$ y $k_y = 100$ a lo largo del eje $x$ para el borofeno $8 - Pmmn$ en (a) el valle $\mathrm{K}_+$ y (b) el valle $\mathrm{K}_-$.

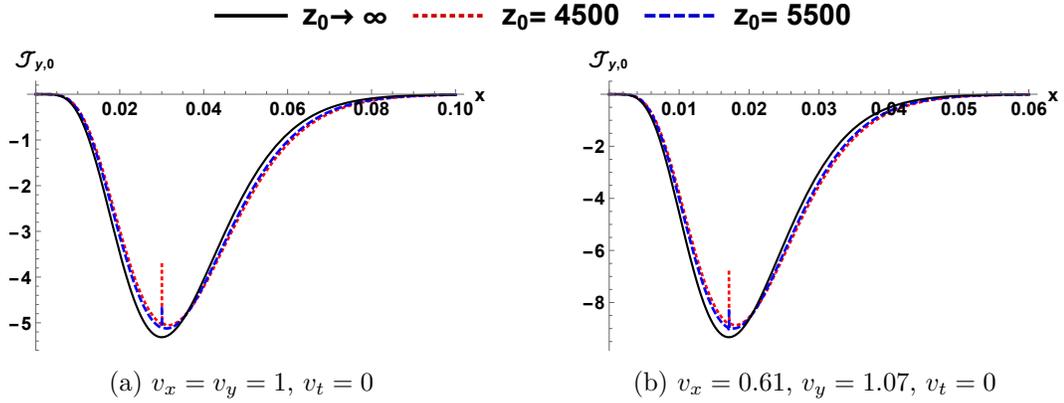

(a) $v_x = v_y = 1$, $v_t = 0$

(b) $v_x = 0.61$, $v_y = 1.07$, $v_t = 0$

**Figura 4.4.7:** Densidad de corriente $\mathcal{J}_{y,0}(x)$ para el caso singular, con $\mathcal{B}_0 = 3$, $\mathcal{E} = 0.5$, $\alpha = 1$ y $k_y = 100$ a lo largo del eje $x$ para (a) el grafeno prístino y (b) el grafeno bajo tensión a lo largo del eje $x$ con $\varepsilon = 0.15$.

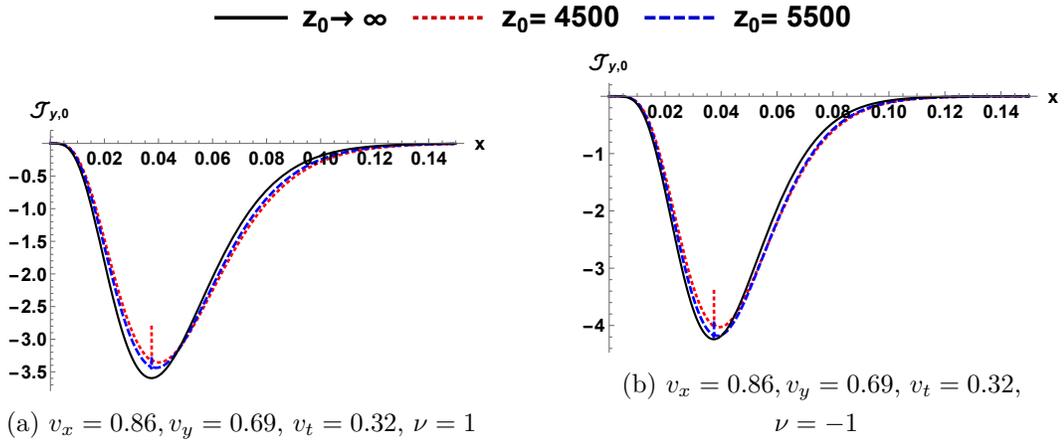

(a) $v_x = 0.86$, $v_y = 0.69$, $v_t = 0.32$, $\nu = 1$

(b) $v_x = 0.86$, $v_y = 0.69$, $v_t = 0.32$,
$\nu = -1$

**Figura 4.4.8:** Densidad de corriente $\mathcal{J}_{y,0}(x)$ para el caso singular, con $B_0 = 3$, $\mathcal{E} = 0.5$, $\alpha = 1$ y $k_y = 100$ a lo largo del eje $x$ para el borofeno $8 - Pmmn$ en (a) el valle $K_+$ y (b) el valle $K_-$.

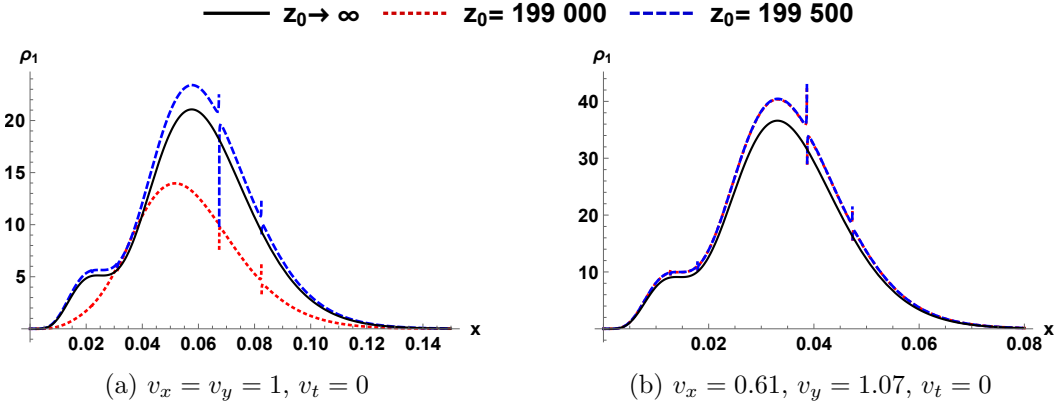

(a) $v_x = v_y = 1$, $v_t = 0$          (b) $v_x = 0.61$, $v_y = 1.07$, $v_t = 0$

**Figura 4.4.9:** Densidad de probabilidad $\rho_1(\mathbf{r})$ para el caso singular, con $\mathcal{B}_0 = 3$, $\mathcal{E} = 0.5$ y $k_y = 100$ a lo largo del eje $x$ para el (a) grafeno prístino y (b) el grafeno bajo tensión a lo largo del eje $x$ con $\varepsilon = 0.15$.

grande para garantizar esta correspondencia.

La corrección al espectro de energía se ha realizado hasta primer orden, donde se observa que el parámetro $z_0$ (o bien $x_0$) influye directamente en este espectro y también sobre las funciones de onda definidas en las ecuaciones (4.2.53) y (4.2.60). Para cada perfil de campo considerado, existe una relación entre $z_0$ y $x_0$. Para el perfil exponencial se tiene:

$$z_0^{\exp}(x_0) = \frac{2D\sqrt{1-\beta_\nu^2}}{\alpha_c}\,\mathrm{e}^{-\alpha_c x_0^{\exp}}, \tag{4.5.1}$$

mientras que para el perfil singular, la expresión es

$$z_0^{\mathrm{sing}}(x_0) = 2x_0^{\mathrm{sing}}\sqrt{k_y^{c\,2} - \bar{E_n}^2}. \tag{4.5.2}$$

Despejando $x_0$ en cada caso, resulta:

$$x_0^{\exp}(z_0) = -\frac{1}{\alpha_c}\ln\left(\frac{\alpha_c z_0^{\exp}}{2D\sqrt{1-\beta_\nu^2}}\right), \tag{4.5.3}$$

$$x_0^{\mathrm{sing}}(z_0) = \frac{z_0^{\mathrm{sing}}}{2\sqrt{k_y^{c\,2} - \bar{E_n}^2}}, \tag{4.5.4}$$

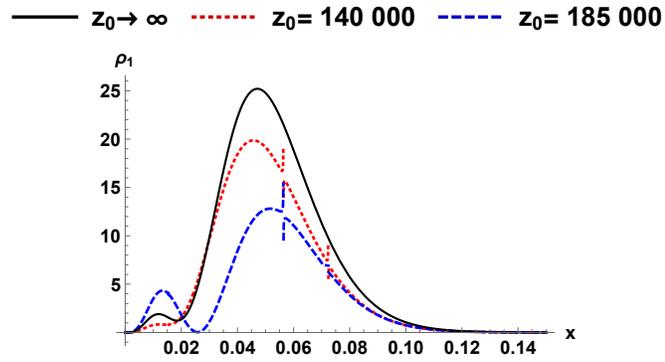

(a) $v_x = 0.86, v_y = 0.69, v_t = 0.32, \nu = 1$

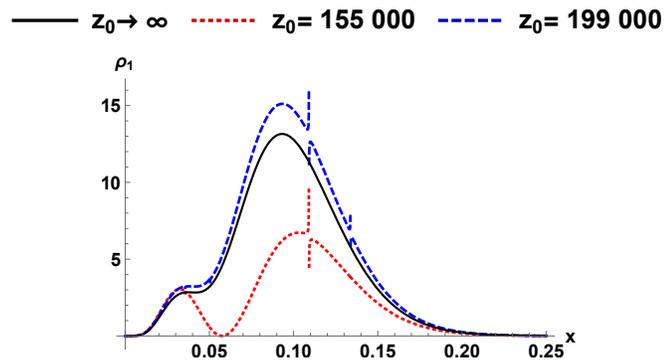

(b) $v_x = 0.86, v_y = 0.69, v_t = 0.32,$
$\nu = -1$

**Figura 4.4.10:** Densidad de probabilidad $\rho_1(\mathbf{r})$ para el caso singular, con $B_0 = 3$, $\mathcal{E} = 0.5$, $\alpha = 1$ y $k_y = 100$ a lo largo del eje $x$ para el borofeno $8 - Pmmn$ en (a) el valle $K_+$ y (b) el valle $K_-$.

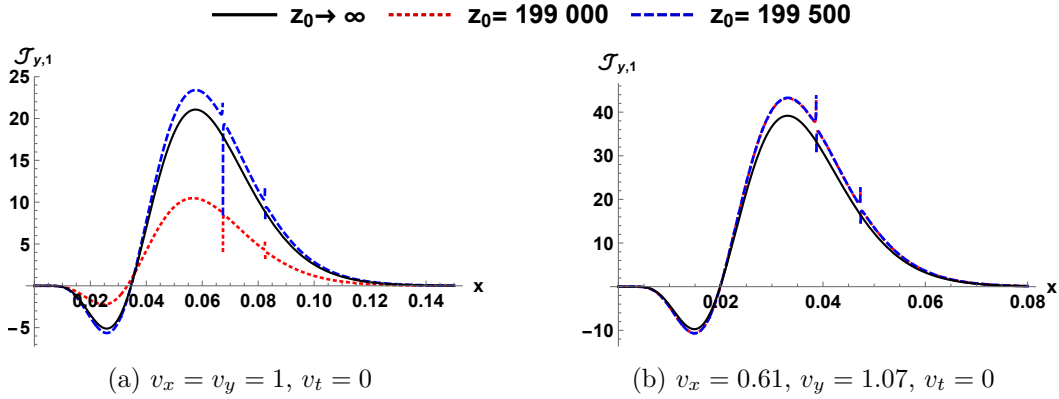

(a) $v_x = v_y = 1$, $v_t = 0$          (b) $v_x = 0.61$, $v_y = 1.07$, $v_t = 0$

**Figura 4.4.11:** Densidad de corriente $\mathcal{J}_{y,1}(x)$ para el caso singular, con $\mathcal{B}_0 = 3$, $\mathcal{E} = 0.5$, $\alpha = 1$ y $k_y = 100$ a lo largo del eje $x$ para (a) el grafeno prístino y (b) el grafeno bajo tensión a lo largo del eje $x$ con $\varepsilon = 0.15$.

donde $\bar{E}_n = (\nu E_n - v_t k_y)/\sqrt{v_x v_y}$, siendo $E_n$ el término dado por (4.4.5) del caso singular.

Con una adecuada elección de parámetros, para ambos perfiles de campos eléctrico y magnético se puede definir una región acotada lo suficientemente representativa para simular condiciones de laboratorio y tener un escenario más realista.

Por otra parte, las divergencias que se observan en las eigenfunciones $\psi_{0,j}^{(1)}(z)$ y $\psi_{1,j}^{(1)}(z)$ y que están asociadas a términos racionales o exponenciales presentes en las soluciones analíticas de las funciones de onda, las cuales están dadas por la ecuación (4.2.54) para el perfil exponencial y por las ecuaciones (4.2.61) y (4.2.62) para el perfil singular, pueden ser expresadas en términos de la variable $x$.

Para el caso $n = 0$, y utilizando (3.2.5), (3.3.13) y (4.2.54), es posible calcular el valor de la singularidad para el estado base $\bar{x}_0$, obteniéndose:

$$\bar{x}_{\exp}^{(0)} = -\frac{1}{\alpha_c} \ln\left(\frac{\alpha_c \lambda_0}{D\sqrt{1 - \beta_\nu^2}}\right), \tag{4.5.5}$$

$$\bar{x}_{\sing}^{(0)} = \frac{F}{\sqrt{k_c^2 - \bar{E}_0^{\,2}}}, \tag{4.5.6}$$

donde $\lambda_0$, de la ecuación (3.2.29), es calculada con el valor de $E_0$ de la ecuación (4.3.5)

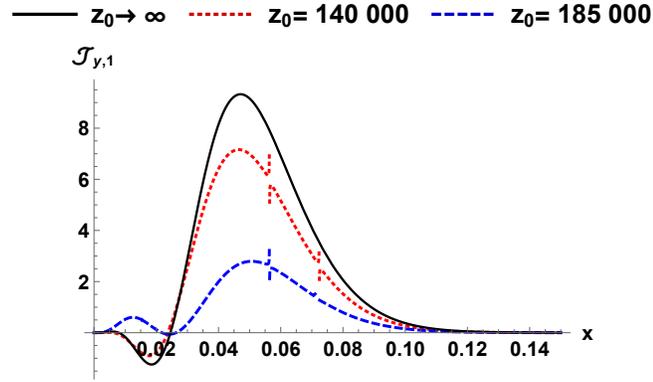

(a) $v_x = 0.86, v_y = 0.69, v_t = 0.32, \nu = 1$

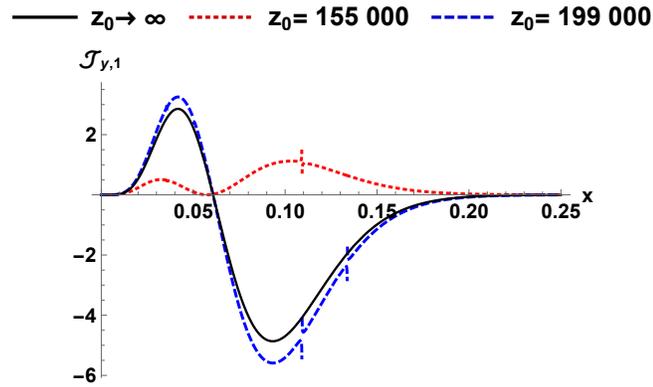

(b) $v_x = 0.86, v_y = 0.69, v_t = 0.32, \nu = -1$

**Figura 4.4.12:** Densidad de corriente $\mathcal{J}_{y,1}(x)$ para el caso singular, con $B_0 = 3$, $\mathcal{E} = 0.5$, $\alpha = 1$ y $k_y = 100$ a lo largo del eje $x$ para el borofeno $8 - Pmmn$ en (a) el valle $K_+$ y (b) el valle $K_-$.

del caso exponencial, mientras que $\bar{E}_0$ viene dado de nuevo por (4.4.5) del caso singular.

Para el primer estado excitado ($n = 1$), y utilizando (3.2.5), (3.3.13) junto con las expresiones (4.2.61) y (4.2.62), también se pueden obtener las expresiones para las singularidades en la variable $x$, las cuales son:

$$\bar{x}_{j,\text{exp}}^{(1,2)} = -\frac{1}{\alpha_c} \ln\left[\frac{\alpha_c\left(4\mu_j + 5 \pm \sqrt{16\mu_j + 17}\right)}{4D\sqrt{1 - \beta_\nu^2}}\right], \quad \text{con } \mu_j = \begin{cases} \lambda_1 - \frac{1}{2} & j = 1, \\ \lambda_1 + \frac{1}{2} & j = 2, \end{cases} \quad (4.5.7)$$

$$\bar{x}_{j,\text{sing}}^{(1,2)} = \frac{4\mu_j + 5 \pm \sqrt{16\mu_j + 17}}{4\sqrt{k_y c^2 - \bar{E}_1^{\,2}}}, \quad \text{donde } \mu_j = \begin{cases} F - \frac{1}{2} & j = 1, \\ F + \frac{1}{2} & j = 2, \end{cases} \quad (4.5.8)$$

donde $\lambda_1$ se obtiene empleando la ecuación (3.2.29) del caso exponencial junto con (4.3.5), y $\bar{E}_1$ se determina mediante la expresión (4.4.5) correspondiente al caso singular.

Del análisis numérico y analítico se observa que las singularidades presentes en la función de onda del estado base y del primer estado excitado para el perfil singular de los campos eléctrico y magnético, ubicadas en los puntos $\bar{x}_{\text{sing}}^{(0)}$, $\bar{x}_{j,\text{sing}}^{(1)}$ y $\bar{x}_{j,\text{sing}}^{(2)}$, se encuentran dentro del dominio de definición de los campos eléctricos y magnéticos, es decir, $\bar{x}_{\text{sing}}^{(0)}, \bar{x}_{j,\text{sing}}^{(1)}, \bar{x}_{j,\text{sing}}^{(2)} \in (0, x_0]$. Estas singularidades se acentúan para valores pequeños de $z_0$ (o bien $x_0$) y tienden a suavizarse a medida que $z_0$ aumenta, como se aprecia en las gráficas de densidad de corriente y densidad de estados. Para valores muy grandes del parámetro $z_0$, algunos factores de las funciones de onda (4.2.53) y (4.2.60) tienden a la unidad, lo que contribuye a la desaparición de las divergencias.

En el caso del perfil exponencial, el análisis numérico y analítico indica que las singularidades de la función de onda del estado base y del primer estado excitado, ubicadas en $\bar{x}_{\text{exp}}^{(0)}$, $\bar{x}_{j,\text{exp}}^{(1)}$ y $\bar{x}_{j,\text{exp}}^{(2)}$, se encuentran fuera del dominio de definición de los campos eléctricos y magnéticos $x \in (-\infty, x_0]$, es decir, $\bar{x}_{\text{exp}}^{(0)}, \bar{x}_{j,\text{exp}}^{(1)}, \bar{x}_{j,\text{exp}}^{(2)} \gg x_0$. Por esta razón, las gráficas de densidad de corriente y densidad de estados presentadas anteriormente son más suaves en comparación con las obtenidas para perfiles de campo singular.

Cabe destacar que la elección del parámetro $z_0$ (o $x_0$) influye directamente en las funciones de onda calculadas en este trabajo y, en consecuencia, en las densidades de probabilidad y corriente. Al modificar su valor, cambia la región espacial donde se definen los estados electrónicos, lo que se refleja en la amplitud, el desplazamiento y la simetría de dichas densidades. Esta sensibilidad refuerza la interpretación de $z_0$ (o
$x_0$) como un parámetro de control espacial dentro del modelo, permitiendo explorar
distintas configuraciones de confinamiento del sistema, como se aprecia en la Figura
4.2.1.

# 5 | Conclusiones

El estudio de los materiales de Dirac se ha convertido en un tema de gran relevancia en la investigación actual, particularmente en sistemas donde los portadores de carga interactúan con campos magnéticos y eléctricos. Motivados por este enfoque, en este trabajo se ha abordado el problema de eigenvalores que surge al considerar a los electrones de materiales de Dirac anisótropos interactuando con campos eléctricos y magnéticos dependientes de la posición. Específicamente, al considerar dos perfiles para los campos externos aplicados al material, uno en el que la intensidad de los campos decrece exponencialmente conforme $x \to \infty$ y otro en el que la intensidad varía como $1/x^2$. En el Capítulo 3 se obtuvieron expresiones analíticas para el espectro de energía y eigenfunciones de onda, así como para las densidades de probabilidad y de corriente, mostrando su dependencia con los parámetros característicos del material: la intensidad de los campos aplicado, $\mathbf{B}(x)$ y $\mathbf{E}(x)$, el llamado índice de valle, $\nu$, y el parámetro $v_t$ relacionado con la anisotropía intrínseca del material.

Las soluciones halladas reproducen consistentemente los casos límite conocidos en la literatura. Por ejemplo, al considerar $\alpha \to 0$ para el caso exponencial, se obtienen los resultados del caso con campos constantes reportados en [Díaz-Bautista, 2022]; en tanto, en el límite $\beta_\nu \to 0$ para ambos casos, se recuperan exactamente las soluciones de [Ş. Kuru et al., 2009] correspondientes al grafeno prístino. Es importante destacar que los resultados para el perfil exponencial ya han sido validados y publicados en [Mojica-Zárate et al., 2024]. Por otro lado, el caso del perfil singular aquí abordado, aunque inédito en la literatura, muestra una congruencia notable con los resultados conocidos para el caso del campo eléctrico nulo, y representa una extensión natural del trabajo presentado en [Mojica-Zárate et al., 2024]. La coherencia de las soluciones obtenidas en este trabajo de tesis, en todos los límites físicos pertinentes, confirma la solidez del enfoque teórico empleado.

Un resultado particularmente significativo es la identificación de un valor crítico $\mathcal{E}_c$ del campo eléctrico que produce el colapso de los niveles de Landau que surgen de la interacción con el campo magnético. Este valor crítico muestra una dependencia fundamental con los parámetros característicos del material: $\nu$, $v_t$ y $v_y$, y la intensidad del campo magnético aplicado, $\mathcal{B}_0$, como se muestra en la ecuación (3.2.41). El análisis del espectro de energía en función del número de onda (o momento transversal) $k_y$ revela la existencia de estados ligados cuyo número depende del valor de $k_y$, siendo esto ilustrado mediante una envolvente característica, como se puede ver, por ejemplo, en las Figuras 3.2.3 y 3.2.4.

Otro aspecto notable de los resultados presentados es el comportamiento de las densidades de probabilidad y de corriente cuando se considera la influencia del campo eléctrico $\mathbf{E}(x)$. A diferencia de los casos donde el campo eléctrico está ausente, se encuentra que el estado fundamental ($n = 0$) genera una corriente de probabilidad claramente distinguible (ver Figuras 3.2.7, 3.2.8, 3.3.7 y 3.3.8). Este fenómeno, presenta interesantes paralelos con el efecto Schwinger [Schwinger, 1951], donde campos eléctricos intensos pueden función crear pares electrón-positrón (o electrón-hueco, en este caso) directamente del vacío cuántico. Estos hallazgos adquieren especial relevancia al considerar recientes observaciones experimentales en grafeno. Estudios reportados en [Allor et al., 2008, Berdyugin et al., 2022, Schmitt et al., 2023], han detectado corrientes eléctricas significativamente mayores a las predichas por modelos convencionales cuando se aplican campos eléctricos intensos mediante estructuras de superred. Los autores atribuyen este efecto a la generación espontánea de pares electrón-hueco en el material. En el modelo usado, la aparición de niveles de Landau con valores negativos y la presencia de una corriente de probabilidad no nula para el estado fundamental ($n = 0$) sugieren la posible manifestación de un análogo del efecto Schwinger en estos sistemas (es decir, la generación de pares electrón-hueco en lugar de electrón-positrón bajo campos eléctricos intensos). Esta interpretación abre nuevas perspectivas para entender los fenómenos de transporte cuántico en materiales bidimensionales bajo campos electromagnéticos intensos y dependientes de la posición.

Por otro lado, el análisis de las densidades de probabilidad y de corriente para los distintos índices de valle revela una clara asimetría entre los estados asociados a $K_+$ y $K_-$. Esta diferenciación, inducida por la anisotropía del sistema, o por campos externos inhomogéneos, se traduce en una ruptura de la simetría en la distribución espacial de los estados electrónicos y en la dirección del flujo de corriente. Además, se observa que la velocidad $v_x$ influye directamente en la localización de los electrones: velocidades menores favorecen su permanencia en las regiones asociadas a los puntos críticos, mientras que velocidades mayores reducen su probabilidad de detección en dichas zonas. Estos resultados no solo evidencian el papel dinámico de los valles en la física del sistema, sino que también apuntan al control efectivo de estos grados de libertad como una vía factible para el diseño de dispositivos *valletrónicos*. En conjunto, estos efectos reflejan cómo el transporte electrónico en materiales de Dirac puede ser modulado mediante ingeniería de anisotropías y perfiles de campo, abriendo nuevas posibilidades para aplicaciones funcionales basadas en el control de los portadores de carga en función de los valles $K_+$ y $K_-$ a los cuales pertenecen.

Ahora bien, el análisis de los sistemas confinados a una región finita mediante la aplicación de campos externos con perfiles dependientes de la posición, como se comentó en el Capítulo 4, revela comportamientos cualitativamente distintos respecto a lo discutido en el Capítulo 3. Cuando $z_0$ toma valores suficientemente grandes, el sistema tiende a recuperar las propiedades del caso en el cual las condiciones de frontera no se han modificado, lo cual se refleja en el espectro de energía, así como en las densidades de probabilidad y de corriente. En el caso de perfiles de la forma $e^{-\alpha x}$, las singularidades asociadas a las soluciones ocurren para valores críticos de $x$ que están fuera del intervalo $x \in (-\infty, x_0]$, por lo que las funciones resultan regulares en el dominio físico considerado.

En contraste, para los perfiles de la forma $x^{-2}$, el dominio está acotado entre $x = 0$ y $x = x_0$, y para ciertos valores de $x_0$, en este intervalo sí pueden aparecer valores críticos de $x$ en los que las eigenfunciones divergen. Esto indica que, aunque las soluciones son físicamente consistentes en la mayor parte del dominio, se vuelven no confiables en la vecindad inmediata de ciertos puntos críticos, donde presentan comportamientos no bien definidos o inestabilidades numéricas. Este contraste resalta la sensibilidad del sistema a la forma del campo aplicado y subraya la importancia de delimitar correctamente el intervalo espacial efectivo para garantizar la regularidad de las soluciones. Sin embargo, a medida que aumenta el valor de $x_0$, dichas singularidades tienden a desaparecer.

Cabe destacar que la modificación de las condiciones de frontera o confinamiento, abordada en el Capítulo 4, constituye un paso importante para vincular el análisis teórico con el entorno experimental. Esta modificación proporciona una base teórica útil para explorar configuraciones factibles en el laboratorio. Asimismo, permite estudiar cómo la geometría del confinamiento y los parámetros del sistema influyen en las propiedades electrónicas, energéticas, en otros aspectos relevantes del comportamiento físico del sistema.

Finalmente, el análisis comparativo entre ambos perfiles de campo pone de manifiesto diferencias cualitativas significativas en el comportamiento del sistema, especialmente en lo referente a la estructura de niveles de energía y la localización espacial de los estados. Estos resultados no solo enriquecen nuestra comprensión de los materiales de Dirac sometidos a campos externos inhomogéneos, sino que también aportan herramientas teóricas útiles para guiar el diseño y la caracterización de dispositivos basados en este tipo de sistemas.

Como perspectivas de este trabajo, se podría profundizar en el estudio detallado de la densidad de corriente en el estado fundamental, dado que esta presenta características particulares bajo la acción del campo eléctrico. De igual forma, el análisis de otras densidades, como la densidad de probabilidad de espín, permitiría comprender mejor los mecanismos de transporte cuántico de los electrones. Asimismo, sería de gran interés explorar otros perfiles de campo más complejos o realistas, que reproduzcan con mayor precisión las condiciones experimentales no homogéneas. Otro camino prometedor consiste en extender el análisis hacia diferentes materiales de Dirac, tales como aislantes topológicos o semimetales topológicos, donde la interacción entre campos electromagnéticos y estados protegidos topológicamente podría dar lugar a fenómenos aún más interesantes en el transporte cuántico. Estos enfoques abrirían nuevas posibilidades tanto en la comprensión teórica como en el desarrollo de aplicaciones tecnológicas basadas en sistemas bidimensionales y de fase topológica.

# A | Aplicación del método de iteración asintótica a la ecuación de Whittaker

Como se puede ver de la forma de las expresiones en (3.2.15) y (3.3.14), ambas coinciden con la ecuación diferencial de Whittaker, por lo cual pueden resolverse con el mismo procedimiento analítico. Sin embargo, ahora se optará por emplear el método de iteración asintótica para hallar las soluciones de dicha ecuación diferencial. Esto se realiza para que el lector se familiarice con el método de iteración asintótica y mostrar que se recuperan las mismas soluciones que las mostradas en el Capítulo 3. Además, esta descripción permitirá abordar los mismos problemas comentados en dicho capítulo pero modificando ligeramente las condiciones de frontera consideradas en cada caso, como se propone en el Capítulo 4.

Para aplicar el método de iteración asintótica, es necesario estudiar primero los comportamientos asintóticos de la ecuación de Whittaker cuando $z$ tiende a cero e infinito, para hallar sus soluciones alrededor de estos puntos.

Sea entonces la ecuación de Whittaker

$$\frac{\mathrm{d}^2\psi_j(z)}{\mathrm{d}z^2} + \left[-\frac{1}{4} + \frac{\mu}{z} + \frac{1/4 - \bar{m}_j^2}{z^2}\right]\psi_j(z) = 0, \quad j = 1, 2, \tag{A.0.1}$$

donde $\bar{m}_j$ y $\mu$ son constantes que cambian dependiendo del perfil de campo externo que se elija.

Entonces, cuando $z \to \infty$, (A.0.1) toma la siguiente forma

$$\frac{\mathrm{d}^2\psi_j(z)}{\mathrm{d}z^2} - \frac{1}{4}\psi_j(z) = 0, \tag{A.0.2}$$

cuya solución es $\psi_j(z) = a\,\mathrm{e}^{\frac{z}{2}} + b\,\mathrm{e}^{-\frac{z}{2}}$, donde $a$ y $b$ son constantes. La condición de frontera en este caso requiere que la función se comporte adecuadamente en esta región ($a = 0$). Por lo tanto, la solución viable es

$$\psi_j(z) = b\,\mathrm{e}^{-\frac{z}{2}}. \tag{A.0.3}$$

Para valores de $z$ alrededor de cero ($z \to 0$), la ecuación (A.0.1) puede expresarse como

$$\frac{\mathrm{d}^2\psi_j(z)}{\mathrm{d}z^2} + \frac{\frac{1}{4} - \bar{m}_j^2}{z^2}\psi_j(z) = 0, \tag{A.0.4}$$

teniendo como solución a la función $\psi_j(z) = c\,z^{\frac{1}{2}+\bar{m}_j} + d\,z^{\frac{1}{2}-\bar{m}_j}$. Análogamente a la condición de frontera anterior, en este caso se requiere que $d = 0$. Así, la solución está dada por:

$$\psi_j(z) = c\,z^{\frac{1}{2}+\bar{m}_j}. \tag{A.0.5}$$

Siguiendo el método de iteración asintótica y usando (A.0.3) y (A.0.5), se propone que la función $\psi_j$ sea de la forma

$$\psi_j(z) = \mathrm{e}^{-\frac{z}{2}}z^{\frac{1}{2}+\bar{m}_j}\bar{\phi}_j(z), \tag{A.0.6}$$

donde $\bar{\phi}_j(z)$ es una función a ser determinada.

Introduciendo (A.0.6) en (3.3.14), se obtiene una ecuación diferencial para $\bar{\phi}_j(z)$:

$$\frac{\mathrm{d}^2\bar{\phi}_j(z)}{\mathrm{d}z^2} - \frac{z - 1 - 2\bar{m}_j}{z}\frac{\mathrm{d}\bar{\phi}_j(z)}{\mathrm{d}z} - \frac{\frac{1}{2} - \mu + \bar{m}_j}{z}\bar{\phi}_j(z) = 0. \tag{A.0.7}$$

Comparando (A.0.7) con la expresión (4.1.1) del Capítulo 4, se pueden obtener las funciones $\lambda_0(z)$ y $s_0(z)$ como:

$$\lambda_0(z) = \frac{z - 1 - 2\bar{m}_j}{z}, \tag{A.0.8}$$

$$s_0(z) = \frac{\frac{1}{2} - \mu + \bar{m}_j}{z}. \tag{A.0.9}$$

Para continuar con el método de iteración asintótica, se requieren calcular las funciones $\lambda_n(z)$ y $s_n(z)$, definidas en las ecuaciones (4.1.8) y (4.1.9) del Capítulo 4. A continuación se muestran únicamente los casos para $n = 1$ y $n = 2$:

$$\lambda_1(z) = \frac{1 + 2\bar{m}_j + (z - 1 - 2\bar{m}_j)^2 + z\left(\frac{1}{2} - \mu + \bar{m}_j\right)}{z^2}, \tag{A.0.10}$$

$$s_1(z) = \frac{(z - 2 - 2\bar{m}_j)\left(\frac{1}{2} - \mu + \bar{m}_j\right)}{z^2}, \tag{A.0.11}$$

$$\lambda_2(z) = \frac{z^3 + (-4\bar{m}_j - 2\mu - 2)z^2 + 8\left(\bar{m}_j + 1\right)\left(\bar{m}_j + \frac{\mu}{2} + \frac{1}{2}\right)z - 8\bar{m}_j^3 - 24\bar{m}_j^2 - 22\bar{m}_j - 6}{z^3}, \tag{A.0.12}$$

$$s_2(z) = \frac{1}{2}\frac{\left(\frac{1}{2} - \mu + \bar{m}_j\right)\left(8\bar{m}_j^2 - 6z\bar{m}_j - 2z\mu + 2z^2 + 20\bar{m}_j - 5z + 12\right)}{z^3}. \tag{A.0.13}$$

Las expresiones (A.0.8)-(A.0.13) permiten obtener los eigenvalores de (3.3.14). Para esto es necesario construir las funciones $\delta_n(z)$ descritas en la expresión (4.1.20) también del Capítulo 4. Las primeras tres expresiones para $\delta_n(z)$ son:

$$\delta_1(z) = -\frac{\left(\bar{m}_j - \mu + \frac{3}{2}\right)\left(\bar{m}_j - \mu + \frac{1}{2}\right)}{z^2}, \tag{A.0.14}$$

$$\delta_2(z) = \frac{\left(\bar{m}_j - \mu + \frac{5}{2}\right)\left(\bar{m}_j - \mu + \frac{3}{2}\right)\left(\bar{m}_j - \mu + \frac{1}{2}\right)}{z^3}, \tag{A.0.15}$$

$$\delta_3(z) = -\frac{\left(\bar{m}_j - \mu + \frac{7}{2}\right)\left(\bar{m}_j - \mu + \frac{5}{2}\right)\left(\bar{m}_j - \mu + \frac{3}{2}\right)\left(\bar{m}_j - \mu + \frac{1}{2}\right)}{z^4}. \tag{A.0.16}$$

Empleando la condición de cuantización, $\delta_n(z) = 0$, los ceros de las funciones delta mostradas para los 3 primeros casos son los siguientes:

$$\delta_1(z) = 0 \quad \Rightarrow \quad \bar{m}_j - \mu + \frac{3}{2} = 0, \quad \bar{m}_j - \mu + \frac{1}{2} = 0, \tag{A.0.17}$$

$$\delta_2(z) = 0 \quad \Rightarrow \quad \bar{m}_j - \mu + \frac{5}{2} = 0, \quad \bar{m}_j - \mu + \frac{3}{2} = 0, \quad \bar{m}_j - \mu + \frac{1}{2} = 0, \tag{A.0.18}$$

$$\delta_3(z) = 0 \quad \Rightarrow \quad \bar{m}_j - \mu + \frac{7}{2} = 0, \quad \bar{m}_j - \mu + \frac{5}{2} = 0, \quad \bar{m}_j - \mu + \frac{3}{2} = 0, \tag{A.0.19}$$

$$\bar{m}_j - \mu + \frac{1}{2} = 0. \tag{A.0.20}$$

La expresión general para los ceros anteriores es:

$$\mu - \bar{m}_j - \frac{1}{2} = n \quad n = 0, 1, 2\ldots. \tag{A.0.21}$$

Ahora bien, para calcular las eigenfunciones $\psi_j(z)$ de (3.3.14), el método de iteración asintótica indica que primero es necesario calcular las funciones auxiliares $\eta_n(z)$ evaluadas en los ceros correspondientes, como se menciona en la ecuación (4.1.12) del Capítulo 4. Entonces, $\eta_1(z)$, $\eta_2(z)$ y $\eta_3(z)$ están dados por

$$\eta_1(z) = \left.\frac{\frac{1}{2} - \mu + \bar{m}_j}{z - 1 - 2\bar{m}_j}\right|_{\bar{m}_j - \mu + \frac{1}{2} = 0} = 0, \tag{A.0.22}$$

$$\eta_2(z) = \left.\frac{-\left(\frac{1}{2} - \mu + \bar{m}_j\right)(2\bar{m}_j - z + 2)}{z^2 - (6\bar{m}_j + 2\mu + 3)\frac{z}{2} + 4(1 + 2\bar{m}_j)(1 + \bar{m}_j)}\right|_{\bar{m}_j - \mu + \frac{3}{2} = 0} = \frac{1}{2\bar{m}_j - z + 1}, \tag{A.0.23}$$

$$\eta_3(z) = \left.\frac{-\left(\frac{1}{2} - \mu + \bar{m}_j\right)\left[z^2 - \left(3\bar{m}_j + \mu + \frac{5}{2}\right)z + 2(3 + 2\bar{m}_j)(1 + \bar{m}_j)\right]}{-z^3 + (4\bar{m}_j + 2\mu + 2)z^2 - (1 + \bar{m}_j)(8\bar{m}_j + 4\mu + 4)z + 2(3 + 2\bar{m}_j)(1 + 2\bar{m}_j)(1 + \bar{m}_j)}\right|_{\bar{m}_j - \mu + \frac{5}{2} = 0}$$

$$= \frac{2(2\bar{m}_j - z + 2)}{z^2 - 4z(1 + \bar{m}_j) + 2(1 + \bar{m}_j)(1 + 2\bar{m}_j)}. \tag{A.0.24}$$

Combinando las funciones $\eta_n(z)$ calculadas previamente junto con la ecuación (4.1.19), es posible calcular las funciones $\bar{\phi}_j(z)$. Por lo tanto, para los tres primeros casos se tiene

$$\bar{\phi}_{0j} = \exp\left[-\int \eta_1(z')\,dz'\right] = 1 \tag{A.0.25}$$

$$\bar{\phi}_{1j} = \exp\left[-\int \eta_2(z')\,dz'\right] = 2\bar{m}_j - z + 1 \tag{A.0.26}$$

$$\bar{\phi}_{2j} = \exp\left[-\int \eta_3(z')\,dz'\right] = z^2 - 4z(1 + \bar{m}_j) + 2(1 + \bar{m}_j)(1 + 2\bar{m}_j). \tag{A.0.27}$$

Comparando con los polinomios asociados de Laguerre, para $n = 0, 1, 2$,

$$L_0^k(z) = 1, \tag{A.0.28}$$

$$1!L_1^k(z) = -z + (k + 1), \tag{A.0.29}$$

$$2!L_2^k(z) = z^2 - 2(k + 2)z + (k + 1)_2. \tag{A.0.30}$$

donde $(k + n)_n$ denota nuevamente al símbolo de Pochhammer, y tras definir a $2m_j = k$, las funciones $\bar{\phi}_{nj}$ se pueden expresar en términos de tales polinomios.

Como se mencionó al comienzo de este apartado, dependiendo del valor de $\bar{m}_j$ y $\mu$ se tendrá una descripción para uno de los perfiles de los campos externos aplicados.

1. **Campos externos con perfil exponencial decreciente**

A partir de las soluciones $\bar{\psi}_j(z)$ para $j = 1, 2$, expresadas en términos de los polinomios asociados de Laguerre, las funciones de onda $\psi_{1,n}(z)$ y $\psi_{2,n}(z)$ del sistema están dadas por:

$$\psi_{1,n}(z) = \mathcal{N}_1 \, e^{-z/2} z^{\lambda_n} L_n^{2\lambda_n-1}(z) \,, \tag{A.0.31}$$

$$\psi_{2,n}(z) = \mathcal{N}_2 \, e^{-z/2} z^{\lambda_{n+1}+1} L_n^{2\lambda_{n+1}+1}(z) \,, \ \ n = 0, 1 \,..., \tag{A.0.32}$$

con $\mathcal{N}_j$, $j = 1, 2$, siendo las constantes de normalización.

Usando (A.0.21), y de la definición de $\mu$ y $\bar{m}_j$, se puede obtener el siguiente espectro de energía (compárese con (3.2.23)):

$$E_n = v_F \beta_\nu \alpha_c n \sqrt{1-\beta_\nu^2} - v_d k_y + \kappa \nu v_F \sqrt{1-\beta_\nu^2} \sqrt{\left(k_y^c + D\right)^2 - \left(k_y^c + D - \alpha_c n \sqrt{1-\beta_\nu^2}\right)^2} \,. \tag{A.0.33}$$

2. **Campos externos singulares** De las soluciones $\bar{\psi}_j(z)$ para $j = 1, 2$, escritas en términos de $L_n^k(z)$, las funciones de onda $\psi_{1,n}(z)$ y $\psi_{2,n}(z)$ del sistema se expresan como:

$$\psi_{1,n}(z) = \bar{\mathcal{N}}_1 \, e^{-z_n/2} z^F L_n^{2F-1}(z_n) \,, \tag{A.0.34}$$

$$\psi_{2,n}(z) = \bar{\mathcal{N}}_2 \, e^{-z_{n+1}/2} z^{F+1} L_n^{2F+1}(z_{n+1}) \,, \tag{A.0.35}$$

donde $\bar{\mathcal{N}}_j$, $j = 1, 2$, son las constantes de normalización.

Como en el caso anterior, de la ecuación (A.0.21) y de la definición de $\mu$ y $\bar{m}_j$, se obtiene el espectro de energía correspondiente (compárese con (3.3.20)):

$$E_n = k_y \nu \, v_t - \frac{\mathcal{B}_0^2 \, \beta_\nu \, k_y^c \, v_F + \kappa v_F \, k_y^c \, (F+n) \sqrt{\mathcal{B}_0^2(\beta_\nu^2 - 1) + (F+n)^2}}{(F+n)^2 + \mathcal{B}_0^2 \beta_\nu^2} \,. \tag{A.0.36}$$

# Bibliografía